\documentclass{article}
\usepackage{amssymb}
\usepackage{fullpage} 
\title{Effective dynamics for solitons in the nonlinear Klein Gordon Maxwell 
system and the Lorentz force law}
\author{Eamonn Long\thanks{Supported by EPSRC} 
and David Stuart\thanks{Supported by EPSRC}
\\{\it{\small Centre for Mathematical Sciences, Wilberforce Road, 
Cambridge, CB3 OWA,
England}}\\{\it\small{dmas2@cam.ac.uk}} \\}
\usepackage{graphicx}
\usepackage{amsmath}
\date{}
\newcommand\dv{\hbox{div\,}}
\newcommand{\myqed}{\hfill $\Box$}
\newcommand{\proof}{{\em Proof\quad}}
\newcommand\beq{\begin{equation}}
\newcommand\eeq{\end{equation}}
\newcommand\ba{\begin{eqnarray}}
\newcommand\ea{\end{eqnarray}}
\newcommand\la{\label}
\newcommand{\mcn}{\mathcal V}
\newcommand{\mbxi}{\mbox{\boldmath$\xi$}}
\newcommand\mbx{\mathbf{x}}
\newcommand{\om}{\omega}

\newcommand{\Real}{\mathbb R}
\newcommand\mbR{\mathbf R}
\newcommand\mbS{\mathbf S}
\newcommand{\R}{\Real}
\newcommand\D{\mathbb D}
\newcommand\A{\mathbb A}
\newcommand\E{\mathbb E}
\newcommand\J{\mathbb J}
\newcommand\N{\mathbb N}
\newcommand\mbA{\mathbf A}
\newcommand\hh{{\dot H}^{1}}

\newcommand\bfu{\mathbf u}
\newtheorem{theorem}{Theorem}

\newtheorem{corollary}[theorem]{Corollary}

\newtheorem{lemma}[theorem]{Lemma}
\newtheorem{notation}[theorem]{Notation}

\newtheorem{proposition}[theorem]{Proposition}
\newtheorem{remark}[theorem]{Remark}

\begin{document}
\maketitle
\thispagestyle{empty}
\vspace{-0.3in}
\begin{abstract}
We consider the nonlinear Klein Gordon Maxwell system derived from the
Lagrangian $\int \bigl(-\frac{1}{4}F_{\mu\nu}F^{\mu\nu}+\frac{1}{2}\langle
(\partial-ie\A)_\mu\phi,(\partial-ie\A)^\mu\phi\rangle-\mcn(\phi)
-e\A^\mu\J^B_\mu\bigr)$ 
on four dimensional Minkowski space-time, where $\phi$ is a complex scalar 
field and $F_{\mu\nu}=\partial_\mu\A_\nu-\partial_\nu\A_\mu$ is the
electromagnetic field. For appropriate nonlinear potentials $\mcn$, the
system admits soliton solutions which are gauge invariant generalizations
of the non-topological solitons introduced and studied by T.D. Lee
and collaborators for pure complex scalar fields. 
In this article we develop a rigorous
dynamical perturbation theory for these solitons in the small 
$e$ limit, where $e$
is the electromagnetic coupling constant. The main theorems assert
the long time stability of the solitons with respect to perturbation
by an external electromagnetic field produced by the background
current $\J^B$, and compute their effective dynamics to $O(e)$. The
effective dynamical equation is the equation of motion for a relativistic
particle acted on by the Lorentz force law familiar from 
classical electrodynamics.
The theorems are valid in a scaling regime in which
the external electromagnetic fields are $O(1)$, but 
vary slowly over space-time scales
of $O(\frac{1}{\delta})$, and $
{{\delta}} =e^{1-k}
$
for $k\in \left( 0,\frac{1}{2}\right) $ as $e\to 0$.
We work entirely in the energy norm, and  the 
approximation is controlled in this norm for times of
$O(\frac{1}{e})$.
\end{abstract}

\tableofcontents

\section{Statement of results}

\subsection{\normalsize Introduction}

In this article, we are interested in the effective dynamics of a class of
solitary wave, or soliton, solutions to the
nonlinear Klein-Gordon-Maxwell (nl-KGM) equations, in the presence of an 
external
electromagnetic field. In this introduction we start by writing down the 
equations and giving a 
heuristic statement of, and motivation for, our results in \S\ref{heu}
and \S\ref{mot}. Then,
in \S\ref{Sec:ExternalField} and \S\ref{Sec:NonTopSol},
we provide the necessary background for a precise formulation of 
the main results - theorems \ref{thm:Stability} and 
\ref{Thm:Lorentz Force Law} - which appear in \S\ref{statmain}.
These theorems are proved in the subsequent sections; a list of
notation appears in \S\ref{not} to facilitate reading of the article.

\subsubsection{The equations}\label{first}
We study the following system of equations, called the
nonlinear Klein-Gordon-Maxwell system, or (nl-KGM) system, which describe 
the interaction of a complex scalar field $\phi$ with an electromagnetic
field $F_{\mu\nu}$ in the presence of an external space-time current $\J^B$:
\begin{equation}
\begin{split}
& \partial^\mu F_{\mu\nu} =e\langle i\phi,\D_\nu\phi\rangle+e\J^B_\nu\\
& \D_\mu\D^\mu\phi+\mcn'(\phi)=0 .\\
\end{split}
\label{nl-KGM}
\end{equation}
Here $\phi$ is a complex function on Minkowski space-time ${\Real}^{1+3}$, and
$\D_\mu=\partial_\mu-ie\A_\mu$ is the covariant derivative associated
to an electromagnetic potential $\A_\mu dx^\mu=\A_0dt+\A_j dx^j$ with
associated field $F_{\mu\nu}=\partial_\mu \A_\nu-\partial_\nu\A_\mu$. 
(The operator $\D$ determines an $S^1$ connection over ${\Real}^{1+3}$
whose curvature is $-iF$.)
We
use standard relativistic notation in which $\{x^\mu\}_{\mu=0}^{\mu=3}$
are co-ordinates, with greek indices running over $\{0,1,2,3\}$, 
$x^0=t$ is the time co-ordinate, and $\{x^j\}_{j=1}^3$ are
space co-ordinates with latin indices running over $\{1,2,3\}$; the Minkowski
metric is 
$$
\eta_{\mu\nu}dx^\mu dx^\nu=dt^2-(dx^1)^2 - (dx^2)^2 - (dx^3)^2,
$$
and is used to raise/lower indices in the usual way.
When the spatial part of a space time vector or 1-form is considered separately
bold face will often be used e.g. ${\bf x}=(x^1,x^2,x^3)$ for clarity.
We refer to $e$ as the (electromagnetic)
coupling constant: for the purposes of this article
it is a small positive parameter.
The current four-vector is of the form 
$$\J^B=\J^{B,\nu}\partial_\nu=\rho_B\partial_t+j_{B}^k\partial_k$$ 
and is conserved, i.e.
$$
\partial_t\rho_B+\dv\mathbf{j}_B=0.
$$
The quantity  $\rho _{B}$ is called the (background) charge
density, while  $\mathbf{j}_{B}$ is referred to as the (background
spatial) current density. Throughout the paper we make the following
hypotheses on 
the nonlinear potential
function $\mcn$:
\begin{enumerate}
\item[(H1)]
Phase invariance: there exists $G:\Real\to\Real$ such that
$\mcn(\phi)=G(|\phi|)$.
\item[(H2)]
Positive mass: $\mcn(\phi)=\frac{m^2}{2}|\phi|^2+\mcn_1(\phi)$
where $m>0$ and $\mcn_1(\phi)=-U(|\phi|)$ is smooth with 
$U(0)=U'(0)=U''(0)=0$.
\item[(H3)]
Sub-criticality: the third derivative $D^{(3)}\mcn_1=\mcn_1^{'''}$
satisfies a growth condition $|\mcn_1^{'''}(\phi )| \leq
c(1+| \phi| ^{p-3})$, for some $p\in (3,6)$.
The significance of $6$ is that it is the critical Sobolev exponent
for the embedding $H^{1}({\Real}^3)\hookrightarrow L^{p}({\Real}^3)$.
\end{enumerate}

The function $\mcn$ is subject to a number of additional 
more specialized hypotheses, which we detail in \S\ref{Sec:Hyp}, 
in particular to ensure existence and uniqueness of solitons solutions
with the properties described in \S\ref{Sec:NonTopSol}.

\subsubsection{Solitons}
\la{solitonsi}

The research in this paper is built upon the existence results for solitons
in semi-linear wave equations given in \cite{Berestycki},\cite{Peletier} and
\cite{Strauss}.
These solitons are time-periodic solutions of the nonlinear Klein-Gordon
equation
$$
 \partial_\mu\partial^\mu\phi+\mcn'(\phi)=0,
$$
which is obtained by putting $e=0$ in \eqref{nl-KGM}
(i.e. when there is no electromagnetic coupling),
and are of the form
$$
\phi(t,\mathbf{x})=e^{i\omega t}f_{\omega }(\mathbf{x}).
$$
T.D. Lee emphasized that solutions of this type, which 
he called non-topological solitons, provide a way of circumventing
the Derrick-Pohozaev non-existence results on static solitons in scalar
field theories; see 
\cite[Chapter 7]{Lee} for a discussion 
of their properties from the physical point of view.

It is proved in the references \cite{Berestycki},\cite{Peletier} and
\cite{Strauss} that, for certain potentials $\mcn$,
solutions of this form exist with $f_\omega$ positive and radial. Also
under further conditions
these solutions
are known to be essentially unique ( \cite{McLeod}) and dynamically stable
( \cite{GSS1,Stuart}); see \S\ref{Sec:NonTopSol} and the appendices for further
details. For non-zero values of the coupling constant $e$ solutions to
\eqref{nl-KGM} of this type have been constructed in 
\cite{Benci,Aprile} directly, using a 
spherically symmetric ansatz, and perturbatively in \cite{Thesis,Paper1}
for small $e$ using the $e=0$ case as a starting point. For small $e$ it 
is possible to use the information on stability for $e=0$ from \cite{Stuart}
to prove modulational stability of the solitons and their Lorentz boosts, see
\S\ref{sab} and \cite{Paper1} for details. 
Much of the same information for the $e=0$
case will also be used in the present article to study the stability of 
the solitons when subjected to external (background) electromagnetic fields.

\subsubsection{Informal statement of results on 
interaction of solitons with electromagnetic field}
\label{heu}
Our main concern in this article is to understand the interaction of 
the solitons just described, with an external electromagnetic field
produced by the space-time current $\J^B$. In order to 
be able prove theorems giving precise information on the effect of
this field on the soliton, we study \eqref{nl-KGM} in a regime determined
by two small parameters:
\begin{itemize}
\item The electromagnetic coupling constant $e=o(1)$.
\item The external electric and magnetic
fields, $\mathbf{E}_{ext}^\delta$ and $\mathbf{B}_{ext}^\delta$,
vary over 
scales which are $O(\frac{1}{{\delta}})$, where ${\delta}=o(1)$. Thus 
the small parameter ${\delta}$ is the ratio of the size of the 
soliton to the length scale over which the external field varies.
\end{itemize}
The following is an informal version of our main theorems:
\begin{quote}
{\em The system \eqref{nl-KGM} has solutions which are close, in energy norm, 
to solitons of the type described above and which, 
in an appropriate scaling regime,
move according to the Lorentz force equation:
\beq\la{lofl}
\frac{d}{dt}\left( \gamma \mathbb{M}_{S}\mathbf{u}\right) =e\mathbb{Q}_{S}\left({{\mathbf E}^\delta_{ext}} 
+\mathbf{u\times {{\mathbf B}^\delta_{ext}} }\right),
\eeq
where the effective mass $\mathbb{M}_{S}$ and charge $\mathbb{Q}_{S}$
of the soliton are as in \eqref{defmass} and \eqref{defcharge}. The
scaling $
{{\delta}} =e^{1-k}
$
for $k\in \left( 0,\frac{1}{2}\right) $ ensures
that this holds for time intervals of length $\frac{T_0}{e}$
as $e\to 0$.
}
\end{quote}
The precise formulation is in the two theorems 
stated in \S\ref{statmain}.

\subsubsection{Motivation and related work}
\la{mot}
Our interest in this problem stems
from the classical, but ongoing, 
controversy surrounding the
classical equation of motion for a point charge in an 
external
electromagnetic field. The difficulty arises in attempts to
account for the ``back reaction'' 
of the charge's own field on itself. Attempts to derive 
an equation of motion lead to modifications of the 
Lorentz force law \eqref{lofl} ,
most notably the  Lorentz-Dirac equation 
(\cite[Equation 9.1]{Spohn}). This equation is third order
in time, and is 
difficult to interpret consistently without some
further constraint on the type of solution allowed, due
to the occurence of runaway solutions and
violations of causality,
(see \cite{Dirac} and \cite[Chapter 28]{feyn}). Recent
discussions of this problem have been given in \cite{zeid}
and the books \cite{Spohn,Yaghjian}. 
One natural and well established approach to the problem 
of making sense of the back reaction
is to start with a 
well-posed system of equations in which the point charge 
is explicitly replaced by a smooth bounded 
charge distribution, the Abraham model, or one of its 
generalizations like the Lorentz model, for example. 
One can then derive an
equation of motion for the charge as an 
expansion, valid when the size of the charge distribution is
small (compared with typical length scales set by
the external fields), 
and show that this agrees with the Lorentz-Dirac equation 
at a certain order of approximation -
see \cite{KS}. In this setting it turns out, however, that 
{\em at the same order} the Lorentz-Dirac equation
can be approximated by a more conventional
equation of motion
which seems to be free of interpretational difficulties,
(see \cite[Equation 9.10]{Spohn}, where the name
Landau-Lifshitz equation is suggested for this
effective equation of motion. The Landau-Lifshitz equation, 
which is second order in time,
can be obtained formally from the Lorentz-Dirac equation by 
substituting for the third derivatives the expression obtained
by differentiating the ordinary Lorentz force law \eqref{lofl}
once in time.)

Our aim in studying soliton motion in the (nl-KGM) system is
to attempt a similar analysis using a solitonic model for the
particle (in place of the Abraham or Lorentz model). 
Our model has the virtue of being, in a very natural 
way, a Lorentz invariant system which is well posed (and so 
free of causality problems). Unfortunately the calculations
required even just to derive the equation of 
motion for the soliton to $O(e)$ (i.e. the Lorentz
force equation \eqref{lofl}) are long, and further work
will be required to calculate additional corrections which 
may be compared with the Lorentz-Dirac equation in 
appropriate regimes. To achieve this, the starting point would
be the equation of motion \eqref{Eq:ModEq} for the soliton
parameters derived from modulation theory. In 
\S\ref{Sec:Lorentz Force Law} this equation is computed to
highest order (i.e. to $O(e)$), and shown to give the
Lorentz force law. A computation to the next order should
give the Landau-Lifshitz equation
(\cite[Equation 9.10]{Spohn}). However it seems that some
renormalization of the soliton mass and charge will have 
to be taken into account in this computation, and it is
possible a refinement of the ansatz \eqref{Eq:Ansatz}
will be needed to achieve this. It is to be hoped that
at least in some simple cases such as 
one dimensional motion of the soliton
in an electric field ${{\mathbf E}^\delta_{ext}}
=(0,0,E(\delta t,\delta x))$ of fixed direction
it will be possible to
carry this through, and make a comparison with the corresponding
specialization of the Landau-Lifshitz equation (\cite[Equation
9.11]{Spohn}).

A corresponding theorem to our main result
was proved for solitons in interaction 
with gravitational fields in the articles 
\cite{Stuart04a,Stuart04b}. The system treated there (Einstein's
equation coupled to a nonlinear klein-Gordon 
equation) is in many ways
more difficult than the one studied here (for example it is 
quasi-linear). Correspondingly, it is possible to carry out
a more general analysis for the Klein-Gordon-Maxwell system 
under consideration here:
in particular we emphasize
that in the present article we are able to work entirely with
the energy norm throughout (whereas for the Einstein system it was
necessary to work with much stronger norms).   There have also
been theorems proved on effective dynamics for 
solitons moving under a potential in the nonlinear 
Schroedinger equation, see \cite{JFGS,BJ,Weinstein}.

\subsubsection{\label{not}Notation}
The following is a list of notations for 
important objects, with the section
in which they are first introduced, for reference.

\begin{itemize}

\item $L^{p}({\Real}^3)$ is the Lebesgue space of 
(equivalence classes of) measurable functions with norm 
$\| f\| _{L^{p}}=\int_{{{\Real}}^{3}}\left| f\right|
^{p}dx<\infty$, and
$H^{k}({\Real}^3)$ is the Sobolev space of 
(equivalence classes of) measurable functions with norm
$\|f\|_{H^k}=\sum_{\left| \alpha \right| =0}^{k}
\| \partial^{\alpha}f\| _{L^{2}}<\infty$, 
where $\partial^\alpha$ means the weak partial derivative determined
by the multi-index $\alpha$. We say $f\in H^k_{loc}$ if $\chi f\in
H^k$ for every smooth, compactly supported $\chi$, and
\beq
\la{defsob}
\hh =\left\{ f\in H^1_{loc}\cap L^6:\left\| \nabla f\right\| _{L^{2}}=\left\|
f\right\| _{\hh}<\infty \right\} \text{.}
\eeq
\noindent
Further we define $H_{r}^{k}$ 
to be the intersection of $H^{k}$ and the space of radial
functions, i.e. functions of $|x|$, 
and similarly define $L_{r}^{p}$ and ${\dot H}^1_r$.

\item Electromagnetic potential $\A_\mu dx^\mu=\A_0dt+\A_j dx^j$, 
electromagnetic field 
$F_{\mu\nu}=\partial_\mu \A_\nu-\partial_\nu\A_\mu$, and covariant derivative
$\D_\mu=\partial_\mu-ie\A_\mu$:\;\S\ref{first}.

\item Complex scalar (soliton) field $\phi$ and its
self-interaction potential $\mcn(\phi)=G(|\phi|)$ subject to hypotheses
(H1)-(H3):\;\S\ref{first}. Additional hypotheses (SOL), (KER) and
(POS):\; \S\ref{Sec:Hyp} and \S\ref{sab}.

\item (nl-KGM) is the nonlinear Klein-Gordon-Maxwell  system:\;
\eqref{nl-KGM} and \S\ref{statmain}.

\item $\Psi =\left( \phi ,\psi ,{\A}_i, 
{\E}_i\right)$ is the dependent
variable in the Hamiltonian formulation:\;\S\ref{statmain} 
(and \S\ref{hamform} for zero external current case).

\item $e$ electromagnetic coupling constant, $\delta$ external
field scaling parameter are both small:\;\S\ref{heu} and 
\S\ref{scex}.

\item Scaled external electromagnetic potentials $a_\mu^\delta$,
and electric and magnetic fields
$
{\mathbf E}_{ext}^{{{\delta}} }$ and ${\mathbf B}_{ext}^{{{\delta}}}$
induced by external current
$
(\rho_B^{\delta},\,
\mathbf{j}_B^{\delta}):
$
\;\S\ref{scex}.

\item $\Psi^{\delta}_{ext}=(0,0,\mathbf{a}^{\delta},\mathbf{E}^{\delta}_{ext})$
represents the external field in the Hamiltonian formulation, and
$\Psi^{{\delta},\chi}_{ext}
=(0,0,\mathbf{a}^{{\delta},\chi},\mathbf{E}^{\delta}_{ext}),$ its gauge
transform by $\chi$:\;\S\ref{statmain} and \S\ref{bp5}.

\item $f_\omega$, $f_{\omega,e}$ are the soliton profile functions
in (respectively) the $e=0$ case and
for non-zero $e$, while $\alpha_{\omega,e}$ is the $A_0$ component
of the electromagnetic potential for soliton solutions: 
\S\ref{Sec:NonTopSol}.

\item $\Psi_{S,e}$ is the set of Lorentz transformed soliton solutions
in Hamiltonian formalism, or
$\Psi_{SC,e}$ in Coulomb gauge:\;
\S\ref{solitonsi} and  \S\ref{Sec:NonTopSol}. Gauge transformation to Coulomb
gauge generated by $\zeta$:\;
\S\ref{sab}.

\item 
${{\lambda}}=
(\lambda_{-1},\lambda_0,\lambda_1,\dots,\lambda_6)
=(\omega,\theta ,\mbxi,\mathbf{u})
$
are parameters for Lorentz transformed solitons:\;
\S\ref{Sec:NonTopSol}.

\item $\gamma,P_{\mathbf{u}}, Q_{\mathbf{u}}, \Theta, \Theta_c,\mathbf{Z},
V_0(\lambda),N_\lambda,\tilde\Xi
$ and $\zeta$:\;
\S\ref{sab}.

\item $\widetilde{O}_{stab}\in\R^8$ 
is the stable region of soliton
parameter space, where Grillakis-Shatah-Strauss stability
condition \eqref{stab} holds and $\tilde\Xi$
is positive on the symplectic normal subspace $N_\lambda$:\;\S\ref{sab}.

\item $({\cal H}_0,\Omega_0)$, $({\cal H},\Omega)$ and $\|\Psi\|_{\cal H}$ 
and $\|\Psi\|^2_{{\cal H}_s}$
are the symplectic phase spaces
and norms:\S\ref{hamform} and \S\ref{Sec:LWP}.

\item $W,\,K$ and $\Xi$ are quadratic forms used in stability
analysis:\;\S\ref{mgre}.

\item $\widetilde{W}$ quantity like $W$ but inluding 
certain nonlinear interaction $\tilde H$ parts of the Hamiltonian:\;
\S\ref{Sec:Proof of dWtildedt}.

\item $T_{loc},T_0,T_1$:\;\S\ref{Sec:LWP},\S\ref{spef} and 
\S\ref{rmodth}, respectively.

\item $\widetilde{\partial_{\mathbf\lambda}}$:\;\S\ref{rmodth}.

\end{itemize}

\subsection{\label{Sec:ExternalField}The External Electromagnetic
Field and Scaling }

The external electromagnetic field $F_{\mu\nu}^{ext}$ is
induced by the space-time current $\J^B=\rho_B\partial_t+\mathbf{j}_{B}^k\partial_k$
according to Maxwell's equations, i.e. the first equation of \eqref{nl-KGM} with $\phi$
set equal to zero. Introducing an external electromagnetic potential,
written in lower case symbols,
$a_\mu dx^\mu=a_{0}dt+{a}_j dx^j$, such that $F_{\mu\nu}^{ext}=
\partial_\mu a_\nu-\partial_\nu a_\mu$, and imposing the Coulomb condition
$\nabla\cdot{\bf a}=0$, these equations can be written:
\begin{equation}
\begin{split}
-\triangle a_{0} =&e\rho _{B}\text{,} \\
\Box \mathbf{a} =&\nabla \partial_ta_{0}-e\mathbf{j}_{B}.
\end{split}
\label{Eq:Maxwell}
\end{equation}
Here $\rho _{B}$ is the background charge density, $
\mathbf{j}_{B}$ is the
background current density. 
The associated electric field, ${{\mathbf E}_{ext}} $, 
and magnetic field, ${{\mathbf B}_{ext}} $,
are given by
\begin{eqnarray}
{{\mathbf E}_{ext}} &=&\frac{\partial\mathbf{a}}{\partial t}
-\nabla a_{0}\text{,} \\
{{\mathbf B}_{ext}} &=&\nabla \times \mathbf{a}\text{.}
\end{eqnarray}
We shall make the following assumptions on the external field:
\begin{description}
\item{(BG)}
The external electromagnetic potentials are smooth
and satisfy:
\beq\label{Eq:ExternalAssumption2}
\mathop{\max_{|\alpha|=j}}_{\mu=0,1,2,3}
\left\|\nabla_{t,x}^\alpha a_\mu\right\| _{L^{\infty }({{\Real}}^{1+3})}
=L_j<\infty,
\eeq
(using multi-index notation
$\nabla_{t,x}^\alpha$ for arbitrary partial 
derivatives of order $|\alpha|$.)
\end{description}

It might appear that these assumptions are restrictive: in particular, the
assumption that $\left\| \mathbf{a}\right\| _{L^{\infty }({{\Real}}%
^{3}\times {{\Real}}^{+})}<\infty $ precludes the consideration of a
constant magnetic field. However, since we shall scale 
so that the external electric and magnetic fields do not change
appreciably over the spread of the soliton, which is exponentially
localized, 
these conditions could probably be
relaxed with some further work.
A more important
restriction in our
study appears to arise in the consideration of the 
scaling of the the external field, which we discuss below,
after presenting results on local well-posedness for the
(nl-KGM) system in the presence of an external field.

\subsubsection{The Cauchy problem for (nl-KGM) in an external field}
\label{Sec:LWP}

Throughout this article we make use of local
well-posedness of the (nl-KGM) system in the energy norm. In the case
that there is no external field and $\mcn\equiv 0$ this was proved
in \cite{Klainerman}. In this section we give conditions under which
this is true in the more general situation of \eqref{nl-KGM} considered here. 
Since our assumptions on the external field do not
require
finite energy it is convenient to subtract off the external field. 
Thus assume
given an external electromagnetic potential $a_0dt+a_j dx^j$
as above, in Coulomb gauge
$\nabla\cdot{\bf a}=0$, which solves 
the inhomogeneous Maxwell
equations \eqref{Eq:Maxwell} and verifies 
\eqref{Eq:ExternalAssumption2}. 
Write the electromagnetic potential appearing in
\eqref{nl-KGM} as $\A_\mu=a_\mu+A_\mu$. Then, requiring
the Coulomb gauge condition $\nabla\cdot{\bf A}=0$, as is always possible,
\eqref{nl-KGM} is equivalent to the following system:
\begin{equation}
\begin{split}
\dot\phi=&\psi+ie(a_0+A_0)\phi,\\
\dot\psi  =&\Delta\phi
-2ie(\mathbf{A}+\mathbf{a})\cdot\nabla \phi 
-e^{2}\left| \mathbf{A}+\mathbf{a}\right| ^{2}\phi 
-\mcn^{\prime }(\phi)
+ie(a_0+A_0)\psi,\\
\Box \mathbf{A}=&\left\langle ie\phi ,(\nabla -ie\mathbf{A})\phi
\right\rangle +\nabla \overset{.}{A_{0}}-e^{2}\left| \phi \right| ^{2}%
\mathbf{a}^{ }\text{,}\\
-\triangle A_{0}=&\langle ie\phi ,{\psi }\rangle
\text{,}
\end{split}\label{Eq:Cauchy}
\end{equation}
where $\mathbf{A}=(A_1,A_2,A_3)$ is the 
spatial part of $A=\A-a$. 
We solve this system in the energy space
${\cal H}\equiv
H^1\times L^2\times\hh\times L^2,
$
which is endowed with the energy norm
$\|\Psi\|_{\cal H}=\|(\phi,\psi,{\bf A},{\bf E})\|_{H^1\times L^2\times\hh\times L^2}$; see \S\ref{not} for notation on standard norms.
We also
define corresponding
higher energy norms indexed by $s\in\N$ by
\beq\la{hsnorm}
\|(\phi,\psi,{\bf A},{\bf E})\|^2_{{\cal H}_s}\equiv
\sum_{|\alpha|=0}^{s-1}
\|\nabla_x^\alpha(\phi,\psi,{\bf A},{\bf E})\|^2_{\cal H},
\eeq
with corresponding space denoted ${\cal H}_s$.
We say that the Cauchy problem for (\ref{Eq:Cauchy}) 
is locally well posed in ${\cal H}$
if the following two conditions hold:
\begin{description}
\item[(WP1)]
 given initial data $%
\bigl( \phi (0),{\psi }(0),\mathbf{A}(0),\overset{.}{\mathbf{A}}%
(0)\bigr) \in {\cal H}$ in
Coulomb gauge (i.e. $\dv\mathbf{A}(0)=0$, $\dv\overset{.}{%
\mathbf{A}}(0)=0$), satisfying
\beq
\Bigl\|\bigl(\phi (0),{\psi }(0),\mathbf{A}%
(0),\overset{.}{\mathbf{A}}(0)\bigr)\Bigr\|_{\cal H}\leq k_0
\label{fe}
\eeq
there exists $T_{loc}=T_{loc}(k_0)>0$
and a unique solution $\bigl( (\phi (t),{\psi }(t),
\mathbf{A}(t),%
\overset{.}{\mathbf{A}}(t)\bigr) $ such that
\begin{eqnarray*}
&&\bigl(\phi (t),{\psi }(t), \mathbf{A}(t),\overset{.}{\mathbf{A}}(t)\bigr) \in C([0,T_{loc});{\cal H})
\text{,}\\
&&\int_0^{T_{loc}}\left(\|\Box A\|_{L^2}+\|\Box \phi\|_{L^2}\right)dt<\infty.
\end{eqnarray*}
\item[(WP2)] the solution is continuous with respect to
the initial data in that, for another set of initial data 
$\bigl( \phi_1 (0),
{\psi }_1(0),\mathbf{A}_1(0),
\overset{.}{\mathbf{A}}_1(0)\bigr) $, 
which are close in ${\cal H}$, and
also satisfy \eqref{fe}, and the Coulomb gauge conditions,
the following holds
on the common domain of definition $[0,T_{loc}]$,
for some constant $c>0$:
\begin{multline}
\max_{[ 0,T_{loc}]}\Bigl\| \bigl(\phi -\phi _{1},
{\psi }-
{\psi }_{1},\mathbf{A}-\mathbf{A}_{1},\overset{.}{\mathbf{A}}-%
\overset{.}{\mathbf{A}}_{1}\bigr)\Bigr\| _{\cal H}
\leq  \notag \\
c \Bigl\| \bigl(\phi (0)-\phi _{1}(0),{\psi }(0)
-{\psi }_{1}(0),\mathbf{A}(0)-\mathbf{A}_{1}(0),
\overset{.}{\mathbf{A}}(0)-
\overset{.}{\mathbf{A}}_{1}(0)\bigr)\Bigr\|_{\cal H}\,.
\end{multline}

\end{description}

As remarked above, in the absence of the external field, and with
$\mcn\equiv 0$ the validity of (WP1)-(WP2) was proved in \cite{Klainerman}. The general
case was addressed in  the thesis \cite{Paper1} where it was 
shown, using in addition Strichartz inequalities from 
\cite{Grillakis,Strichartz}, that
(WP1)-(WP2) hold if $\mcn$ is a smooth sub-critical nonlinearity:

\begin{proposition}
\label{Prop:WpCond} Suppose $\mcn$ is smooth and that there exists a positive number
$\kappa\in(0,4)$ such that, for all $\phi ,\varphi $,
\begin{equation}
\left| \mcn^{\prime }(\phi) -\mcn^{\prime }(\varphi )\right| \leq C\left| \phi
-\varphi \right| \left( 1+\left| \phi \right| ^{4-\kappa }+\left| \varphi
\right| ^{4-\kappa }\right)  \label{Eq:WpCond}
\end{equation}
and that $\mcn^{\prime }(0)=0$. Assume that the external
potential  is smooth and verifies 
\eqref{Eq:ExternalAssumption2} for every non-negative integer $j$. 
Then the Cauchy problem for 
\eqref{Eq:Cauchy} is well-posed in the sense
of (WP1) and (WP2).
Further, if the initial data lie in ${\cal H}_s$ 
for some $s\geq 2$ then the solution exists for all
time, and remains in ${\cal H}_s$, and is smooth if the 
initial data are smooth.
\end{proposition}
\begin{remark}\la{rgi}
{\em 
The Coulomb condition leaves a residual gauge invariance
by functions $\chi(t,\mbx)$ which are harmonic in $\mbx$. 
(These are either constant or unbounded.) In particular
the system \eqref{Eq:Cauchy} is invariant under the
transformation $a_\mu\mapsto a_\mu+\partial_\mu\chi$,
$(\phi,\psi)\mapsto e^{ie\chi}(\phi,\psi)$ if
$\chi=\alpha_0(t)+\alpha_j(t) x^j$ is linear in $\mbx$
and smooth in $t$.
In this case the map $(\phi,\psi)\mapsto 
e^{i\chi}(\phi,\psi)=(\tilde\phi,\tilde\psi)$ is 
Lipschitz on $H^1\times L^2$. It follows that proposition
\ref{Prop:WpCond} remains valid if the external potential is obtained
from one satisfying \eqref{Eq:ExternalAssumption2}
by gauge transformation by $\chi=\alpha_0(t)+\alpha_j(t) x^j$
.}
\end{remark}
\begin{remark}
{\em 
Notice that when the nonlinearity is determined by 
a smooth function
$\mcn$ whose third derivative satisfies:
\begin{equation}
\left| D^{(3)}\mcn(\phi )\right| \leq
c(1+\left| \phi \right| ^{3-\kappa}),\quad\hbox{for all }\,\phi
\label{Eq:U2prime1}
\end{equation}
for some $c>0,0<\kappa<3$ the conditions of proposition
\ref{Prop:WpCond}
hold, and the Cauchy problem is well-posed. This  assumption is also
sufficient to estimate the nonlinear terms in the perturbation theory
developed in \S\ref{Sec:Stability},\S\ref{modth} and
\S\ref{Sec:Proof of dWtildedt} of this article. 
Introduce ${\mathcal F}(\phi)=\mcn'(\phi)-m^2\phi=
\mcn_1'(\phi)=\beta(|\phi|)\phi$ 
as the nonlinear part of $\mcn'(\phi)$, 
with $\mcn$ as in the introduction.
Then \eqref{Eq:U2prime1} implies the inequality
\beq
\left| {\mathcal F}^{
\prime }(f+v)-{\mathcal F}^{\prime }(f)\right| \leq
c(1+|f|^3)(|v|+|v|^{4})\quad\hbox{where}\;c\;\hbox{is a positive constant}\text{,} \label{Eq:U2prime2}
\eeq
which is convenient for our use.
In fact, for our purposes
it would be sufficient to make the following slightly more general
assumption on ${\mathcal F}$:
\begin{equation}
\text{For all }f>0\text{ and for any }v\text{, }\left| {\mathcal F}^{
\prime }(f+v)-{\mathcal F}^{\prime }(f)\right| \leq
c(f^{r-1}+f^3)(|v|+|v|^{4})\text{,} \label{Eq:U2prime2r}
\end{equation}
where $r,c$ are positive constants, see \cite{Thesis}.
Of course
given a smooth potential $\mcn$ satisfying
(\ref{Eq:U2prime1}), let ${\mathcal F}$ be as just defined,
then (\ref{Eq:U2prime2r}) will also hold with $r=1$.}
\end{remark}

\subsubsection{Scaling the external fields}
\la{scex}

As already mentioned, we require that the external electric and magnetic
fields are approximately constant over the soliton. To ensure this, we
introduce a scaled version of the external fields. Thus, we have
\beq
a_{0}^{{{\delta}} }(t,\mathbf{x}) =\frac{1}{{{\delta}} }a_{0}({{\delta}} t,{{\delta}} \mathbf{x})\text{%
,} \qquad
\mathbf{a}^{{{\delta}} }(t,\mathbf{x}) =\frac{1}{{{\delta}} }\mathbf{a}({{\delta}} t,{{\delta}}
\mathbf{x})\text{,}
\eeq
with the scaled external electric and magnetic fields given by:
\beq
{\mathbf E}_{ext}^{{{\delta}} } ={{\mathbf E}_{ext}}({{\delta}} t,{{\delta}} \mathbf{x})\text{,} \qquad
{\mathbf B}_{ext}^{{{\delta}} } ={{\mathbf B}_{ext}}({{\delta}} t,{{\delta}} \mathbf{x})\text{.}
\eeq
Clearly these fields correspond to the following rescaled
charge and current densities:
\beq
\rho_B^{\delta}(t,x)={\delta}\rho_B({\delta} t,{\delta} x),\qquad
\mathbf{j}_B^{\delta}(t,x)={\delta}
\mathbf{j}_B^{\delta}({\delta} t,{\delta} x).
\eeq
Henceforth, we shall almost always refer exclusively to 
the scaled fields. It remains to choose the length scale, $\frac{1}{{{\delta}} }$, over which the external fields change: this is determined by the analysis
in \S\ref{Sec:Proof of dWtildedt} which bounds the deviation of the solution 
from the modulated soliton. This analysis seems to require two main conditions
on the scaling of $\delta$ and $e$:
\begin{itemize}\item
From lemma \ref{lem:W Wtilde
equivalence}, it seems that we need
\begin{equation}
\lim_{e\rightarrow 0}\frac{e}{{{\delta}} }=0\text{,}
\end{equation}
to bound the effect of the scaled external electromagnetic potential.
\item
Treatment of the last term in \eqref{Eq:dWtildedt}, seems 
to suggest that we need
\begin{equation}
\lim_{e\rightarrow 0}\frac{{{\delta}} ^{2}}{e}=0\text{.}
\end{equation}
This condition is used
to ensure the deviation from the Lorentz force law
is small for times of order $\frac{1}{e}$.
\end{itemize}
We will consider the limit
$e\to 0$ with
\begin{equation}
{{\delta}} =e^{1-k}
\end{equation}
for some constant $k\in \left( 0,\frac{1}{2}\right) $, so that
both of these conditions hold. It remains to be seen what are the
optimal conditions for scaling $e,\delta$
under which the results of this paper hold.

\subsection{Non-Topological Solitons 
\label{Sec:NonTopSol}}
We now discuss existence and stability properties of
non-topological solitons as solutions of (nl-KGM) in the absence
of external fields. This means we are here concerned with
the (nl-KGM) system with $\rho_B=0=\mathbf{j}_B$. We first discuss
the Hamiltonian formulation of (nl-KGM), since that gives the
appropriate context in which to introduce non-topological solitons.

\subsubsection{Hamiltonian formalism}
\la{hamform}

It
is useful to present the Hamiltonian formalism for the (nl-KGM), not least
because it will give us a language which we shall use in proving 
the existence and 
long-time stability of the non-topological solutions. Indeed,
as we shall see, from the 
Hamiltonian point of view, non-topological solitons are relative 
equilibria, and recognizing this fact leads to the identification
of the appropriate quantities with which to work. 

In order to 
define the phase space we recall the
standard function spaces defined in \S\ref{not}.

To start with, consider the nonlinear wave equation in isolation
\beq
\Box \phi  + \mcn^{\prime }(\phi)=0. 
\la{NLW}\eeq
This can be written as a Hamiltonian system on the phase space
$
{\cal H}_0\equiv\{(\phi,\psi)\in H^1\times L^2\},
$
with symplectic form
$
\Omega_0((\phi',\psi'),(\dot\phi,\dot\psi))
=\int\langle\phi',\dot\psi\rangle-
\langle\psi',\dot\phi\rangle\,dx,
$
and Hamiltonian 
\begin{equation}
H_0(\phi ,\psi)=\frac{1}{2}\int
\left| \nabla\phi \right| ^{2}+2\mcn(\phi).
\label{HAMNLW}
\end{equation}
The corresponding Hamiltonian evolution equations, equivalent to \eqref{NLW}, 
are :
\begin{equation}
\partial_t\left(
\begin{array}{c}
\phi \\
\psi
\end{array}
\right) =\left(
\begin{array}{c}
\psi\\
\triangle\phi -\mcn^{\prime }(\phi )
\end{array}
\right).  \label{Eq:HNLW}
\end{equation}

Next, for (nl-KGM),  introduce the phase space
\beq\la{enorm}
{\cal H}\equiv\{\Psi=(\phi,\psi,{\bf A},{\bf E})\in
H^1\times L^2\times\hh\times L^2\},
\eeq
which is endowed with the norm
$\|\Psi\|_{\cal H}=\|(\phi,\psi,{\bf A},{\bf E})\|_{H^1\times L^2\times\hh\times L^2}$
and the (densely defined, weak) symplectic form
\beq
\Omega(\Psi',\dot\Psi)=\int\langle\phi',\dot\psi\rangle-
\langle\psi',\dot\phi\rangle+{\bf A}'\cdot\dot{\bf E}
-{\bf E}'\cdot\dot{\bf A}dx,
\eeq
where $\Psi'=(\phi',\psi',{\bf A}',{\bf E}')$ and similarly
for $\dot\Psi$.
The (nl-KGM)
equations with $\rho_B=0=\mathbf{j}_B$ 
arise formally as the Hamiltonian flow on ${\cal H}$
associated to the Hamiltonian
\begin{equation}
H(\phi ,\psi ,\mathbf{A},\mathbf{E})=\frac{1}{2}\int \bigl(\left| \mathbf{E}%
\right| ^{2}+\left| \nabla \times \mathbf{A}\right| ^{2}+\left| \psi \right|
^{2}+\left| \nabla _{\mathbf{A}}\phi \right| ^{2}+2\mcn(\phi)\bigr),
\label{HAM}
\end{equation}
and subject to the constraint:
\begin{equation}
\mathcal{C}_{0}\equiv\dv\mathbf{E}-\left\langle ie\phi ,\psi \right\rangle
=0\text{.\label{Eq:Gauss2}}
\end{equation}
Here $\nabla _{\mathbf{A}}\phi $ is the covariant derivative of $\phi $
given by $\nabla _{\mathbf{A}}\phi =\nabla \phi -ie\mathbf{A}\phi $ and $%
\mathbf{A}$ is the spatial part of the gauge field. The equations of motion
for the augmented Hamiltonian $H_{1}=H-\int A_{0}\mathcal{C}_{0}$ are:
\begin{equation}
\partial_t\left(
\begin{array}{c}
\phi \\
\psi \\
\mathbf{A}_{i} \\
\mathbf{E}_{i}
\end{array}
\right) =\left(
\begin{array}{c}
\psi +ieA_{0}\phi \\
\triangle _{\mathbf{A}}\phi -\mcn^{\prime }(\phi )+ieA_{0}\psi \\
\mathbf{E}_{i}+\nabla _{i}{A}_{0} \\
\triangle \mathbf{A}_{i}-\nabla _{i}(\dv\mathbf{A})+\left\langle ie\phi
,\nabla _{\mathbf{A}}\phi \right\rangle 
\end{array}
\right)  \label{Eq:HKGM}
\end{equation}
where the ``Lagrange multiplier'' 
$A_{0}$ is identifiable with the temporal part of the gauge field, $%
\triangle _{\mathbf{A}}\phi =\triangle \phi -2ie\mathbf{A}\cdot\nabla \phi
-ie \dv\mathbf{A}\phi+e^{2}\left| \mathbf{A}\right| ^{2}\phi $, $%
i=1...3 $, and we have not yet imposed any gauge condition.

\subsubsection{Existence of non-topological solitons: the $e=0$ case}

\label{Sec:Hyp}

The class of solitary wave solutions of interest is that of 
non-topological solitons discussed in \cite[Chapter 7]{Lee}. 
These are examples of a special type of
solution to a Hamiltonian system with symmetry called relative
equilibrium: this means that the time evolution is given by an orbit
of a one parameter subgroup of the symmetry group. For
\eqref{Eq:HNLW}
the Hamiltonian is invariant under the action of $S^1$ by phase
rotation, as long as $\mcn(\phi)=G(|\phi|)$ is a function of $|\phi|$
only; the charge corresponding to this $S^1$ action is
$$Q(\phi,\psi)=\int\langle i\psi,\phi\rangle\,dx.$$ 
A relative equilibrium is then a solution of the form
$(\phi,\psi)=e^{i\omega t}(f_\omega(x),i\omega f_\omega)$, 
where $f_\omega$ is a real valued function
which satisfies an elliptic equation. These solutions are critical
points of the functional $H_0+\omega Q$, often called the
{\it augmented Hamiltonian} in this context.
We consider $G$ of the form
$$G(f)=\frac{m^{2}}{2}f^{2}-U(f),\quad \hbox{with}\quad
U(f)=\int_{0}^{|f|}t\beta (t)dt.$$ then the equation satisfied by
$f_\omega$ is $-\triangle f_\omega+(m^2-\omega^2)
f_\omega=\beta(f_\omega)f_\omega$.  This equation typically has many
solutions (see \cite{Berestycki} and references therein), but we are
only interested in positive, radially symmetric solutions because it
is these which are dynamically stable: these are sometimes called the
{\em ground state solitons}. Thus, crucial to our analysis 
is the following hypothesis on existence and uniqueness of
the $e=0$ ground state soliton:
\begin{multline*}
\text{(SOL)}\qquad \text{For }\omega ^{2}<m^{2}\text{, there exists a unique
positive radial function }
f_{\omega}\in H^{4}({{\Real}}^{3})
\text{ which solves}\\
\left( -\triangle +m^{2}
-\omega ^{2}\right) f_{\omega}=\beta
(f_{\omega})f_{\omega}\text{.}\\
\end{multline*}

\begin{theorem}
The existence part of ${\rm (SOL)}$ holds under the following conditions:
\begin{equation}
U^{\prime }(f)=-U^{\prime }(-f)\quad \text{and}\quad U^{\prime }\in C^{1}(%
{{\Real}})\cap C^{2}((0,\infty )),  \label{Eq:Exist1}
\end{equation}
\begin{equation}
U^{\prime }(0)=U^{\prime \prime }(0)=0\quad \text{and}\quad \exists s\in
(0,1):\lim_{f\rightarrow 0}f^{s}U^{\prime \prime \prime }(f)=0,
\label{Eq:Exist2}
\end{equation}
\begin{equation}
\exists \zeta >0:U(\zeta )>\frac{m^{2}-\omega ^{2}}{2}\zeta ^{2},
\label{Eq:Exist3}
\end{equation}
\begin{equation}
\lim_{f\rightarrow \infty }\frac{U^{\prime }(f)}{f^{5}}=0.  \label{Eq:Exist4}
\end{equation}
The uniqueness part of ${\rm (SOL)}$ holds under the additional conditions:
\begin{equation*} 
\text{{\rm (U1)}\qquad \qquad }\exists l_1 >0:0<f<l_1 \Longrightarrow
U^{\prime }(f)<(m^{2}-\omega ^{2})f\text{\quad \qquad \qquad }
\end{equation*}
\begin{equation*}
\text{and\quad }l_1 <f<\infty \Longrightarrow U^{\prime
}(f)>(m^{2}-\omega ^{2})f
\end{equation*}
\begin{equation*}
\text{and }U^{\prime \prime }(l_1 )-(m^{2}-\omega ^{2})>0\text{,}
\end{equation*}
and that
\begin{eqnarray*}
\text{{\rm (U2)}\qquad For }
l_2 &>&l_1 ,\exists \lambda =\lambda (l_2 )\in
C[(l_1 ,\infty ),{{\Real}}^{+}]\text{ } \\
&&\text{such that }2(m^{2}-\omega ^{2})f+\lambda fU^{\prime }(f)-(\lambda
+2)U^{\prime }(f) \\
&&\text{is non-negative on }(0,l_2 )\text{ and non-positive on }(l_2
,\infty )\text{.}
\end{eqnarray*}
\end{theorem}
\proof
The existence part of this hypothesis 
was proved in \cite{Berestycki} under the given conditions on the 
nonlinearity. It was shown in several
articles (see for example, \cite{McLeod}, where further references
are given), that these solutions are unique under the given 
additional conditions.
\myqed

The following two operators, $L_\pm(\omega)$, which appear on linearizing  
\eqref{Eq:HNLW} about the soliton solution,
are crucial to an understanding of the stability and dynamical
properties of the $e=0$ soliton:
\begin{align}
\begin{split}
L_+(\omega)&=-\triangle +m^{2}-\omega ^{2}-\beta (f_{\omega})-\beta ^{\prime
}(f_{\omega})f_{\omega},\\
L_-(\omega)&=-\triangle +m^{2}-\omega ^{2}-\beta (f_{\omega}).
\end{split}
\la{lpm}
\end{align}

We make the following 
hypothesis on $L_+(\omega)$:
\begin{multline*}
\text{{\rm (KER)}}\qquad \qquad \text{The kernel of } L_{+}(\omega) \text{ is empty in }
H^2_r({{\Real}}^{3}).
\end{multline*}
 (Recall
that $H^s_r$ was defined as the space of radial Sobolev $H^s$ functions,
immediately following \eqref{defsob}.)

\begin{theorem}
The hypothesis $\hbox{\rm (KER)}$ is valid under the conditions 
{\rm (U1)-(U2)}.
\end{theorem}
\proof 
See \cite{McLeod}: establishing
$\hbox{{\rm (KER)}}$ is a crucial step in proving uniqueness of the
positive function $f_{\omega}$. \myqed

The operators $L_\pm$ also determine stability properties
of the soliton.
For proving stability the
following spectral assumption is used:

\begin{equation*}
\text{(S1)\qquad\qquad The subspace in which }L_{+}\text{ is strictly negative is
one dimensional,}
\end{equation*}

This assumption is valid for the ground state solitons
$f_{\omega}$ obtained by the constrained minimisation technique
of \cite{Berestycki}, because they are minimizers subject to 
a single constraint, see \cite{Stuart} (where a direct proof in the
pure power case is also given).

\noindent
Some additional more technical
results on the solitons can be found in appendix \ref{Sec:Sol}.

\subsubsection{Existence of non-topological solitons: the general case}

We now show that for small values of the coupling constant $e$ the 
ground state solitons just discussed can be continued (via the
implicit function theorem) to give soliton
solutions of \eqref{Eq:HKGM}. The properties of the $e=0$ soliton needed 
to achieve this were detailed 
already in \S\ref{Sec:Hyp}. As shown in \cite{Aprile,Benci} it is 
also possible to obtain soliton solutions for systems like \eqref{Eq:HKGM}
by variational techniques {\em applied within the class of radial functions}, 
but for present purposes 
we prefer to use the implicit function theorem so that we can carry over
stability information from the $e=0$ case, which seems to be hard to
obtain otherwise.

Generalizing the class of non-topological solitons
to the case of the gauge invariant system \eqref{Eq:HKGM}
leads us to search for solutions to \eqref{Eq:HKGM} of the form
\begin{equation}
\left(
\begin{array}{c}
\phi \\
\psi \\
\mathbf{A} \\
\mathbf{E}
\end{array}
\right) =\left(
\begin{array}{c}
Exp[i\omega t]f_{\omega ,e} \\
Exp[i\omega t]i(\omega -e\alpha _{\omega ,e})f_{\omega ,e} \\
0 \\
-\nabla \alpha _{\omega ,e}
\end{array}
\right) \text{,}  \label{Eq:radialsoliton}
\end{equation}
where we have emphasized the dependence on the parameters $\omega $
and $e$; we will assume the functions  $f_{\omega ,e}$ and
$\alpha_{\omega ,e}$ to be radially symmetric.
It can easily be checked that this gives a solution to \eqref{Eq:HKGM}
with
$A_0=\alpha_{\omega ,e}$ as long as
the functions $f_{\omega ,e}$ and $\alpha_{\omega ,e}$ 
satisfy
\begin{eqnarray}
&&-\triangle \alpha _{\omega ,e}+e^{2}f_{\omega ,e}^{2}\alpha _{\omega ,e}
-e\omega f_{\omega ,e}^{2}=0\text{,} \\
&&-\triangle f_{\omega ,e}-U^{\prime }(f_{\omega ,e})+(m^{2}-\left( \omega
+e\alpha _{\omega ,e}\right) ^{2})f_{\omega ,e} =0.
\end{eqnarray}
The first of these equations implies $\mathcal{C}_{0}=0.$ It can
readily be checked that if a gauge transformation is made to put the
solution thus obtained into temporal gauge, $A_0=0$, then its time dependence
amounts to the action of the one parameter group of gauge
transformations $e^{i(\omega-e\alpha_{\omega ,e})t}$, so that it is 
indeed a relative equilibrium solution as defined above.

\begin{theorem}[\cite{Paper1}]
\label{thm:Exist} Assume that the hypotheses 
$(SOL)$ and 
$(KER)$ 
hold for $\omega_0$ with $ {\omega_0 }^{2}<m^{2}$.
Then, there exists a
neighbourhood $U$ of $ {\omega }_0$ such that for $\omega\in U$, 
there is a number $e( {\omega })>0 $  
such that for $\omega\in U,|e|<e( {\omega })$, there exists $f_{%
 {\omega },e}\in H_{r}^{2}({{\Real}}^{3})$ such that
\begin{equation}
-\triangle f_{ {\omega },e}+m^{2}f_{ {\omega },e}-\left(
 {\omega }-e\alpha _{ {\omega },e}\right) ^{2}f_{%
 {\omega },e}=\beta (f_{ {\omega },e})f_{ {\omega
},e}\text{,}  \label{Eq:ef}
\end{equation}
where $\alpha _{ {\omega },e}\in {\dot H}_{r}^{1}({{\Real}}%
^{3})$ is a non-local function of $f_{ {\omega },e}$ uniquely
determined by
\begin{equation}
-\triangle \alpha _{ {\omega },e}+e^{2}f_{ {\omega }%
,e}^{2}\alpha _{ {\omega },e}= {\omega }ef_{ {%
\omega },e}^{2}\text{.}  \label{Eq:alph}
\end{equation}
In addition the map $ {\omega }\mapsto f_{%
 {\omega },e}$ is $C^{2}$ from $U$ to $H_{r}^{2}$.
\end{theorem}

\subsubsection{Stability in the absence of an external field}
\la{sab}
The stability of the solutions  to \eqref{Eq:HNLW} of the form
$e^{i\omega t}f_\omega(x)$ was first considered in \cite{Shatah,GSS1}
where it was proved that the positive radial solution was stable,
with respect to radially symmetric perturbations of the initial data,
as long as
\beq
\partial_\omega\left( \omega \left\|
f_{\omega}\right\| _{L^{2}}^{2}\right) <0.
\la{stab}\eeq
It was also shown that the solutions are unstable when this quantity
is positive. In \cite{Stuart} an alternative, modulational, approach
to stability was adopted along the lines of \cite{Weinstein}, with the
aim, both of generalizing previous stability results
to prove stability of uniformly moving solutions
with respect to arbitrary (non-symmetric) perturbations, and also of 
providing techniques which could provide useful information in
dynamically non-trivial settings. The presence of external fields
is an example of the latter circumstance, and so the analysis in this
article is based on that in \cite{Stuart}, which we will now
summarize. It turns out that the condition \eqref{stab} implies the
strict positivity of the Hessian of the augmented Hamiltonian on the
symplectic normal space to the space of solitons. To explain this
properly in the generality needed it is necessary to consider the
action of the Poincare (or inhomogeneous Lorentz) group  
\\[1ex]
\noindent{\em Action of the Poincare group on the solitons.}
The equations (\ref{Eq:HKGM}) are Poincare
covariant. The action of the Poincare group on the
radial soliton (\ref{Eq:radialsoliton}) gives
a family of functions depending smoothly on eight parameters
 $\{\lambda_A\}_{A=-1}^6$, with
\beq
\la{dl}
{{\lambda}}=
(\lambda_{-1},\lambda_0,\lambda_1,\dots,\lambda_6)
=(\omega,\theta ,\mbxi,\mathbf{u})
\eeq
determining (respectively) the frequency, the  phase, 
the centre and the velocity
 of the soliton. Explicitly:
\begin{equation}\Psi_{S,e}(\mathbf{x};{{\lambda}})
=\left(
\begin{array}{c}
Exp[i\Theta ](f_{\omega ,e}(\mathbf{Z})) \\
Exp[i\Theta ](i\gamma (\omega -e\alpha _{\omega ,e}(\mathbf{Z}))f_{\omega
,e}(\mathbf{Z})-\gamma \mathbf{u}\cdot\nabla _{\mathbf{Z}}f_{\omega ,e}(\mathbf{Z%
})) \\
-\gamma \mathbf{u}\alpha _{\omega ,e}(\mathbf{Z}) \\
-(\frac{1}{\gamma }P_{\mathbf{u}}+\gamma Q_{\mathbf{u}})\nabla _{\mathbf{Z}%
}\alpha _{\omega ,e}(\mathbf{Z})
\end{array}
\right) .  \label{Eq:Sol}
\end{equation}
Here the projection operators $P_{\mathbf{u}}:{{\Real}}^{3}\rightarrow {{\Real}}%
^{3}$ and $Q_{\mathbf{u}}:{{\Real}}^{3}\rightarrow {{\Real}}^{3}$ are
defined by $(P_{\mathbf{u}})_{ij}=\frac{u_{i}u_{j}}{\left| \mathbf{u}\right|
^{2}}$ and $Q_{\mathbf{u}}=1-P_{\mathbf{u}}$, and 
\beq\la{defz}
\mathbf{Z}(\mathbf{x},\lambda)
=\gamma P_{\mathbf{u}}(\mathbf{x}-\mbxi)
+Q_{\mathbf{u}}(\mathbf{x}-
\mbxi),\eeq

\beq\la{dth}\Theta(\mathbf{x},\lambda) =\theta -\omega \mathbf{u}\cdot\mathbf{Z},\eeq 
with $\gamma (%
\mathbf{u})=\frac{1}{\sqrt{1-\left| \mathbf{u}\right| ^{2}}}$.
The parameters are required to lie in the set
$\widetilde{O}\subset {{\Real}}^{8}$ defined by
\begin{equation}
\widetilde{O}\equiv
\{(\omega,\theta ,\mbxi,\mathbf{u})
\subset {{\Real}}%
^{8}:|\mathbf{u|}<1\text{ and }\omega ^{2}<m^{2}\}\text{.}
\end{equation}
The parameter range corresponding to stable solitons is
\begin{equation}
\widetilde{O}_{stab}
\equiv
\{
(\omega,\theta ,\mbxi,\mathbf{u})
\subset \widetilde{O}:\hbox{ condition \eqref{stab} holds}\}.
\la{spr}
\end{equation}
The Poincare covariance of the equations
of motion (\ref{Eq:HKGM}) implies that the 
solitons given by (%
\ref{Eq:Sol}) form an eight parameter family of solutions 
$t\mapsto\Psi_{S,e}(\mathbf{x};{{\lambda}}(t))$
of (\ref{Eq:HKGM}) as long as $\frac{d}{dt}{{\lambda}}=
V_0(\lambda)$,
where $V_0$ is the vector field on $\widetilde{O}$ defined by
\beq
\la{vf}
V_0(\lambda)\equiv \bigl(0,\frac{\omega }{\gamma },\mathbf{u},0\bigr),
\eeq
for ${{\lambda}}=(\omega,\theta ,\mbxi,\mathbf{u})$.

The case of the
nonlinear wave equation \eqref{Eq:HNLW} can be obtained 
by putting $e=0$ in the first two components of the formulae just given.
Simplifying to this case
we obtain an eight parameter family of functions,
\beq
\bigl(\phi_{S,0},\psi_{S,0})(\mathbf{x};{{\lambda}}\bigr)
\equiv e^{i\Theta}\bigl(f_{\omega}(\mathbf{Z}),
(i\gamma\omega f_{\omega}(\mathbf{Z})
-\gamma \mathbf{u}\cdot\nabla _{\mathbf{Z}}f_{\omega}(\mathbf{Z})) \bigr)
\la{es}\eeq
such that
$$t\mapsto (\phi_{S,0},\psi_{S,0})(\mathbf{x};{{\lambda}}(t)),$$ 
solves \eqref{Eq:HNLW},
as long as $\frac{d}{dt}{{\lambda}}=V_0(\lambda)$, with $V_0$
as above.
\\[1ex]
\noindent
{\em Stability for $e=0$ (nonlinear Klein-Gordon).} 
The starting point for stability analysis
is the observation that
$(\phi_{S,0},\psi_{S,0})$ is a
critical point of the {\it augmented Hamiltonian}
\beq\la{augh}
F_0(\phi,\psi;{\mathbf\lambda})=H_0(\phi,\psi)+u^i\Pi_i(\phi,\psi)
+\frac{\omega}{\gamma}Q(\phi,\psi)
\eeq
where $H_0,Q$ are the functionals defined above, and $\Pi_i$
are the momenta $\Pi_i(\phi,\psi)=\int\langle\psi,\partial_i\phi\rangle\,dx$.
The Hessian of $F_0$ at $(\phi_{S,0},\psi_{S,0})$ is a quadratic form
depending upon ${\mathbf\lambda}$:
$$
\tilde\Xi(\tilde\phi,\tilde\psi;{\mathbf\lambda})\equiv
D^2F_0(\phi_{S,0},\psi_{S,0};{\mathbf\lambda})
((\tilde\phi,\tilde\psi),(\tilde\phi,\tilde\psi)).
$$
Introduce the subspace
\beq
N_{\mathbf\lambda}\equiv\{(\tilde\phi,\tilde\psi)\in H^1\times L^2:
\Omega_0\bigl((\tilde\phi,\tilde\psi),
\partial_{\mathbf\lambda}(\phi_{S,0},\psi_{S,0})({\mathbf\lambda})\bigr)=0\}
\la{sns}\eeq
then the following hypothesis is crucial for stability:

\begin{multline*}
\text{{\rm (POS)}}\qquad
\text{For each compact }
\mathbf{K}\subset \widetilde{O}_{stab}\,
\exists\tau_*=\tau_*(\mathbf{K})>0\text{ such that}\\
\shoveleft{\phantom{\text{{\rm (POS)}}\qquad}
\tilde\Xi(\tilde\phi,\tilde\psi;{\mathbf\lambda})\geq
\tau_*\|(\tilde\phi,\tilde\psi)\|_{H^1\times L^2}^2\,\text{ for all }
(\tilde\phi,\tilde\psi)\in N_{\mathbf\lambda}.}\\
\end{multline*}
\begin{remark}{\rm
$\widetilde{O}_{stab}$ is the set of parameter values
corresponding to stable solitons, which are obtained as
Poincare transforms of solitons $e^{i\omega t}f_{\omega }$
with $\omega$ such that \eqref{stab} holds.
}
\end{remark}

\begin{theorem}[\cite{Stuart}]
\la{modstab}
If the nonlinearity satisfies the conditions given
in \S\ref{Sec:Hyp} then (POS) is true.
Furthermore, solitons of \eqref{Eq:HNLW} 
corresponding to frequencies $\omega$ such that 
\eqref{stab} holds are modulationally stable with respect to 
small, arbitrary perturbation in energy norm. To be precise,
consider the initial value problem for \eqref{Eq:HNLW} with
initial data close to a soliton 
$\bigl(\phi_{S,0},\psi_{S,0})(\cdot;{{\lambda}(0)}\bigr)$
with $\lambda(0)\in\widetilde{O}_{stab}$, in the 
sense that
$$\epsilon=
\bigl\|
\bigl(\phi(0,\cdot),\psi(0,\cdot)
\bigr)-
\bigl(\phi_{S,0},\psi_{S,0})(\cdot;{{\lambda}(0)}\bigr)
\bigr\|_{{\cal H}_0}
$$
is sufficiently small.
Then there
exists a global solution which satisfies:
\beq\sup_{t\in\R}\bigl\|
\bigl(\phi(t,\cdot),\psi(t,\cdot)
\bigr)-
\bigl(\phi_{S,0},\psi_{S,0})(\cdot;{{\lambda}(t)}\bigr)
\bigr\|_{{\cal H}_0}
\leq
c\epsilon,
\la{eqmodtsab}\eeq
for some $C^1$ curve $t\mapsto\lambda(t)
\in\widetilde{O}_{stab}$.
\end{theorem}

\noindent
{\em Stability for small $e$ 
(nonlinear Klein-Gordon-Maxwell).} 
It was shown in  \cite{Paper1}, that stability holds also
for solitons in \eqref{Eq:HKGM} under the condition
\eqref{stab}, for sufficiently small values of the 
electromagnetic coupling constant $e$. This was proved using
the Coulomb condition, so we first write down
the soliton solutions \eqref{Eq:Sol} in Coulomb gauge. 
(The Coulomb condition is not
invariant under Lorentz boosts, therefore, it is necessary 
to perform a gauge transformation to move the Lorentz 
boosted solitons into the Coulomb gauge).
The Lorentz boosted solitons 
$\Psi_{SC,e}$
in the Coulomb gauge have the form
\begin{equation}\la{lbs}
\left(
\begin{array}{c}
\phi _{SC,e}(\mathbf{x}) \\
\psi _{SC,e}(\mathbf{x}) \\
\mathbf{A}_{SC,e}(\mathbf{x}) \\
\mathbf{E}_{SC,e}(\mathbf{x})
\end{array}
\right) =\left(
\begin{array}{c}
Exp[i{\Theta_C}](f_{\omega ,e}(\mathbf{Z})) \\
Exp[i{\Theta_C}](i\gamma (\omega -e\alpha _{\omega ,e}(\mathbf{Z}%
))f_{\omega ,e}(\mathbf{Z})-\gamma \mathbf{u}.\nabla _{\mathbf{Z}}f_{\omega
,e}(\mathbf{Z})) \\
-\gamma \mathbf{u}\alpha _{\omega ,e}(\mathbf{Z})+\nabla \zeta \\
-(\frac{1}{\gamma }P_{\mathbf{u}}+\gamma Q_{\mathbf{u}})\nabla _{\mathbf{Z}%
}\alpha _{\omega ,e}(\mathbf{Z})
\end{array}
\right)
\end{equation}
where ${\Theta_C}=\Theta +ie\zeta $, and $\zeta(x;\lambda) $ is a
solution of
\beq
\la{cc0}
-\triangle \zeta =-\gamma \mathbf{u}\cdot\nabla \alpha _{\omega
,e}(\mathbf{Z}).
\eeq
It is a smooth function of $x$ and
also depends smoothly on $\lambda$; requiring that
$\nabla\zeta\in L^p,p>3$ fixes it up to a constant. Some estimates
for $\zeta$ are given in appendix \ref{Sec:alpha Estimates}.
The temporal part of the gauge field is
given by 
$$(A_{SC,e})_{0}=\gamma \alpha _{\omega ,e}(Z)+\overset{.}{\zeta }%
=\gamma \alpha _{\omega ,e}(Z)
+V_0(\lambda)\cdot\partial_\lambda{\zeta }.
$$

\begin{theorem}[\cite{Paper1}]
\la{modstabe}
In the situation of the previous theorem 
the solitons \eqref{lbs} of \eqref{Eq:HKGM} 
corresponding to frequencies $\omega$ such that \eqref{stab} 
holds are, for sufficiently small $|e|$, 
modulationally stable in Coulomb gauge 
with respect to small, 
arbitrary perturbation of the initial data
in energy norm $\|\,\cdot\|_{\cal H}$
defined in \eqref{enorm}. The stability 
is in the same sense as in the previous theorem,
see \cite{Paper1} for full details.
\end{theorem}

\subsection{The main theorems}
\label{statmain}
We now explain and state our main results on the interaction 
of the solitons of \S\ref{Sec:NonTopSol}
with the scaled external electromagnetic field 
of \S\ref{Sec:ExternalField}. We write
the total electromagnetic potential as $\A=\A_\mu dx^\mu$ (as described in \S\ref{first})
with corresponding electric field
$\E_j=\partial_t \A_j-\partial_j\A_0$. The potential $\A$ 
will be formed from three constituents:
\begin{enumerate}
\item
the external field, produced by a background charge 
$\rho_B^{\delta}$ and current $\mathbf{j}_B^{\delta}$, and scaled as 
described in \S\ref{Sec:ExternalField}, 
\item
the soliton contribution, 
as described in \S\ref{Sec:NonTopSol} but with parameters
$\lambda(t)$ varying in a dynamically determined way,
\item an additional component produced by interaction of the
initial data with the two previous components. This component
is not explicitly given, and must be estimated.
\end{enumerate}
Similarly, the solitonic field will be made up of a component which 
is the moving soliton, and a remainder produced by 
interactions, which must be estimated. 

It is convenient to 
write the (nl-KGM) equations in first order form. Including
the scaled background current density, the equations read:
\begin{equation}
\partial_t\left(
\begin{array}{c}
\phi \\
\psi \\
\A_{i} \\
\E_{i}
\end{array}
\right) =\left(
\begin{array}{c}
\psi +ie\A_0\phi \\
\triangle _{\A}\phi -\mcn^{\prime }(\phi )+ie\A_0\psi \\
\E_{i}+\nabla _{i}{\A}_{0} \\
\triangle \A_{i}+\left\langle ie\phi
,\nabla _{\A}\phi \right\rangle-e{\mathbf j}_{B}^{\delta} 
\end{array}
\right),  \label{Eq:HKGMB}
\end{equation}
with the Coulomb gauge condition imposed. These equations
are to be solved with the Gauss law 
\beq
\dv{\E}-\left\langle ie\phi ,\psi \right\rangle
=\rho_B^{\delta},\label{Eq:Gauss3}
\end{equation}
as a constraint.
We shall abbreviate a general solution by making use of the 
following definition:
\begin{equation}
\Psi =\left( \phi ,\psi ,{\A}_i, {\E}_i\right) \text{,}
\end{equation}
with $i\in\{1,2,3\}$.

Using this Hamiltonian formulation with $\Psi$ as dynamical variable
we write the external field
$$\Psi^{\delta}_{ext}=(0,0,\mathbf{a}^{\delta},\mathbf{E}^{\delta}_{ext}).$$ 
It will be convenient also to have the freedom of 
applying a gauge transformation $\chi(t,x)$ to this:
$$\Psi^{{\delta},\chi}_{ext}
=(0,0,\mathbf{a}^{{\delta},\chi},\mathbf{E}^{\delta}_{ext}),$$ 
with $\mathbf{a}^{{\delta},\chi}=\mathbf{a}^{{\delta}}+\mathbf{d}\chi$.
The aim is now to construct a solution $\Psi$ to \eqref{Eq:HKGMB}
consisting of $\Psi^{{\delta},\chi}_{ext}$ with a soliton
$\Psi _{SC,e}({{\lambda}})$ superimposed. We choose the
gauge transformation $\chi$ 
so that the transformed external 
electromagnetic potentials vanish along the 
world-line of the
soliton 
$\mbx=\mbxi(t)$; in particular, at $t=0$ we will choose
$$
\chi(0,\mbx)=\chi_0(\mbx)=-
(\mathbf{x}-\mbxi(0))\cdot\mathbf{a}^{{{\delta}}}(0,\mbxi(0))
$$ 
so that  $\mathbf{a}^{{\delta},\chi_0}(0,\mbx)=
\mathbf{a}^{\delta}(0,\mbx)-\mathbf{a}^{\delta}(0,\mbxi(0))$.
\subsubsection{Stability in the presence of an external field}
\la{spef}

The following theorem asserts the long time stability, under 
the influence of an external field, of stable solitons
to \eqref{Eq:HKGMB}. Recall that the stable solitons are 
those parametrised
by $\lambda\in\widetilde{O}_{stab}$, so that \eqref{stab}
and hence (POS) hold, 
and they are stable by theorems \ref{modstab} and \ref{modstabe}
in the absence of an external field.
\begin{theorem}
\label{thm:Stability} 
Assume that the nonlinearity satisfies the hypotheses 
(H1)-(H3), and also is such that the hypotheses (SOL),\,(KER) and (POS)
in \S\ref{Sec:Hyp} hold.
In addition, assume that the external field 
satisfies the assumptions 
in \S\ref{Sec:ExternalField}.
Suppose further that the scaling parameters
satisfy ${\delta}^2=o(e), 
e=o(1)$ and $e=o({\delta})$. 

(i) Consider initial data of the form
$$
\Psi(0) =\left(\phi_{SC,e}(\lambda(0)),\psi_{SC,e}(\lambda(0)) ,{\A}_i(0), {\E}_i(0)\right)
$$
where $\lambda(0)
\in\widetilde{O}_{stab}$ corresponds 
to a stable soliton (which verifies (POS)).
It follows
that, if $e$ is sufficiently small and
\begin{equation}
\bigl\| \Psi (0)-\Psi_{ext}^{{{\delta}},\chi_0 }(0)-\Psi _{SC,e}
(\lambda(0))\bigr\|_{\cal H}^2=o(e),
\end{equation}
there exists 
\begin{itemize}
\item a positive number $T_0>0$, independent of $e$,
\item a $C^1$ gauge transformation $\chi(t,\mbx)$ defined in
\eqref{defg}, linear in $\mbx$ 
at each time $t$, satisfying $\chi(0,\mathbf{x})=\chi_0(\mathbf{x})$
\item a curve ${{\lambda}}(t)\in C^{1}
(\mathbb[0,\frac{T_0}{|e|}],\widetilde{O}_{stab})$, and
\item a distributional solution $\Psi (t)$ of (\ref{Eq:HKGMB}), 
\end{itemize}
such that
$$\Psi (t)-\Psi_{ext}^{{{\delta}},\chi }(t)
\in C([0,\frac{T_0}{|e|}];{\cal H})$$
and
\begin{equation}
\sup_{t\in [0,\frac{T_0}{|e|}]}\bigl\| \Psi (t)
-\Psi_{ext}^{{{\delta}},\chi}(t)
-\Psi _{SC,e}({{\lambda}}(t))\bigr\|_{\cal H}^2
=o(e)
\text{,}
\end{equation}
Furthermore, ${\lambda }(t)$ 
satisfies a system of ordinary differential equations given by
(\ref{Eq:ModEq}) with
$
|\partial_t\lambda-V_0(\lambda)|=O(e)\text{.}
$
The time component of the potential $\A_0$ is determined by the
Coulomb condition and the Gauss law, \eqref{Eq:Gauss2m}, 
and has properties detailed in \S\ref{Sec:Stability}.

(ii) More generally,  the same conclusions hold 
for initial data sufficiently close to a
stable soliton in an appropriate sense: 
see \S\ref{mg} for a precise statement.
\end{theorem}
This theorem is proved in \S\ref{Sec:Stability}.

\begin{remark}{\rm 
As explained in \ref{Sec:Hyp}, 
if the nonlinear potential satisfies (\ref
{Eq:Exist1})-(\ref{Eq:Exist4}), 
$U(1)$, $U(2)$, $S(1)$ above 
in addition to (H1)-(H3) then the conditions
(SOL),\,(KER) and (POS) all hold.
}
\end{remark}

\subsubsection{Motion in the presence of an external field: the Lorentz force}

The previous theorem provides ordinary differential equations 
(\ref{Eq:ModEq}) which determine the evolution of the soliton
parameters. A detailed investigation of these
equations allows us to deduce an equation of motion
for the soliton, which is expected to be
the Lorentz force law for a moving charge, at least to highest
order in $e$. As remarked earlier, if the analysis were carried
out explicitly to higher order in $e$, corrections would be
expected to appear, in particular  due to the back reaction of 
the soliton's electromagnetic field on itself. However, these
are not expected to appear in the $O(e)$ force law, and the
following theorem validates this:
\begin{theorem}
\label{Thm:Lorentz Force Law}
Assume the hypotheses and conclusions of
theorem \ref{thm:Stability} hold, and let
$
{{\lambda}}
=(\omega,\theta ,\mbxi,\mathbf{u})$ 
be the
parameters of the soliton $\Psi_{SC,e}({{\lambda}})$. 
Then, on the interval
$[0,\frac{T_{0}}{|e|}]$, the 
centre and velocity of 
the soliton evolve according to the equations:
\ba
\frac{d}{dt}\mbxi&=&\mathbf{u}+o(e)\\
\frac{d}{dt}\left( \mathbb{M}_{S}\gamma 
(\mathbf{u})\mathbf{u}\right) &=&
e\mathbb{Q}_{S}\left( {\mathbf E}_{ext}^{{{\delta}} }(t,\mbxi)+\mathbf{u}\times
{\mathbf B}_{ext}^{{{\delta}} }(t,\mbxi)\right) 
+o(e)\text{,}
\ea
where the mass of the soliton, $\mathbb{M}_{S}$, is given by
\begin{equation}\la{defmass}
\mathbb{M}_{S}=\frac{1}{3}\left\| \nabla f_{\omega}\right\|
_{L^{2}}^{2}+\omega ^{2}\left\| f_{\omega}\right\| _{L^{2}}^{2}\text{,}
\end{equation}
and the charge of the soliton is given by
\begin{equation}\la{defcharge}
\mathbb{Q}_{S}=\int \left( \omega -e\alpha \right) f_{\omega ,e}^{2}\text{.}
\end{equation}
\end{theorem}

This theorem is proved in \S\ref{Sec:Lorentz Force Law}.

\begin{remark}
{\em 
Observe that, since we have scaled the external field 
so that 
${\mathbf{E }^{\delta}_{ext}}$ and 
${{\mathbf B}^{\delta}_{ext}}$ 
are independent of $e$, the soliton undergoes $O(1)$ motion
on the time interval $[0,\frac{T_0}{|e|}]$
according to the Lorentz force law.}
\end{remark}

\section{Stability: proof of theorem \ref{thm:Stability}}

\label{Sec:Stability}

In this section we explain the proof of theorem
\ref{thm:Stability}, making use of results 
which are proved separately in \S\ref{modth} and
\S\ref{Sec:Proof of dWtildedt}.
Throughout this section the hypotheses of theorem
\ref{thm:Stability} are understood to hold without
explicit mention. 
Also we may assume, without loss of generality, that the
solution is smooth in the course of the following calculations:
since finite energy solutions
can be approximated by smooth ones by $(WP2)$ in 
\S\ref{Sec:LWP}, 
and all the bounds we use depend only on the energy norm, 
this implies the result for finite energy initial data as 
in theorem \ref{thm:Stability}.

\subsection{Beginning of proof of theorem 
\ref{thm:Stability}}\label{bp5}
 
\subsubsection{Ansatz for the solution}

We make an ansatz for a solution 
$
\Psi (t)
=\Psi_{ext}^{{{\delta}},\chi}(t)
+\Psi _{SC,e}({{\lambda}}(t))+Perturbation,
$
which is close to a soliton with
time varying (modulating) parameters $\lambda(t)$,
in the background external field $\Psi_{ext}^{{{\delta}},\chi}(t)$.
Explicitly the ansatz reads:
\begin{equation}
\left(
\begin{array}{c}
\phi (t,\mathbf{x}) \\
\psi (t,\mathbf{x}) \\
{\A}_\mu(t,\mathbf{x}) \\
{\E}_j(t,\mathbf{x})
\end{array}
\right) =\left(
\begin{array}{c}
\phi _{SC,e}\left( {{\lambda}}(t)\right) +Exp[i\Theta_C]v \\
\psi _{SC,e}\left( {{\lambda}}(t)\right) +Exp[i\Theta_C]w \\
\left({A}_{SC,e}\right)_\mu\left( {{\lambda}}(t)\right) 
+{a}^{{{\delta},\chi}
}_\mu+{\tilde A}_\mu\\
\left({E}_{SC,e}\right)_j\left( {{\lambda}}(t)\right) 
+\left({E}_{ext}^{{{\delta}}}\right)_j
+{\tilde E}_j
\end{array}
\right) .\label{Eq:Ansatz}
\end{equation}
Notice that we have included here an
ansatz for the temporal part of the potential $\A_0$.
Since we have imposed the Coulomb gauge 
throughout, it follows that $\dv\mathbf{{\tilde A}}
=0$. The choice of the
gauge transformation $\chi$ is:
\beq
\chi(t,\mbx) =-(\mathbf{x}-\mbxi)\cdot\mathbf{a}^{{{\delta}}
}(t,\mbxi)
-\int_{0}^{t}a_{0}^{{{\delta}} }(s,\mbxi(s))
+\overset{.}{\mbxi}(s)
\cdot\mathbf{a}^{{{\delta}} }(s,\mbxi)ds.
\la{defg}\eeq
This is chosen so that the gauge transformed external 
potentials vanish along the world line of the soliton:
\begin{align}
\la{da}
\begin{split}
&a_\mu^{{\delta},\chi} =a_\mu^{\delta}+\partial_\mu\chi,\\
&a_\mu^{{\delta},\chi}\bigl(t,{\mbxi}(t)\bigr)=0.
\end{split}
\end{align}
These imply
\begin{align}
\begin{split}
a_0^{{\delta},\chi} (t,x)&=a_0^{\delta} (t,x)
-a_0^{\delta} (t,\mbxi(t))
-(x-\mbxi(t))\cdot{\dot {\mathbf a}}^{\delta}(t,\mbxi(t)),\\
\mathbf{a}^{{\delta},\chi}(t,x)&=\mathbf{a}^{\delta}(t,x)
-\mathbf{a}^{\delta}(t,\mbxi(t)),
\end{split}
\la{dad}
\end{align}
exhibiting the claimed vanishing of $a_\mu^{{\delta},\chi}$ along the soliton's
world line.
This allows certain quantities to be proved to be bounded
in the course of the proof. Notice that $\chi$ is linear (and so
harmonic) in $\mathbf{x}$, and so preserves the 
Coulomb condition (see remark \ref{rgi}).

There is clearly a
redundancy in our ansatz, in that $\lambda(t)$ is
so far completely undetermined.
The appropriate choice of ${{\lambda}}(t)$ is dictated by the requirement
that the solution be close to a soliton determined by the 
parameters $\lambda(t)$, i.e. by the requirement that we
have good bounds for  field
perturbation $(v,w,\mathbf{\tilde A},\mathbf{\tilde E})$.
This is carried out in \S\ref{modth}, with
the main results summarized next in
\S\ref{rmodth}. First we write explicitly the equations for the 
$(v,w,\mathbf{\tilde A},\mathbf{\tilde E})$, and give 
some bounds for
the inhomogeneous terms in these equations.

\subsubsection{Equations for the perturbations of the fields}

\begin{equation}
\partial_t{v}+
i\left( \omega  \gamma+h\right) v
=w+\jmath_1,
\label{de1}\end{equation}

\begin{equation}
\partial_tw+i( \omega\gamma +h)w=-M_{{{\lambda}}}v+
\jmath_2+{\mathcal N},\label{de2}
\end{equation}

\begin{equation}
\partial_t\mathbf{\tilde A}=\mathbf{\tilde E}+\jmath_3,
\label{de3}\end{equation}

\begin{equation}
\partial_t\mathbf{\tilde E}=\Delta\mathbf{\tilde A}
+\jmath_4,
\label{de4}\end{equation}
where the inhomogeneous terms 
$h,\jmath_1,\dots ,\jmath_4$ and ${\mathcal N}$ are defined 
in \S\ref{defin},
and $M_\lambda$ is the operator
\begin{equation}
M_{{{\lambda}}}v=(-\triangle _{x}+m^{2}+\gamma ^{2}\omega
^{2}|u|^{2})v+2i\omega\gamma \mathbf{u}\cdot\nabla _{x}v
-\beta (f_{\omega})v
-f_{\omega}\beta ^{\prime }(f_{\omega}){\Re} v\text{.}
\label{Eq:Mlambda}
\end{equation}
\noindent
The last two terms have been chosen to 
depend on the $e=0$ profile function $f_\om$,
rather than $f_{\om, e}$, so that
it is possible to make direct use of the stability assumption (POS)
in \S\ref{sab}. (This choice is reflected in the expression for
the inhomogeneous term ${\mathcal N}$ in \eqref{Eq:N(f,f,v)} and
its corresponding estimate in \eqref{estn}).

In addition to these evolution equations, 
the fields are constrained to satisfy the
Gauss law (\ref{Eq:Gauss2}), which takes the form:
\begin{equation}
\dv\mathbf{\tilde E}=-\triangle {\tilde A}_0
=e\left\langle i Exp[-i\Theta_C]\phi _{SC,e}
,w\right\rangle
+e\left\langle iv, Exp[-i\Theta_C]\psi_{SC,e}
+w\right\rangle \text{.}
\la{Eq:Gauss2m}
\end{equation}
Under finite energy assumptions this equation has a unique
solution with ${\tilde A}_0\in \hh$; this defines uniquely
${\tilde A}_0$ as a nonlocal function of $v,w,\lambda$ at each time.
Estimates for ${\tilde A_0}$ are given in lemma \ref{lem:rbarLp}.

\subsubsection{Inhomogeneous terms in the field perturbation equations 
\eqref{de1}-\eqref{de4}}
\label{defin}

The following quantity appears in both \eqref{de1} and 
\eqref{de2}:
\begin{equation}  
\label{Eq:h}
h=\dot\Theta_c-\omega\gamma-e(A_{SC,e})_0-ea_0^{{\delta},\chi}-e\tilde A_0.
\end{equation}

The inhomogeneous term in 
\eqref{de1} is $\jmath_1=\jmath_1^I+\jmath_1^{II}+\jmath_1^0$, where
\ba
\la{dj1}
\jmath_1^I
&=&
-(\dot\lambda-V_0(\lambda))\cdot e^{-i\Theta_c}{\partial}_\lambda\phi_{SC,e}\\
\jmath_1^{II}&=& ie a_0^{{\delta},\chi}f_{\omega,e}\\
\jmath^0_1&=&ie\tilde A_0f_{\omega,e}.
\ea

The inhomogeneous terms in \eqref{de2} are
\ba
\la{Eq:N(f,f,v)}
{\mathcal N}(f_{\omega ,e},f_{\omega},v)
&=&\beta(|f_{\omega ,e}+v|)(f_{\omega ,e}+v)
-\beta(|f_{\omega ,e}|)f_{\omega,e}\\
&&\;\quad -\beta (|f_{\omega}|)v-f_{\omega}\beta ^{\prime }(|f_{\omega}|){\Re} v,
\notag\ea
and $\jmath_2=\jmath_2^{I}+\jmath_2^{II}+\jmath_2^{III}+\jmath_2^{IV}+\jmath_2^0$
where
\ba
\la{dj21}
\jmath_2^I&=&
-(\dot\lambda-V_0(\lambda))\cdot e^{-i\Theta_c}{\partial}_\lambda\psi_{SC,e}\notag\\
\jmath_2^{II}&=&e \mbR f_{\omega,e}+ie a_0^{{\delta},\chi}e^{-i\Theta_c}\psi_{SC,e},\la{dj22}\\
\jmath_2^{III}&=& e\mbR v+\mbS v\la{dj23}\\
\jmath_2^{IV}&=&e^{-i\Theta_c}\left[
\Delta_{\A}(e^{i\Theta_c}v)+
(\Delta_{\A}-\Delta_{A_{SC,e}})\phi_{SC,e}\right]-\jmath_2^{II}
-\jmath_2^{III}\la{dj24}\\
\jmath_2^0&=& ie\tilde A_0e^{-i\Theta_c}\psi_{SC,e}.
\ea
Here, the operators $\mbR,\mbS$ are given by
\ba
\mbR v&=&2i\left( \mathbf{a}^{{{\delta}},\chi }\right) .(i\gamma
\left( \omega -e\alpha _{\omega ,e}\right) \mathbf{u}-\nabla )v-e\left|
\mathbf{a}^{{{\delta}},\chi }\right| ^{2}v\text{,}  \label{Eq:E0}\\
\mbS v&=&2ie\alpha _{\omega ,e}\gamma \mathbf{u}\cdot\nabla v+ie\gamma
\left( \mathbf{u}\cdot\nabla \alpha _{\omega ,e}\right) v
+2e\gamma^{2}|\mathbf{u}|^{2}\omega \alpha _{\omega ,e}v
-e^2\left( \gamma \alpha _{\omega,e}
|\mathbf{u}|\right) ^{2}v\text{.} \notag
\ea
\noindent
(In verifying these formulas, it is helpful to
note that by the exact solutions in \S\ref{sab}
$$e^{-i\Theta_c}(\nabla-ieA_{SC,e}
-ie{\mathbf a}^{{\delta},\chi})e^{i\Theta_c}v
=\nabla v
-i(\gamma(\omega-e\alpha_{\omega,e}){\mathbf u}
+e{\mathbf a}^{{\delta},\chi})v,
$$
and a similar formula for the second derivatives.)

The inhomogeneous term in \eqref{de3} is
$\jmath_3=\jmath_3^I+\jmath_3^0$ where
\beq
\begin{split}
\jmath_3^I&=-(\dot\lambda-V_0(\lambda))\cdot
\partial_\lambda A_{SC,e},\\
\jmath_3^0&=\nabla\tilde A_0.
\end{split}
\la{dj3}
\eeq
and in \eqref{de4} we have  $\jmath_4=\jmath_4^{I}+\jmath_4^{II}
+\jmath_4^{III}+\jmath_4^{IV}+\jmath_4^0$, with $\jmath_4^0=0$ and
\ba
\la{dj41}
\jmath_4^I&=&-(\dot\lambda-V_0(\lambda))\cdot
\partial_\lambda E_{SC,e},\\
\jmath_4^{II}&=&
-e^2|f_{\omega ,e}|^2{\mathbf a}^{{\delta},\chi}
\la{dj42}\\
\jmath_4^{III}&=&e\langle ie^{i\Theta_c}v,
\nabla_{\A}(\phi_{SC,e}+e^{i\Theta_c}v)\rangle
-e^2|f_{\omega ,e}|^2\mathbf{\tilde A},\la{dj43}\\
\jmath_4^{IV}&=&e\langle i\phi_{SC,e},
\nabla_{\A}(e^{i\Theta_c}v)\rangle.\la{dj44}
\ea

To clarify the structure of these terms it is helpful to insert
the ansatz \eqref{Eq:Ansatz} into the Hamiltonian \eqref{HAM}
and write $H-\int\mcn=\sum_{n=0}^4 \hat H^{(n)}$, where 
$\hat H^{(n)}$ has 
homogeneity $n$ in $(v,\tilde A)$.  (The terms of degree larger than
two arise solely from $\frac{1}{2}\int|\nabla_{\A}\phi|^2$.) Then
the pieces of $\jmath_2$, (resp. $\jmath_4$), which are of degree 
$n\in\{1,2,3\}$
in $(v,\tilde A)$ arise, respectively, as the Frechet derivatives
$-D_v \hat H^{(n+1)}$, (resp. $-D_{\tilde A} \hat H^{(n+1)}$). 
The nonlinear potential $\mcn$ only appears through $M_\lambda v$ and
${\mathcal N}$ in \eqref{de2}.
With this 
understood we now introduce notation for the various terms arising
in \eqref{de2} and \eqref{de4}, organized according to their 
homogeneity. Let $\hat H=\sum_{n=2}^4\hat H^{(n)}$, then
in \eqref{de2} the corresponding terms are
$$
-D_v \hat H=-M^0_\lambda v+\jmath_2^{III}+\jmath_2^{IV},
$$
where
\begin{equation}
M^0_{{{\lambda}}}v=(-\triangle _{x}+\gamma ^{2}\omega
^{2}|u|^{2})v+2i\omega\gamma \mathbf{u}\cdot\nabla _{x}v
=-e^{-i\Theta}\triangle\bigl(e^{i\Theta}v\bigr)
\end{equation}
with $\Theta$ the soliton phase factor in \eqref{dth}.
Notice that
the operator $M^0_{{{\lambda}}}$ consists of those terms
in (\ref{Eq:Mlambda}), which do not arise from the $\mcn$
term in the energy, because we have so far excluded this term 
in our expansion (which is of $H-\int\mcn$). However, it is convenient
to put back in the quadratic parts of the Taylor expansion of 
$\mcn$, but expanded around $f_\om$ (the $e=0$ soliton), so as to obtain
the $M_\lambda$ operator which appears in \eqref{de2}.
Thus we let
$$
\tilde H=\hat H+\frac{1}{2} D^2\mcn(f_\om)(v,v)=
\hat H+\frac{1}{2}\Bigl[m^2|v|^2-\beta(f_\omega)|v|^2-f_\om\beta'(f_\om)(\Re v)^2\Bigr],
$$
so that, using the same notation for the homogeneous components of 
$\tilde H$ as for $\hat H$, we have $\tilde H^{(n)}=\hat H^{(n)}$ for
$n>2$ and $\tilde H^{(2)}-\hat H^{(2)}=\frac{1}{2} D^\mcn(f_\om)(v,v)$.
In \eqref{de4} the corresponding terms are
$$
-D_{\tilde A} \tilde H=\Delta\mathbf{\tilde A}+\jmath_4^{III}+
\jmath_4^{IV}.
$$
To write these terms explicitly we introduce a multilinear
notation as follows.
\beq
\la{ml}
(-D_v \tilde H,
-D_{\tilde A} \tilde H)
=\mathbb{B}^{(1)}(
v,
\mathbf{{\tilde A}}
) +\mathbb{B}^{(2)}(
v,
\mathbf{{\tilde A}}
) +\mathbb{B}^{(3)}(
v,
\mathbf{{\tilde A}}
),
\eeq
where $\mathbb{B}^{(n)}(v,\mathbf{\tilde A})$ 
is a homogeneous degree $n$ function
of $(v,\mathbf{\tilde A})$, as indicated by the superscript.
We will define 
${\mathbb{B}^{(1)}}:H^{1}( {{\Real}}^{3};\mathbb{C})
\oplus {\dot H}^{1}( {{\Real}}^{3};{{\Real}}^{3}) \mapsto
H^{-1}( {{\Real}}^{3};\mathbb{C}) \oplus {\dot H}%
^{-1}( {{\Real}}^{3};{{\Real}}^{3}) $, where by 
$H^{-1}$ (resp. ${\dot H}^{-1}$) 
we mean the dual space of $H^1$ (resp. ${\dot H}^1$.)
Explicitly:
\begin{equation}
{\mathbb{B}^{(1)}}(
v,
\mathbf{{\tilde A}}
) =(
{\mathbb{B}}_{11}v+{\mathbb{B}}_{12}\mathbf{{\tilde A}},
{\mathbb{B}}_{21}v+{\mathbb{B}}_{22}\mathbf{{\tilde A}}
) \text{,\label{Eq:A}}
\end{equation}
where
\begin{equation}
{\mathbb{B}}_{11}v=-M_{{{\lambda}}}v+e\mbR v+\mbS v\text{,}
\end{equation}
and the
operators $\mbR$ and $\mbS$ are as just defined.
Next
\begin{align}
{\mathbb{B}}_{12}\mathbf{{\tilde A}}&=-2ef_{\omega ,e}\left( \gamma \left( \omega
-e\alpha _{\omega ,e}\right) \mathbf{u}
+e\mathbf{a}^{{{\delta},\chi} }\right)
\cdot\mathbf{{\tilde A}} 
-2ie\mathbf{\tilde A}\cdot\nabla f_{\omega ,e}
\text{,}
\notag\\
{\mathbb{B}}_{21}\,v&=-2ef_{\omega ,e}\left( \gamma \left( \omega -e\alpha
_{\omega ,e}\right) \mathbf{u}+e\mathbf{a}^{{{\delta}},\chi } 
\right) \Re v +\langle iev,\nabla f_{\omega ,e}\rangle+
\langle ief_{\omega ,e},\nabla v\rangle\text{,}
\notag\end{align}
and finally,
$
{\mathbb{B}}_{22}\mathbf{{\tilde A}}=\triangle \mathbf{{\tilde A}} -e^{2}f_{\omega ,e}
\mathbf{{\tilde A}} \text{.}
$ Since $\dv\tilde{\mbA}=0$ integration by parts
yields $\langle \tilde{\mbA},{\mathbb{B}}_{21}\,v\rangle_{L^2}=
\langle v,{\mathbb{B}}_{12}\,\tilde{\mbA}\rangle_{L^2}$, and
\begin{align}
\tilde H^{(2)} &=
-\frac{1}{2}\left\langle(
v ,
\mathbf{{\tilde A}}),
{\mathbb{B}^{(1)}}(
v,
\mathbf{{\tilde A}}) 
\right\rangle _{L^{2}},\\
&=\frac{1}{2}\int \Bigl[|\nabla\tilde\mbA|^2
+e^2|f_{\omega ,e}\tilde\mbA|^2
+|\nabla v-i( \gamma 
\left( \omega -e\alpha_{\omega ,e}\right) 
\mathbf{u}+e\mathbf{a}^{{{\delta}},\chi })v|^2\notag\\
&\qquad\quad
+\frac{1}{2}\bigl(m^2|v|^2-\beta(f_\omega)|v|^2
-f_\om\beta'(f_\om)(\Re v)^2\bigr)
\notag\\
&\qquad\qquad-2\langle ie\tilde\mbA f_{\omega ,e},
\nabla v-i( \gamma 
\left( \omega -e\alpha_{\omega ,e}\right) 
\mathbf{u}+e\mathbf{a}^{{{\delta}},\chi })v\rangle\notag\\
&\qquad\qquad\qquad-2\langle ie\tilde\mbA v,
\nabla  f_{\omega ,e}-i( \gamma 
\left( \omega -e\alpha_{\omega ,e}\right) 
\mathbf{u}+e\mathbf{a}^{{{\delta}},\chi }) f_{\omega ,e}\rangle\Bigr]\,dx.
\notag\end{align}

Next, the quadratic terms in the equations can be expressed in terms of a 
rank three symmetric tensor 
$$\mathbb{B}^{(2)}:\left( H^{1}\left(
{{\Real}}^{3};\mathbb{C}\right) \oplus {\dot H}^{1}\left( {{\Real}}%
^{3};{{\Real}}^{3}\right) \right) ^{2}\mapsto H^{-1}\left( {{\Real}}^{3};%
\mathbb{C}\right) \oplus {\dot H}^{-1}\left( {{\Real}}^{3};{{\Real}}%
^{3}\right) $$ which is given explicitly by
\begin{equation}
\label{Eq:B}
\mathbb{B}^{(2)}(
v,
\mathbf{{\tilde A}}
) =\left(\begin{array}{c}
\mathbb{B}_{111}[v,v]+\mathbb{B}_{112}[v,\mathbf{{\tilde A}}]
+\mathbb{B}_{121}[\mathbf{{\tilde A}},v]
+\mathbb{B}_{122}[\mathbf{{\tilde A}},\mathbf{{\tilde A}}],\\
\mathbb{B}_{211}[v,v]+\mathbb{B}_{212}[v,\mathbf{{\tilde A}}]
+\mathbb{B}_{221}[\mathbf{{\tilde A}},v]
+\mathbb{B}_{222}[\mathbf{{\tilde A}},\mathbf{{\tilde A}}]
\end{array}
\right)
\text{,}\notag
\end{equation}
where $\mathbb{B}_{111}=\mathbb{B}_{222}=0$, and
$$
\mathbb{B}_{112}[v,\mathbf{{\tilde A}}]
=-ev\left( \gamma \left( \omega -e\alpha
_{\omega ,e}\right) \mathbf{u}+e( \mathbf{a}^{{{\delta}},\chi }
) \right) \cdot\mathbf{{\tilde A}}
-ie\nabla v\cdot\mathbf{{\tilde A}}\text{,}
$$
$$
\mathbb{B}_{121}[\mathbf{{\tilde A}},v]
=-ev\left( \gamma \left( \omega -e\alpha
_{\omega ,e}\right) \mathbf{u}+e( \mathbf{a}^{{{\delta}},\chi }
) \right) \cdot\mathbf{{\tilde A}}
-ie\nabla v\cdot\mathbf{{\tilde A}}\text{,}
$$
$$
\mathbb{B}_{122}[\mathbf{{\tilde A}},\mathbf{{\tilde A}}]
=-e^{2}f_{\omega ,e}|\mathbf{{\tilde A}}|^{2}\text{,}
$$
and
\begin{equation*}
\mathbb{B}_{211}[v,v]=-e\left( \gamma \left( \omega -e\alpha _{\omega
,e}\right) \mathbf{u}+e( \mathbf{a}^{{{\delta}},\chi }) 
\right) |v|^{2}+\left\langle iev,\nabla v\right\rangle \text{,}
\end{equation*}
along with
\begin{equation*}
\mathbb{B}_{221}[\mathbf{{\tilde A}},v]
=\mathbb{B}_{212}[v,\mathbf{{\tilde A}}]
=-e^{2}\left\langle f_{\omega ,e},v\right\rangle \mathbf{{\tilde A}}\text{.}
\end{equation*}
These terms are obtained by differentiation of the cubic part of the
expanded Hamiltonian, which is 
\ba
\tilde H^{(3)}&=&
-\frac{1}{2}\left\langle (
v,\mathbf{{\tilde A}}) ,
\mathbb{B}^{(2)}(
v,
\mathbf{{\tilde A}})\right\rangle _{L^{2}}\notag\\
&=&\langle\nabla
v-i\gamma\mathbf{u}(\omega-e\alpha_{\omega,e})v-ie\mathbf{a}^{{\delta},\chi} v,
-ie\mathbf{\tilde A}v\rangle_{L^2}+e^2\langle\mathbf{\tilde A}f_{\omega,e},\mathbf{\tilde A}v\rangle_{L^2}.
\notag\ea

Finally the cubic terms in the equations arise by differentiation of
the quartic part of the Hamiltonian
$$
\tilde H^{(4)}=
-\frac{1}{2}\left\langle (
v,\mathbf{{\tilde A}}
) ,\mathbb{B}^{(3)}(
v,
\mathbf{{\tilde A}}
)\right\rangle _{L^{2}}
=
\frac{e^2}{2}\int |\mathbf{\tilde A}|^2|v|^2,
$$
and
are determined by a rank four tensor, 
 $$\mathbb{B}^{(3)}:\left( H^{1}\left( {{\Real}}^{3};%
\mathbb{C}\right) \oplus {\dot H}^{1}\left( {{\Real}}^{3};{{\Real}}%
^{3}\right) \right) ^{3}\mapsto H^{-1}\left( {{\Real}}^{3};\mathbb{C}%
\right) \oplus {\dot H}^{-1}\left( {{\Real}}^{3};{{\Real}}%
^{3}\right) $$ 
which, using an identical notation to the rank three case, has as its
only non-zero entries
\begin{equation}
\mathbb{B}_{1122}=\frac{-e^{2}}{3}
\label{Eq:C}
\end{equation}
and the other entries obtained by permuting the indices. 

\subsubsection{Some bounds for the inhomogeneous terms} 
\la{boundin}
We record here
some simple bounds for the quantities defined above:

\begin{itemize}
\item $\|\jmath_1^{II}\|_{L^p}+\|\jmath_2^{II}\|_{L^p}=O(e)$ and
$\|\jmath_4^{II}\|_{L^p}=O(e^2)$ for every $p\in[1,\infty]$ by
\eqref{Eq:a0mu},\eqref{Eq:amu},
\item $\|hf_{\omega, e}\|_{L^p}
=O(e+|\dot\lambda-V_0|+e|\tilde A_0|_{L^q})$, for any
$q>3$, 
which can be read off from \eqref{Eq:h}, using 
results from appendices \ref{Sec:alpha Estimates} \ref{a0est} and
\ref{eref}, and the assumptions on the applied fields.
${\tilde A}_0$ can be bounded in $L^q,\,q>3$ by 
appendix \ref{a0est}.
\item It is possible to write $h=h_1-e{\tilde A}_0$ with
$\|\nabla h_1\|_{L^\infty}=O(e+|{\dot\lambda}-V_0|)$
and $\nabla{\tilde A}_0$ bounded in $L^p,\,p\in (3/2,3]$, 
by appendix \ref{a0est}.
\end{itemize}
\noindent
Finally, consider ${\mathcal N}$: by lemma \ref{lem:different}
we can write
\ba
{\mathcal N}(f_{\omega ,e},f_{\omega},v)
&=&\beta(|f_{\omega ,e}+v|)(f_{\omega ,e}+v)
-\beta(|f_{\omega ,e}|)f_{\omega,e}\notag\\
&&\;\quad -\beta (|f_{\omega,e}|)v
-f_{\omega,e}\beta ^{\prime }(|f_{\omega,e}|){\Re} v+O(e^2|v|)\notag\\
&=&{\mathcal N}(f_{\omega ,e},f_{\omega,e},v)+O(e^2|v|)
\label{rrr}\ea
Using the condition \eqref{Eq:U2prime1}, or more generally
\eqref{Eq:U2prime2}, and the fundamental theorem of calculus,
we can estimate 
\beq
|{\mathcal N}(f,f,v)|
\leq
c
\left( 1+|f|^3\right)
\left( |v|^{2}+|v|^{5}\right),
\la{faf}\eeq
for any $f$. Therefore,
choosing $f=f_{\omega,e}$, which is bounded, 
and using \eqref{rrr} we have
\beq
|{\mathcal N}(f_{\omega ,e},f_{\omega},v)|
\leq
c_1
\left( |v|^{2}+|v|^{5}\right)+c_2 e^2|v|.
\la{estn}
\eeq

\subsection{Results from modulation theory}
\label{rmodth}

The assumptions on the nonlinearity under which we are
working ensure that the Cauchy problem for 
\eqref{Eq:HKGMB} is locally well-posed in the sense of
$(WP1)$ and $(WP2)$, see \S\ref{Sec:LWP}.
Since so far $\chi$ is unknown 
(since $\lambda(t)$ and hence $\mbxi(t)$ 
are not yet determined) we cannot solve directly 
for $\Psi=(\phi,\psi,\A_j,\E_j)$
in the  background potential $a_\mu^{{\delta,\chi}}$.
Instead we exploit gauge invariance and solve for 
\beq\la{gte}
\hat\Psi=\bigl(\hat\phi,\hat\psi,\hat\A_j,\E_j\bigr)
\equiv
\bigl(e^{-ie\chi}\phi,e^{-ie\chi}\psi,
\A_j-\partial_j\chi,\E_j\bigr)=e^{-ie\chi}\cdot\Psi
\eeq
in the potential $a_\mu^\delta$, which is known. (Since
$\chi(t,\mbx)$ is harmonic in $\mbx$ this gauge 
transformation preserves both the equations 
\eqref{Eq:HKGMB} and 
the Coulomb gauge condition (see remark \ref{rgi})).
By 
proposition \ref{Prop:WpCond} on local well-posedness,
there exists a time $T_{loc}>0$
and unique solution to \eqref{Eq:HKGM}) 
with 
\beq\la{areg}
\left(\hat\Psi -\Psi_{ext}^{{{\delta}}}\right)
\in C([0,T_{loc}];{\cal H}),
\eeq
with initial data
\beq\la{aregi}
\hat\Psi(0)=
\bigl(e^{-ie\chi_0}\phi(0),e^{-ie\chi_0}\psi(0),
\A_j(0)-\partial_j\chi_0,\E_j(0)\bigr).
\eeq
\\
Once ${{\lambda}}(t)
=( \omega(t), \theta(t) ,\mbxi(t),\mathbf{u}(t))$,
and hence $\chi(t)$,
is determined, then $\Psi(t)$ is obtained from 
$\hat\Psi(t)$ by the above relation. As remarked previously, 
by proposition
\ref{Prop:WpCond} these solutions can be approximated
in energy norm by smooth solutions evolving in any of the
spaces ${\cal H}_s$ of \eqref{hsnorm} (after subtracting off
the background field). Thus, although the statement and
proof of  theorem \ref{thm:Stability} involve only the 
energy norm, it is permissible to assume smoothness of the
solutions throughout the proof.

We now state a theorem which asserts that it is 
possible to choose the soliton parameters $\lambda(t)$
in such a way that the quantity $W$ defined in 
\eqref{defw} is equivalent to the energy norm.
This is achieved by choosing $\lambda(t)$ in such 
a way that the pair $(v,w)$
satisfies some conditions which are equivalent to 
those in \eqref{sns} (after adjusting the phase). 
\begin{theorem}
\label{rModEq} 

(a) Let $\hat\Psi $ 
be a solution to the Cauchy problem for (\ref
{Eq:HKGMB}) satisfying \eqref{areg} with initial
data \eqref{aregi} with $\Psi(0)$
as described in theorem \ref{thm:Stability}.
Then, for sufficiently small $e$, 
there exists $T_1>0$ and ${{\lambda}}
\in C^{1}([0,T_{1}];\widetilde{O}_{stab})$
with the following properties.
On the interval $[0,T_1]$ define 
$\Psi(t)=(\phi(t),\psi(t),\A_j(t),\E_j(t))$ by \eqref{defg}
and \eqref{gte}. Then
it is possible to write $\Psi$ 
in the form \eqref{Eq:Ansatz} where
$v,w$ are constrained to satisfy
\beq
\Omega_0\bigl((v,w),
\widetilde{\partial_{\mathbf\lambda}}(\phi_{S,0},\psi_{S,0})\bigr)=0,
\label{cc}\eeq
where we define
\begin{equation}
\widetilde{\partial_{{\lambda}}}\phi _{S,0}=Exp[-i\left( \Theta
\right) ]\partial_{{\lambda}}\phi _{S,0}
\text{,}
\end{equation}
and likewise for $\widetilde{\partial_{{\lambda}}}\psi
_{S,0}$. Furthermore, the function $t\mapsto\lambda(t)$ solves
a system of differential equations \eqref{Eq:ModEq}. 
The condition \eqref{cc} is equivalent to requiring
$\bigl(\phi-\phi_{SC,e},\psi-\psi_{SC,e})\in
N_{\lambda}.$

(b) If $e$ and $\| (v,w,\mathbf{{\tilde A}},\mathbf{{\tilde E}})\|
_{\cal H}$ are sufficiently small, then
\beq
|\dot\lambda-V_0(\lambda)|=O\left(e+\| (v,w,\mathbf{{\tilde A}},\mathbf{{\tilde E}}%
)\| _{\cal H}^2
\right),
\la{ss}
\eeq
so that, in particular, if 
$\| (v,w,\mathbf{{\tilde A}},\mathbf{{\tilde E}})\| _{\cal H}^2=O(e)$
then
\beq
|\dot\lambda-V_0(\lambda)|=O\left(e\right).
\la{bv}
\eeq
\end{theorem}

\noindent\proof
This is a consequence of the lemmas 
in \S\ref{modth}.
\myqed

\subsection{The main growth estimate}\la{mgre}
As discussed in \S\ref{sab}, the natural quantity 
for stability
and perturbation analyses of the solitons \eqref{lbs} is the Hessian
of the augmented Hamiltonian. Here we modify this quantity to take 
account of the phase shifts in \eqref{Eq:Ansatz}, and discard terms
which are formally $O(e)$, leading us to the introduction of the
following quadratic form:
\begin{equation}
\la{defw}
W(v,w,\mathbf{{\tilde A}},\mathbf{{\tilde E}};{{\lambda}})
=K+\Xi \text{,}
\end{equation}
where
\begin{equation}
K(\mathbf{{\tilde A}},\mathbf{{\tilde E}};{{\lambda}})=\frac{1}{2}
\left(\| \mathbf{\tilde E}\| _{L^{2}}^{2}+\| \nabla \times \mathbf{{\tilde A}}\|
_{L^{2}}^2+2\langle \mathbf{\tilde E},
\left( \mathbf{u}\cdot\nabla \right) 
\mathbf{{\tilde A}}%
\rangle _{L^{2}}\right) \text{,}
\end{equation}
and
\begin{equation}
\Xi (v,w;{{\lambda}})=\frac{1}{2}\left( \left\| w-i\gamma \omega
v\right\| _{L^{2}}^{2}+\left\langle v,M_{{{\lambda}}}-\gamma
^{2}\omega ^{2})v\right\rangle _{L^{2}}
+2\left\langle w,\mathbf{u}\cdot\nabla
v\right\rangle _{L^{2}}\right) \text{,}
\end{equation}
where $M_{{{\lambda}}}$ is as defined in (\ref{Eq:Mlambda}).

\begin{theorem}[Equivalence of $W$ and energy norm]
\label{thm:WNormEq} Suppose that the nonlinearity
is such that (H1)-(H3) and {\em (SOL), (KER)} and 
{\em (POS)} hold.
Suppose further that $\mathbf{%
\lambda }$ lies in a compact subset, $\mathbf{K}$, of 
$\widetilde{O}_{stab}$. 
Then the quadratic form 
$W$ just defined,
is equivalent uniformly on $\mathbf{K}$ to 
$\| (v,w,\mathbf{{\tilde A}},
\mathbf{{\tilde E}})\| _{\cal H}^2$
provided that $( v,w) $ satisfy the 
constraints \eqref{cc}.
\end{theorem}

\noindent\proof
This is essentially theorem $2.7$ in \cite{Stuart}. 
Since there is no coupling in $W$ between $(v,w)$ and
$(\mathbf{{\tilde A}},\mathbf{{\tilde E}})$, it is 
only necessary to show separately the equivalence of 
$\Xi$ and $K$ to the corresponding parts of 
$\| (v,w,\mathbf{{\tilde A}},
\mathbf{{\tilde E}})\| _{\cal H}^2$. For $K$ this
can be achieved by completing the square (since
$\| \nabla \times \mathbf{{\tilde A}}\|
_{L^{2}}=\|\mathbf{{\tilde A}}\|_{\hh}$ by the 
Coulomb condition), while for $\Xi$ it is an
immediate consequence of (POS).
\myqed

\begin{theorem}[Main growth estimate]
\label{thm:Wgrowth} Assume given a solution to the Cauchy problem
for (\ref{Eq:HKGM}) for which theorem \ref{rModEq} applies on an
interval $[0,\frac{T_2}{|e|}]$ for some 
fixed positive $T_2$. Assume that $\lambda(t)\in \mathbf{K}$,
a compact subset of $\widetilde{O}_{stab}$, so that
by theorem \ref{thm:WNormEq} there exists
$c_1>0$ such that, 
\beq
\frac{1}{c_1}W\leq \| (v,w,\mathbf{{\tilde A}},
\mathbf{{\tilde E}})\| _{\cal H}^2\leq c_1 W,
\la{sto}
\eeq
on $[0,\frac{T_2}{|e|}]$.
Assume further that there exist $c_2>0,c_3>0$ such that
that ${\delta}^2\leq c_2 |e|$ and and $W\leq c_3|e|$, and that
$e=o({\delta})$.
It follows that, for sufficiently small $e$,
there exists $c_{4}>0$ such that, on $[0,\frac{T_2}{|e|}]$
\begin{equation}
W(t)\leq c_{4}(W(0)+e^{2}+{\delta}^2)\exp (c_{4}|e|t)
\text{.}
\end{equation}
\end{theorem}

\noindent\proof
See \S\ref{Sec:Proof of dWtildedt}.
\myqed

\subsection{Completion of the proof of theorem \ref{thm:Stability}}

\subsubsection{Local solution verifying constraints}

For simplicity of exposition 
we first prove part (i) of the theorem, i.e.
we consider initial data $\Psi(0)$
consisting of an exact soliton as in \eqref{lbs} determined by
parameters $\lambda(0)=(\theta(0) ,\omega(0) ,\mathbf{u}(0),
\mbxi(0))\in\widetilde{O}_{stab}$,
with $\omega(0)$ satisfying the stability condition. 
On account of the applied fields there will be a non-trivial evolution
starting from this initial value.
Applying the local existence theorem \ref{Prop:WpCond}, and theorem
\ref{rModEq} as in \S\ref{rmodth},
we deduce the existence a positive time $T_1>0$ such that
on the interval $[0,T_1]$ there is a solution 
to the Cauchy problem
which can be written as in \eqref{Eq:Ansatz} where $v(0)=0=w(0)$,
and $\bigl(v(t),w(t)\bigr)$ satisfy the constraints 
\eqref{Eq:Consa}
(or \eqref{cc}), and $t\mapsto\lambda(t)$ solves \eqref{Eq:ModEq}.
We may assume that $\lambda(t)\in\mathbf{K}$, a fixed 
compact subset of $\widetilde{O}_{stab}$, 
so that \eqref{sto} holds.

\subsubsection{Growth of the energy norm}\la{gen}
Since we have a local solution satisfying the constraints \eqref{cc}
we can assume that the conclusions of
theorem \ref{thm:WNormEq} hold.
Furthermore, by continuity we may assume (making $T_1$ smaller if need
be) that on this interval $W(t)\leq c_3|e|$, and \eqref{sto} holds.
Now apply the growth estimate in theorem \ref{thm:Wgrowth}:
$$
W(t)\leq c_{4}(W(0)+e^{2}+{\delta}^2)\exp (c_{4}|e|t)
\text{,}
$$
to deduce by a standard continuation argument, 
since $W(0)=0$ and ${\delta}^2=o(e)$, that there exists
an interval $[0,\frac{T_0}{|e|}]$, with $T_0>0$ 
fixed (independent of $e,{\delta}$), on which
$$
W(t)\leq c_5(e^2+{\delta}^2)=o(e)
$$
which completes the proof of theorem \ref{thm:Stability} for the case
of exact soliton initial data - part (i) of theorem 
\ref{thm:Stability}.

\subsubsection{General initial data}\la{mg}
Part (ii) of theorem \ref{thm:Stability} says that the behaviour
described in part (i) also holds for nearby initial data: for a
precise formulation it is necessary to consider the initial data
for the gauge transform $\hat\Psi$:

\begin{theorem}\la{thm:Stabilityg} Under the same
assumptions as theorem \ref{thm:Stability},
let $\hat\Psi $ 
be a solution to the Cauchy problem for (\ref
{Eq:HKGMB}) with $(\hat\Psi -\Psi_{ext}^{{{\delta}}})
\in C(\R;{\cal H})$ and initial data
$
\hat\Psi(0)
=(\hat\phi(0),\hat\psi(0),\hat\A_j(0),\hat\E_j(0))
$
having the following property.
There exists
$\widetilde{{{\lambda}}}=(\widetilde{\theta },
\widetilde{\omega },\widetilde{\mathbf{u}},
\widetilde{\mbxi})
\in\widetilde{O}_{stab}
$
such that if we define
$\widetilde\chi(\mbx)=-
(\mathbf{x}-\widetilde\mbxi)\cdot\mathbf{a}^{{{\delta}}
}(0,\widetilde\mbxi)$, then
\begin{equation}
\kappa_0\equiv
\left\| e^{-ie\widetilde\chi}\cdot
\hat\Psi (0)-\Psi_{ext}^{{{\delta}},\widetilde\chi }(0)
-\Psi _{SC,e}
(\widetilde\lambda)\right\| _{\cal H}=o(e^{\frac{1}{2}}).
\la{te}\end{equation}
It follows
that, if $e$ is sufficiently small 
there exists 
$T_0>0$, 
$\chi(t,\mbx)$ 
and ${{\lambda}}(t)\in C^{1}
(\mathbb[0,\frac{T_0}{|e|}],\widetilde{O}_{stab})$,
all as in theorem \ref{thm:Stability}, such that if
$\Psi (t)$ is defined as in \eqref{gte}
it satisfies all the conclusions of part (i) of theorem
\ref{thm:Stability}.
\end{theorem}
\noindent\proof
It is only necessary to
argue, as in the proof of lemma \ref{lem:InitDat},
that under the stated conditions
there exists $\lambda(0)\in \widetilde{O}_{stab}$ 
with $|\lambda(0)-\widetilde\lambda|=o(e^{\frac{1}{2}})$
such that
$
\Psi(0) =(\phi(0),\psi(0),\A_j(0),\E_j(0))\equiv
e^{-ie\chi_0}\cdot
\hat\Psi (0)$ can be written as
$$\Psi(0)=
\left(\phi_{SC,e}(\lambda(0))+\tilde\phi(0),
\psi_{SC,e}(\lambda(0))+\tilde\psi(0) ,{\A}_i(0), {\E}_i(0)\right),$$
with $$\bigl(\tilde\phi(0),\tilde\psi(0)\bigr)
\in N_{\mathbf\lambda(0)}$$ 
where $N_{\mathbf\lambda(0)}$ is 
the symplectic normal subspace, of codimension eight,
defined in \eqref{sns}. This is a simple consequence of the
implicit function theorem, as is lemma \ref{lem:InitDat}. There
is only a slight modification required in that
$\phi(0)=e^{-ie\chi_0}\hat\phi(0)$ depends on $\lambda(0)$,
and so does $\psi(0)$,
unlike the case considered in that lemma. However for small $e$
this has no effect on the non-degeneracy  condition required to 
apply the implicit function theorem. (Also the fact that $\chi_0$
grows linearly in $\mathbf{x}$ can easily be handled using the 
exponential decay in $\mathbf{x}$ of $\phi_{SC,e},\psi_{SC,e}$ 
and their derivatives.)

Now using $|\lambda(0)-\widetilde\lambda|=o(e^{\frac{1}{2}})$ we can deduce
from \eqref{te} that
$W(0)=o(e)$. Indeed for the electromagnetic components this is
immediate since the gauge transformation leaves the electric field
unchanged, and only shifts $\A_j$ by $\partial_j\chi_0$, and this
shift is put onto the background potential (and so does not
contribute to $W(0)$ since $\tilde\mbA$ is unchanged). The change of the
electromagnetic components of the soliton induced by the change
of $\widetilde\lambda$ to $\lambda(0)$ are easily estimated
in energy norm as $O(|\widetilde\lambda-\lambda(0)|)$ by 
lemmas \ref{lem:DalphaLambda} and \ref{lem:Laplacezeta}.
For the other
components we just use phase invariance to estimate, e.g.
\begin{align*}
\|e^{-ie\chi_0}\hat\phi(0)-\phi_{SC,e}(\lambda(0))\|_{L^2}
&=\|\hat\phi(0)
-e^{ie\chi_0}\phi_{SC,e}(\lambda(0))\|_{L^2}\\
&\leq 
\|\hat\phi(0)-e^{ie\widetilde\chi}\phi_{SC,e}
(\widetilde\lambda)\|_{L^2}
+
\|e^{ie\widetilde\chi}\phi_{SC,e}(\widetilde\lambda)
-e^{ie\chi_0}\phi_{SC,e}(\lambda(0))\|_{L^2}\\
&\leq \kappa_0+O(|\lambda(0)-\widetilde\lambda|)=o(e^{\frac{1}{2}}).
\end{align*}

From this point on, the argument can be completed as before: 
since
$\bigl(\tilde\phi(0),\tilde\psi(0)\bigr)
\in N_{\mathbf\lambda(0)}$ is equivalent to the conditions 
\eqref{cc}, theorems \ref{rModEq} and \ref{thm:Wgrowth}
can be applied to produce a local solution satisfying the growth
estimate in \S\ref{gen}.

\section{Modulation theory}
\label{modth}

In this section we state and prove some theorems which imply theorem
\ref{rModEq}, which is needed in the proof of the main results 
(theorems \ref{thm:Stability} and
\ref{Thm:Lorentz Force Law}). The proofs are a direct application of the developments
in \cite{Stuart}, and so the presentation will be brief and reference made to 
\cite{Stuart,Thesis} for some of the calculations. The crucial point is that the conditions
\eqref{cc} are equivalent to a locally well-posed set of 
ordinary differential equations.
Recall from \eqref{es} that, for $e=0$, the soliton solutions are of the form
$
\bigl(\phi_{S,0},\psi_{S,0})(\mathbf{x};{{\lambda}}\bigr)
\equiv e^{i\Theta}\bigl(f_{\omega}(\mathbf{Z}),
(i\gamma\omega f_{\omega}(\mathbf{Z})
-\gamma \mathbf{u}\cdot\nabla _{\mathbf{Z}}f_{\omega}(\mathbf{Z})) \bigr)
$
with $\lambda(t)$ an integral curve of the vector field $V_0$. 
Explicitly, the conditions \eqref{cc} read

\begin{equation}
\left\langle v,\widetilde{\partial_{\lambda }}\psi _{S,0}(%
\mathbf{\lambda })\right\rangle _{L^{2}}-\left\langle w,\widetilde
{\partial_{\lambda_A }}\phi _{S,0}(\mathbf{\lambda })\right\rangle _{L^{2}}=0
\label{Eq:Consa}
\end{equation}
for $A=-1,0,...,6$.

In the next two subsections we state two lemmas which prove that these 
constraints can be enforced thoroughout a time interval:
\begin{itemize}
\item
The first shows that by an appropriate choice of
$\lambda(0)$, they can be assumed to hold in an 
open neighbourhood of the set of stable solitons in 
the phase space ${\cal H}$. This shows that the class
of initial data considered in part (ii) of theorem 
\ref{thm:Stability} forms an open set containing the stable
solitons.
\item
The second
shows that an appropriate choice of 
$\partial_t\lambda$ implies that they are preserved
for later times.
\end{itemize}

\subsection{Preparation of the initial data}

\begin{lemma}
\label{lem:InitDat} Suppose that there exists $\widetilde{\mathbf{\lambda }}%
=(\widetilde{\theta },\widetilde{\omega },\widetilde{\mathbf{u}},\widetilde{%
\mbxi})\in\widetilde{O}_{stab}$ 
(so that \eqref{stab} holds with
${\omega} =\widetilde{\omega }$). Then, there exists $%
e( \widetilde{\mathbf{\lambda }}) $, 
$\kappa ( \widetilde{%
\mathbf{\lambda }},e) $, such that, if $|e|<e(
\widetilde{\mathbf{\lambda }}) $ and
\begin{equation}
\widetilde{\kappa_1 }=\| \phi (0)-\phi _{SC,e}( \widetilde{%
\mathbf{\lambda }}) \| _{H^{1}}+\| \psi (0)-\psi
_{SC,e}( \widetilde{\mathbf{\lambda }}) \|
_{L^{2}}<\kappa \text{,}
\end{equation}
there exists $\mathbf{\lambda }(0)\in\widetilde{O}_{stab}$ 
depending differentiably upon $%
(\phi (0),\psi (0))$ such that \newline $(v(0),w(0))$, determined by 
the first two equations of \eqref{Eq:Ansatz} at $t=0$,
satisfy the constraints \eqref{Eq:Consa} with $\lambda=\lambda(0)$.
Furthermore there exists  $c_{1}>0$ such that
\begin{equation}
|\lambda(0)-\widetilde{\lambda}|
+\left\| \phi (0)-\phi _{SC,e}( \mathbf{\lambda }(0))
\right\| _{H^{1}}+\left\| \psi (0)-\psi _{SC,e}( \mathbf{\lambda
}(0)) \right\| _{L^{2}}<c_{1}{\widetilde{\kappa}_1 }\text{.}
\end{equation}
\end{lemma}

\noindent\proof
The condition in \eqref{stab} allows this to be deduced from the implicit function theorem,
see \cite[\S2.3]{Stuart} or \cite{Thesis} for details.
\myqed

\subsection{Modulation equations and constraints}

\begin{lemma}
\label{lem:ModEq} 
Let
$\mathbf{\lambda }(0)\in \widetilde{O}_{stab}$ 
and $(v(0),w(0))$ be as given in the
conclusions of lemma \ref{lem:InitDat}. Let 
$\hat\Psi $ be a solution to the
Cauchy problem for (\ref{Eq:HKGMB}) on the time interval 
$[0,T_{loc}]$ 
with regularity as in \eqref{areg}, and such that
\begin{equation}
\sup_{\lbrack 0,T_{loc}]}
\left\| \hat\Psi (t)-\Psi_{ext}^{{{\delta}} }(t)
\right\| _{\cal H}<N_{0}\text{.}
\end{equation}
Fix a compact subset $\mathbf{K}$ of the stable parameter set
$\widetilde{O}_{stab}$, which is the closure of an open
neighbourhood of $\lambda(0)$.
Then, there exists $\kappa _{2}>0$ and $T_{1}>0$ such that,
if $\left\| (v(0),w(0))\right\| _{H^{1}\oplus L^{2}}<\kappa _{2}$,
there exists $\lambda(t)
\mathbf{\in }C^{1}([0,T_{1}];\mathbf{K})$
such that the constraints (\ref{Eq:Consa}) are satisfied for
$0\le t\le T_1$, where $v,w$ are as in 
\eqref{Eq:Ansatz} with
$\Psi$ obtained from $\hat\Psi$ via 
\eqref{defg} and \eqref{gte}.
The function $t\mapsto\lambda(t)$ is a solution of a 
system of
ordinary differential equations \eqref{Eq:ModEq}.
\end{lemma}

\noindent\proof
The proof of this is essentially the same as \cite[\S\,2.5]{Stuart}. 
For clarity it is divided into three stages.

\subsubsection{Beginning of proof of lemma \ref{lem:ModEq}}
Equations \eqref{de1} and \eqref{de2}
define a linear operator $\tilde {\cal M}_\lambda$ in an obvious way:
\begin{equation}
\widetilde{\cal M}_\lambda(v,w)=\bigl(
-\partial_t v-i\omega\gamma v+w,
-\partial_t w-i\omega\gamma w-M_\lambda v
\bigr).
\la{dm}
\end{equation}
and let $\widetilde{\cal M}_\lambda^*$ be the formal $L^2(dxdt)$ adjoint of this
operator. Then, by \cite[\S\,2.5]{Stuart}, 
there exists an $8\times 8$ matrix $D_{AB}$ such that
\begin{equation}
\widetilde{\cal M}_\lambda^*(-\widetilde{\partial_{\lambda_A }}\psi_{S,0},
\widetilde{\partial_{\lambda_A }}\phi _{S,0})
=\sum_B D_{AB}(-\widetilde{\partial_{\lambda_B }}\psi _{S,0},
\widetilde{\partial_{\lambda_B }}\phi _{S,0})
+({\tilde{\bf I}}^1_{A},{\tilde{\bf I}}^2_{A}).
\label{m21}
\end{equation}
where the inhomogeneous terms ${\tilde{\bf I}}^j_{A}$ are 
proportional to $\dot\lambda-V_0(\lambda)$:
$$
{\tilde{\bf I}}^j_{A}={\tilde{I}}^j_{AB}(\dot\lambda-V_0(\lambda))_B
$$
with ${\tilde{I}}^j_{AB}$ smooth functions of $x$, which are
exponentially decreasing as $|x|\to\infty$; the precise formulae, which
are unimportant here, can be found in \cite[\S\,2.5]{Stuart}.
A simple integration by parts then shows that the constraints
in \eqref{Eq:Consa}
are satisfied 
on an interval containing the initial time, if they hold
at that initial time and if the following is true
\ba
\label{r2m}
&&
\langle -\widetilde{\partial_{\lambda_A }}\psi _{S,0},j_1\rangle_{L^2} 
+\langle \widetilde{\partial_{\lambda_A }}\phi _{S,0},j_2+{\cal N}\rangle_{L^2} \\
&&\quad+\langle {\tilde{\bf I}}^1_{A}-i h \widetilde{\partial_{\lambda_A }}\psi
_{S,0},
v\rangle_{L^2} +\langle
{\tilde{\bf I}}^2_{A}+i h \widetilde{\partial_{\lambda_A }}\phi
_{S,0},w\rangle_{L^2} =0,
\notag
\ea
for all $A=-1,0,\dots 6$, and at each time in the interval.
A calculation as in \cite{Stuart}, which is reviewed in the
next stage of the proof in \S\ref{modcalc},
shows that these latter conditions are equivalent to the following
system of differential equations
\begin{equation}
\bigl( \mathbb{M}(e)_{AB}+\mathbb{j}_{AB}(v,w,{{\lambda}})
\bigr)\bigl({\dot\lambda-V_0(\lambda)}\bigr)_B
=\mathbf{F}_{A}(e,\Psi_{ext}^{{{\delta}} },\Psi ,\mathbf{\lambda )}\text{,}
\label{Eq:ModEq}
\end{equation}
where $\mathbb{M}(e)_{AB}$ is defined in (\ref{Eq:Mass}), 
$\mathbb{j}_{AB}$ is defined in (\ref{Eq:j1}),
$\mathbf{F}_{A}$ is given by (\ref{Eq:Force}) and where
the indices $A,B\in \{-1,0,1,...,6\}$, and we sum over the repeated index $B$.

\subsubsection{Explicit computation of the modulational equation 
\eqref{Eq:ModEq}}
\la{modcalc}
We write out explicitly the various terms in the conditions
\eqref{r2m}. The first thing to note is that the overall expression is
affine in $(\dot\lambda-V_0(\lambda))$ so we divide into the {\it inertial} terms, which are 
proportional to this quantity (and give rise to the left hand side of \eqref{Eq:ModEq}),
and the remaining {\it force} terms, which give rise to the right hand side of 
\eqref{Eq:ModEq}. The dominant contribution to the inertial terms
arises from $\jmath^I_1,\jmath^I_2$, while that to the force terms
arises from $\jmath^{II}_1,\jmath^{II}_2$.

To describe the inertial terms 
we need the following matrix, which, to highest order,
describes the mass of the soliton:
\begin{equation}
\mathbb{M}_{AB}(e)=\left\langle \widetilde{\partial_{{\lambda}_{A}}}%
\psi _{S,0},e^{-i\Theta_c}{\partial_{{\lambda}_{B}}}
\phi_{SC,e}\right\rangle_{L^2} -\left\langle 
\widetilde{\partial_{{\lambda}_{A}}}\phi _{S,0},
e^{-i\Theta_c}{\partial_{{\lambda}_{B}}}
\psi_{SC,e}\right\rangle_{L^2} \text{.}  \label{Eq:Mass}
\end{equation}
Then the dominant inertial term is
$$
\langle -\widetilde{\partial_{\lambda_A }}\psi _{S,0},{\jmath}_1^I\rangle_{L^2} 
+\langle \widetilde{\partial_{\lambda_A }}\phi
_{S,0},{\jmath}_2^I\rangle_{L^2}
=\mathbb{M}_{AB}(e)(\partial_t\lambda-V_0(\lambda))_B.
$$
Next, we have the following matrices, which may be thought of as corrections
- owing to the presence of the perturbations $v$ and $w$ - to the
``inertia'' matrix above :
\begin{equation}  \label{Eq:j1}
\mathbb{j}_{AB}
=\left\langle v,\left(\tilde{I}^1_{AB} -
i\partial_{{\lambda}_{B}}\Theta_c\widetilde{\partial_{\lambda_A }}
\psi_{S,0}
\right)
\right\rangle_{L^2} -
\left\langle w,\left( \tilde{I}^2_{AB}+i\partial_{{\lambda}_{B}}\Theta_C \widetilde{\partial_{\lambda_A }}\phi_{S,0}\right)
\right\rangle_{L^2}.
\end{equation}

We now present the abbreviations for the force terms appearing in the
modulational equation. Firstly, we have what is effectively the Lorentz
force term.
\begin{equation}
\begin{split}
\mathbf{F}_{A}^{L}&=
\langle \widetilde{\partial_{\lambda_A }}\psi _{S,0},{\jmath}_1^{II}\rangle_{L^2} 
-\langle \widetilde{\partial_{\lambda_A }}\phi
_{S,0},{\jmath}_2^{II}\rangle_{L^2}\\
&=\left\langle \widetilde{\partial_{{\lambda}_{A}}}
\psi _{S,0},ie a_0^{{\delta},\chi}f_{\omega ,e}\right\rangle_{L^2} \\
&\quad -\left\langle \widetilde{\partial_{{\lambda}_{A}}}\phi _{S,0},
ie a_0^{{\delta},\chi}\left( i\gamma (\omega -e\alpha _{\omega ,e})-%
\mathbf{u}\cdot\nabla \right) f_{\omega ,e}+e\mbR f_{\omega ,e}
\right\rangle_{L^2} \text{%
.}  
\end{split}
\label{Eq:FL}
\end{equation}
We also have a force $\mathbf{F}_A^n+\mathbf{F}_{A}^{p}$
due to the nonlinear interactions, where
\ba
\mathbf{F}_A^n &=&-\left\langle \widetilde{\partial_{{\lambda}_{A}}}
\phi_{S,0},{\mathcal N}
\right\rangle_{L^2},\\
\mathbf{F}_{A}^{p} &=&\left\langle \widetilde{\partial_{{\lambda}_{A}}}%
\psi _{S,0},
\jmath_1^{0}+
ie\left(
\gamma\alpha_{\omega,e}+a_0^{{\delta},\chi}+\tilde A_0
\right) v\right\rangle_{L^2}  \label{Eq:Fp} \\
&&\quad-\left\langle \widetilde{\partial_{{\lambda}_{A}}}
\phi_{S,0},
\jmath_2^{III}+\jmath_2^{IV}+\jmath_2^0
+ie\left( 
\gamma\alpha_{\omega,e}+a_0^{{\delta},\chi}+\tilde A_0
\right) w\right\rangle_{L^2}.\notag
\ea

We abbreviate the total force as follows:
\begin{equation}
\mathbf{F}_{A}=\mathbf{F}_{A}^{L}+\mathbf{F}_A^n+\mathbf{F}_{A}^{p}.
\label{Eq:Force}
\end{equation}
\noindent
{\bf Bound for the inertia matrix.} It follows from the
definition of $\mathbb{j}_{AB}$ that
\beq
|\mathbb{j}_{AB}|=O\left(\| (v,w,\mathbf{{\tilde A}},
\mathbf{{\tilde E}})\| _{\cal H}\right).
\la{bj}\eeq

\noindent
{\bf Bounds for the forces.}
Firstly, the main force term can be bounded as
\begin{equation}
\mathbf{F}_{A}^{L}=O(e)\text{,}
\end{equation}
because of (\ref{Eq:a0mu}), (\ref{Eq:amu}) and \eqref{Eq:eff}.
For some values of $A$ there are better bounds:
\begin{equation}\la{bb1}
\mathbf{F}_{0}^{L}=O(e^3)\text{.}
\end{equation}
Referring to \eqref{Eq:FL}, and using lemmas \ref{different}
and \ref{lem:Laplacezeta}, we deduce that
\ba
\mathbf{F}_{0}^{L}&=&O(e^3)
-\langle if_\om,e\mbR f_\om\rangle_{L^2}\notag\\
&&+\langle \widetilde{\partial_{\theta }}\psi _{S,0},
iea_0^{{\delta},\chi}f_{\om}\rangle_{L^2} 
-\langle \widetilde{\partial_{\theta }}
\phi_{S,0},iea_0^{{\delta},\chi}(i\om\gamma f_\om-\mathbf{u}\cdot\nabla f_\om)
\rangle_{L^2}.\notag
\ea
By the reality of $f_\om$ and the Coulomb condition the last three
terms vanish, proving the bound \eqref{bb1}.
Also, for $A=3+j$ we have an improvement:
\begin{equation}
\mathbf{F}_{3+j}^{L}=O(e^2+e{\delta}).
\la{bb2}\end{equation}
To establish this, we first argue as above that
\ba
\mathbf{F}_{3+j}^{L}&=&O(e^3)
-\langle \widetilde{\partial_{u^j }}
\phi_{S,0},e\mbR f_\om\rangle_{L^2}\notag\\
&&+\langle \widetilde{\partial_{u^j }}\psi _{S,0},
iea_0^{{\delta},\chi}f_{\om}\rangle_{L^2} 
-\langle \widetilde{\partial_{u^j }}
\phi_{S,0},iea_0^{{\delta},\chi}
(i\om\gamma f_\om-\mathbf{u}\cdot\nabla f_\om)
\rangle_{L^2}.\notag
\ea
Now referring to the formulae in \ref{Sec:Identities} we see
that $\widetilde{\partial_{u^j }}\phi_{S,0}=even + i odd$,
while $\widetilde{\partial_{u^j }}\psi_{S,0}=odd + i even$
where $even$ (resp. $odd$) means a real valued function which
is even (resp. odd ) as a function of $\mathbf{Z}$. The bound
asserted then follows by inspection and use of lemma \ref{err}.

Next, \eqref{estn}
implies, by \eqref{Eq:3 Derivatives of f}),
\eqref{eq:3 Derivatives of dfdomega} ,
\eqref{Eq:eff} and by the 
H\"{o}lder and Sobolev inequalities,
that
$$
|{\mathbf F}_A^n| =O\left(e^2\| (v,w,\mathbf{{\tilde A}},
\mathbf{{\tilde E}})\|_{\cal H}\right) +
O\left(\| (v,w,\mathbf{{\tilde A}},\mathbf{{\tilde E}}%
)\| _{\cal H}^2+
\| (v,w,\mathbf{{\tilde A}},\mathbf{{\tilde E}}%
)\| _{\cal H}^{5}\right).
$$

Finally 
\beq
|{\mathbf F}_A^p| =O\left(e\| (v,w,\mathbf{{\tilde A}},
\mathbf{{\tilde E}})\|_{\cal H}
+e\| (v,w,\mathbf{{\tilde A}},
\mathbf{{\tilde E}})\| _{\cal H}^2
+e^2\| (v,w,\mathbf{{\tilde A}},
\mathbf{{\tilde E}})\| _{\cal H}^{3}\right).
\la{bfp}\eeq
This is obtained directly from the formula above by means
of the Sobolev and H\"older inequalities and using the bounds in 
\S\ref{eref} and \S\ref{a0est}.

\subsubsection{Completion of proof of lemma \ref{lem:ModEq}}
The matrix $\mathbb{M}(e)_{AB}$ is invertible for small $e$ on account
of the stability condition \eqref{stab} and lemma
\eqref{lem:different}. Also the matrix $\mathbb{j}_{AB}$ is small when
$(v,w)$ is small, so that in this case the system of 
evolution equations (\ref{Eq:ModEq})
can be manipulated - as in the proof of theorem 2.6 in \cite{Stuart} - to
form a system of equations of the form
$$
\dot\lambda=V_0(\lambda)+
V_1(e,\Psi_{ext}^{{{\delta}} },\hat\Psi ,\mathbf{\lambda )}.
$$
This is almost a locally well-posed system of ordinary differential equations - there is a slight modification of the standard proof from \cite{Stuart} required: 
$\hat\Psi$ is
known to exist already, 
but $(v,w)$, determined as in the statement, 
depend on $\lambda(t)$ 
through the gauge transformation \eqref{defg}, which is 
nonlocal in the $\mbxi$ component of $\lambda$, and so 
$V_1$ is similarly nonlocal. To allow for
this it is necessary to augment $\lambda$ by the nonlocal 
quantity appearing in \eqref{defg}, which is in fact
$\chi(t,\mbxi)$. Call $\Lambda=(\lambda,\chi(t,\mbxi(t)))$, 
then there is a locally well-posed system of 
ordinary differential equations of the form
$\dot\Lambda=W({\Lambda )},e,\Psi_{ext}^{{{\delta}} },\hat\Psi)$,
allowing the proof of 
lemma \ref{lem:ModEq} to be completed in the
same way in \cite{Stuart}.
\myqed

\subsection{A bound for $\dot\lambda$}

\begin{lemma}
\label{Thm:Control of dLambdadt}
In the situation of the previous lemma,
$$
|\dot\lambda-V_0(\lambda)|=O\left(e+\| (v,w,\mathbf{{\tilde A}},\mathbf{{\tilde E}}%
)\| _{\cal H}^2
+e\| (v,w,\mathbf{{\tilde A}},\mathbf{{\tilde E}}%
)\| _{\cal H}
\right)
$$
in the limit of $e$ going to zero.
\end{lemma}
\noindent\proof
The function $\lambda(t)$ is obtained as a solution
of the modulation equations \eqref{Eq:ModEq}. Referring
to the bounds for the inertial matrix and forces in 
\S\ref{modcalc}, it is immediate that for
$e,\| (v,w,\mathbf{{\tilde A}},
\mathbf{{\tilde E}})\| _{\cal H}$ sufficiently small
the bound claimed holds.
\myqed

\section{\label{Sec:Lorentz Force Law}The Lorentz force law: proof of 
theorem \ref{Thm:Lorentz Force Law}}

The starting point is (\ref{Eq:ModEq}). Define
\begin{equation}
\mathbb{M}_{AB}(0)=
\left\langle 
\widetilde{\partial_{{\lambda}_{A}}}\psi _{S,0},
\widetilde{\partial_{{\lambda}_{B}}}
\phi_{S,0}\right\rangle _{L^{2}}-
\left\langle \widetilde{\partial_{{\lambda}_{A}}}
\phi _{S,0},\widetilde{\partial_{{\lambda}_{B}}}
\psi_{S,0}\right\rangle _{L^{2}}\text{,}
\end{equation}
and observe that by lemmas \ref{lem:different} and
\ref{lem:Laplacezeta}
$\mathbb{M}_{AB}(e)-\mathbb{M}_{AB}(0)=O(e^2)$. 
Using this, and referring
to the decomposition of $\mathbf{F}_A$ in equation \eqref{Eq:Force}, and the
associated bounds following it, we infer that
\begin{equation}\la{infer}
\Bigl( \mathbb{M}(0)_{AB}+O\bigl( e^{{2}}+ 
\widetilde{W}^{\frac{1}{2}}\bigr) 
\Bigr) \bigl(\dot\lambda-V_0\bigr)_{B}=\mathbf{F}_{A}^L+
O\bigl( e\widetilde{W}^{\frac{1}{2}}
+\widetilde{W}\bigr) \text{,}
\end{equation}
where $\mathbf{F}_{A}^L$ is as in \eqref{Eq:FL}. Since the right hand side
is known, up to the stated error term,
it is now just a matter of calculation
to obtain explicit forms for the left hand side of these 
equations, and thence to deduce theorem
\ref{Thm:Lorentz Force Law}. The calculation is done in 
\cite[\S A.7]{Stuart}, using a set of functions defined in 
\S\ref{Sec:Identities} which are convenient
linear combinations of the $\widetilde{\partial_{{\lambda}_{A}}}
(\phi_{S,0},\psi_{S,0})$. We now record the conclusions.

Using \eqref{bv}, 
the $A=0$ component of \eqref{infer} reads:
$$
\partial_\omega (\omega \left\| f_{\omega}\right\| _{L^{2}}^{2})%
\overset{.}{\omega }=\mathbf{F}_{0}^{L}+O(e^2)
+O( \widetilde{W}) \text{,}
$$
with a formula for $\mathbf{F}_{0}^{L}$ given in \eqref{Eq:FL}
which indicates that
$
\mathbf{F}_{0}^{L}=O(e^3)
$
(see \S\ref{modcalc}),
and all together:
\begin{equation}\la{omeq}
\partial_\omega (\omega \left\| f_{\omega}\right\| _{L^{2}}^{2})%
\overset{.}{\omega }=O(e^2)+O( \widetilde{W}) \text{.}
\end{equation}
Similarly, the bound \eqref{bb2} for $\mathbf{F}_{3+j}^{L}$ implies
the following equation for the centre of the soliton:
\beq\la{xieq}
\dot\mbxi=\mathbf{u}+O(e^2)+O( \widetilde{W})+O(e{\delta}).
\eeq

Next, using \eqref{omeq} and \eqref{bv}, 
the $A=i\in\{1,2,3\}$ component of \eqref{infer} reads
\begin{equation}\la{prec}
\partial_t\Bigl[\bigl( \frac{1}{3}\left\| \nabla f_{\omega}\right\|
_{L^{2}}^{2}+\omega ^{2}\left\| f_{\omega}\right\| _{L^{2}}^{2}\bigr) 
\gamma\mathbf{u}^{i}\Bigr]
=\mathbf{F}_{i}^{L}+O( \widetilde{W})
+O(e^{2})\text{,}
\end{equation}
again with $\mathbf{F}_{i}^{L}$ given in \eqref{Eq:FL} as:
\begin{equation}
\begin{split}
\mathbf{F}_{i}^{L}&=
\left\langle \widetilde{\partial_{{\xi}_{i}}}
\psi _{S,0},ie a_0^{{\delta},\chi}f_{\omega ,e}\right\rangle_{L^2} \\
&\quad -\left\langle \widetilde{\partial_{{\xi}_{i}}}\phi _{S,0},
ie a_0^{{\delta},\chi}\left( i\gamma (\omega -e\alpha _{\omega ,e})-%
\mathbf{u}\cdot\nabla \right) f_{\omega ,e}+e\mbR f_{\omega ,e}
\right\rangle_{L^2} \text{%
,}  
\end{split}
\end{equation}
where the operator $\mbR$ is defined in \eqref{Eq:E0}.
Here, on the left hand side, $\|f_\om\|_{L^2}^2
=\int f_\om({\mathbf Z})^2 d^3{\mathbf Z}$  and
by the Lorentz transformation \eqref{defz}
$d^3{\mathbf Z}=\gamma d^3{\mathbf x}$. The inner products
on the right hand side are in $L^2(d^3\mathbf{x})$.
It remains to simplify this expression for 
$\mathbf{F}_{i}^{L}$: firstly,
\begin{align*}
\Bigl\langle \widetilde{\partial_{\xi_i} }
\psi_{S,0},ie a_0^{{\delta},\chi}f_{\omega ,e}
\Bigr\rangle_{L^2}
&-
\Bigl\langle \widetilde{\partial_{{\xi }_{i}}}
\phi_{S,0},ie  a_0^{{\delta},\chi}
\bigl( i\gamma ( \omega -e\alpha _{\omega ,e}) 
-\mathbf{u}\cdot\nabla \bigr) f_{\omega ,e}
\Bigr\rangle_{L^2} \\
&\;=\Bigl\langle \widetilde{\partial_{\xi_i} }
\psi_{S,0},ie a_0^{{\delta},\chi}f_{\omega}
\Bigr\rangle_{L^2}
-
\Bigl\langle \widetilde{\partial_{{\xi }_{i}}}
\phi_{S,0},ie  a_0^{{\delta},\chi}
( i\gamma\omega 
-\mathbf{u}\cdot\nabla ) f_{\omega}
\Bigr\rangle_{L^2}+O(e^3)\\
\intertext{by lemma \ref{lem:different},}
&\;=\Bigl\langle 
( i\gamma\omega 
-\mathbf{u}\cdot\nabla ) f_{\omega},
ie  \nabla a_0^{{\delta},\chi}
f_{\omega}
\Bigr\rangle_{L^2}+O(e^3)\\
\intertext{by integration by parts,}
&\; =e\om\|f_\om\|_{L^2}^2\Bigl[
\nabla_{i}a_{0}^{{{\delta}} }(t,{\mbxi})
-\mathbf{\dot a}^{{{\delta}} }(t,\mbxi)
-\mathbf{u}\cdot\nabla \mathbf{a}^{{{\delta}} }
(t,\mbxi)\Bigr]+O(e{\delta}+e^3),
\intertext{by \eqref{dad} and lemma \ref{absurd}.
(Again, $\|f_\om\|_{L^2}^2
=\int f_\om({\mathbf Z})^2 d^3{\mathbf Z}$.)
But also, referring to \eqref{Eq:E0},}
\Bigl\langle 
-\widetilde{\partial_{{\xi }^{j}}}
\phi_{S,0},e \mbR f_{\omega ,e}\Bigr\rangle_{L^2} 
&=\gamma \omega e\int
f_{\omega}^{2}(\mathbf{Z})\nabla 
\mathbf{u\cdot a}^{{{\delta}}}
(t,\mathbf{x})dx\text{,}\notag\\
&=\omega e
\|f_{\omega}\|_{L^2}^{2}
 \mathbf{u}_l\nabla_j
{\cdot a}_l^{{{\delta}}}(t,{\mbxi})+O(e{\delta}),\notag
\end{align*}
again using lemma \ref{absurd}.
Adding together these contributions, we end up with
$$
\mathbf{F}^{L}=e\omega 
\| f_{\omega}\|_{L^{2}}^{2}
\Bigl( \nabla a_{0}^{{{\delta}} }
-(\partial_t\mathbf{a}^{{{\delta}} })
+
\mathbf{u}\times \bigl( \nabla \times 
\mathbf{a}^{{{\delta}} }\bigr)\Bigr) (t,{\mbxi})
+O(e^{3}+e{\delta})\text{,}
$$
which is the required form of the Lorentz force law, as
given in theorem \ref{Thm:Lorentz Force Law},
once we note that
$
\int \omega f_{\omega}^{2}=\int \left( \omega -e\alpha \right) f_{\omega
,e}^{2}+O(e^{2}).
$
\myqed

\section{Proof of the main growth estimate\la{Sec:Proof of dWtildedt}}

In this section we are concerned with the proof of 
theorem \ref{thm:Wgrowth}. 
In order to control $W$ it is helpful to introduce a quantity 
$\widetilde{W}$ which allows us to take advantage of certain
cancellations occuring in the energy identity to handle some of
the nonlinear interaction terms which would otherwise be difficult
to estimate directly. The direct nonlinear interactions between
$v$ and $\mathbf{\tilde A}$ arise from terms in the Hamiltonian obtained
by expanding the expression $\frac{1}{2}\int|(\nabla-ie\A)\phi|^2$
in terms of $v,\mathbf{\tilde A}$ by means of \eqref{Eq:Ansatz}. (There 
are also indirect interactions mediated by ${\tilde A}_0$ via the
Gauss law, but these are easier to estimate.) In \S\ref{defin}
this expansion of $\frac{1}{2}\int|(\nabla-ie\A)\phi|^2$ is carried
out explicitly, and, including also the quadratic
part of the Taylor expansion of the potential $\mcn$,
leads to the introduction of the quantity:
\ba
\tilde H(v,\mathbf{\tilde A})&=&\sum_{n=2}^4\tilde H^{(n)}\notag\\
&=&
-\frac{1}{2}\sum_{n=2}^4
\left\langle(
v ,
\mathbf{{\tilde A}}),
{\mathbb{B}^{(n-1)}}(
v,
\mathbf{{\tilde A}}) 
\right\rangle _{L^{2}},
\notag\ea
where the superscript $n$ (resp. $n-1$) indicates the homogeneity
in $v,\mathbf{\tilde A}$ of the term $\tilde H^{(n)}$ in the expanded 
Hamiltonian (resp. of the term $\mathbb{B}^{(n-1)}$ in the expanded
evolution equations \eqref{de2},\eqref{de4});
see \S\ref{defin} for explicit expressions and explanations. 
Using these definitions we have an alternative form for the 
expanded evolution:
equations \eqref{de1},\eqref{de3} can be written in the form
\begin{equation}
\partial_t\bigl(
v,
\mathbf{{\tilde A}}
\bigr) =\big(
w,
\mathbf{{\tilde E}}
\bigr) -\bigl(
i(\gamma \omega +h)v,
0
\bigr) 
-\bigl(\partial_t\lambda-V_0(\lambda)\bigr)\cdot\bigl(
\widetilde{\partial_{\lambda}}\phi _{SC,e},
\partial_{\lambda}\mathbf{A}_{SC,e}
\bigr) +\bigl(
\jmath_1^0,\jmath_3^0
\bigr) +\bigl(\Phi _{11},0\bigr)
\text{,\label{Eq:d/dt(v,q)}}
\end{equation}
with $\Phi_{11}=\jmath_1^{II}$. The remaining two equations 
\eqref{de2},\eqref{de4} can be written:
\begin{multline}
\partial_t\bigl(
w ,
\mathbf{{\tilde E}}
\bigr) =
\bigl(-D_v \tilde H,
-D_{\tilde A} \tilde H\bigr)
-\bigl(
i(\gamma \omega +h)w,
0
\bigr) \\
-\bigl(\partial_t\lambda-V_0(\lambda)\bigr)\cdot\bigl(
\widetilde{\partial_{\lambda}}\psi _{SC,e},
\partial_{\lambda}\mathbf{E}_{SC,e}
\bigr) 
+\bigl(\jmath_2^0,
0\bigr) +\bigl(\Phi _{21},\Phi_{22}\bigr)
\text{,\label{Eq:d/dt(w,s)}}
\end{multline}
where $h$ is defined in (\ref{Eq:h}), 
and $\Phi _{21}=\jmath_2^{II}+{\mathcal N}$, 
and $\Phi_{22}=\jmath_4^{II}$
are given in terms of the inhomogeneous terms defined in
\S\ref{defin}; notice that the inhomogeneous terms
$\jmath_2^{III},\jmath_2^{IV},\jmath_4^{III},\jmath_4^{IV}$
are included in the first term on the right hand side of 
\eqref{Eq:d/dt(w,s)}.

To study these equations it will turn out that the following 
quantity is useful:

$$
\widetilde{W}=\frac{1}{2}\bigl\| w-i\gamma \omega v\bigr\| _{L^{2}}^{2}-%
\frac{1}{2}\gamma ^{2}\omega ^{2}\bigl\| v\bigr\| _{L^{2}}^{2}+\frac{1}{2}%
\bigl\| \mathbf{{\tilde E}}\bigr\| _{L^{2}}^{2}+\Bigl\langle \bigl(
w,
\mathbf{{\tilde E}}
\bigr) ,\mathbf{u}\cdot\nabla \bigl(
v,
\mathbf{{\tilde A}}
\bigr) \Bigr\rangle_{L^{2}}
+{\tilde H}(v,\mathbf{\tilde A}).
$$
\noindent
We can think of $\widetilde{W}$ as follows:
it is formed by adding to the Hessian of the augmented Hamiltonian $W$
those terms arising in the expanded
Hamiltonian (when we input the perturbed solution ansatz (\ref
{Eq:Ansatz})) 
which describe the interactions of the fields
$(v,\mathbf{\tilde A})$ with themselves and with the 
external electromagnetic field.
An important reason for introducing $%
\widetilde{W}$ is that the following two lemmas 
imply  a long time bound for 
$W$, and hence a stability estimate in energy norm.

\begin{lemma}
\label{lem:W Wtilde equivalence} In the situation of
theorem \ref{thm:WNormEq}, so that
\begin{itemize}
\item
$\lambda$ lies in a compact subset, 
$\mathbf{K}\subset\widetilde{O}_{stab}$,
\item
$( v,w) $ satisfy the constraints \eqref{cc},
and 
\item
$W$ is equivalent (uniformly on $\mathbf{K}$) to 
$\| (v,w,\mathbf{{\tilde A}},\mathbf{{\tilde E}}%
)\| _{\cal H}^2$,
\end{itemize}
assume that $W<1$, and that
$e=o(1)$ and $e=o({\delta})$.
Then, there exists a constant $c({\mathbf{K}})>0$ such
that, for all ${{\lambda}}\in \mathbf{K}$,
\begin{equation}
cW\leq \widetilde{W}\leq \frac{1}{c}W.
\end{equation}
\end{lemma}
\noindent\proof
Referring to the formulae 
in \S\ref{defin} for
the $\tilde H^{(n)}$ which occur in the definition of $\tilde H$, 
it is a straghtforward consequence of the H\"older inequality
that
$$
W=\widetilde{W}+O(\frac{e}{{{\delta}} }W)
+O(\frac{e^2}{{{\delta}} ^{2}}W)
+O(\frac{e^2}{{{\delta}} }W^{\frac{3}{2}}) 
+O(e^{2}W)+O(e W^{\frac{3}{2}})+O(e^{2}W^{2})\text{,}
$$
lemma \ref{lem:Expdec} and the 
assumptions on the external field in \S\ref{Sec:ExternalField}.
The lemma follows immediately.
\myqed

\begin{notation}\la{noto}
In the following we write, $f=\frac{d}{dt}\left( O(A) +o(B)\right) $
if there exist $C^1$ functions $g,h$ such that 
$f=\frac{d}{dt}(g+h)$ and
$g=O(A)$ and $h=o(B)$.
\end{notation}

\begin{lemma}
\label{lem:dWtildedt}Assume the hypotheses of theorem
\ref{thm:Wgrowth}. It follows that,
\begin{equation}
\left| \frac{d}{dt}\widetilde{W}\right| 
=\frac{d}{dt}\left(O\left(e\widetilde{W}^{\frac{1}{2}}\right) 
+o(\widetilde{W}) \right)
+O\left(e^4+(e+\frac{e^2}{{\delta}})\widetilde{W}+(e^{2}+e{{\delta}})
\widetilde{W}^{\frac{1}{2}}
\right)
\text{,\label{Eq:dWtildedt}}
\end{equation}
in the limit of $e$ and $\widetilde{W}$ going to zero. 
\end{lemma}
\noindent\proof
See \S\ref{lem:Proof of dWtildedt}.
\myqed

\subsection{Proof of theorem \ref{thm:Wgrowth}, assuming lemma \ref{lem:dWtildedt}}
\label{Sec:Proof of Wgrowth}

\noindent\proof
Integrating up equation (\ref{Eq:dWtildedt}), and using
the Cauchy-Schwarz inequality,
$2e{\delta}\widetilde{W}^{1/2}\leq+e{\delta}^2+e\widetilde{W}$,
we infer the existence of a 
constant $c>0$ such that, for $t\in \lbrack 0,T_2/e]$,
\begin{equation}
\left| \widetilde{W}(t)-\widetilde{W}(0)\right| \leq c
\left(e^{2}
+{{\delta}}^{2}+|e|\int_{0}^{t}\widetilde{W}(s)ds\right) \text{,}
\end{equation}
as long as $e=O({\delta})$.
By Gronwall's inequality and lemma \ref{lem:dWtildedt}, for $|e|$ sufficiently
small there exists a constant $c>0$ such that, on $[0,T_2/e]$,
\begin{equation}
\widetilde{W}(t)\leq c\left( \widetilde{W}(0)+e^{2}+{\delta}^2\right)
\exp [c|e|t]\text{.}
\end{equation}
By lemma \ref{lem:W Wtilde equivalence}, the result is proved.
\myqed

\subsection{Proof of lemma \ref{lem:dWtildedt}}
\label{lem:Proof of dWtildedt}

\subsubsection{Beginning of proof of lemma \ref{lem:dWtildedt}}

By the assumptions of theorem \ref{thm:Wgrowth} we have a solution
of equations \eqref{Eq:d/dt(v,q)},\eqref{Eq:d/dt(w,s)}
satisfying the conclusions of theorems \ref{rModEq} and
\ref{thm:WNormEq}, so that the
constraints \eqref{Eq:Consa} hold and $W=O(e)$. Then, by
lemma \ref{lem:W Wtilde equivalence} and theorem \ref{thm:WNormEq},
there exists $c>0$ such that
$$
\frac{1}{c}\widetilde{W}\leq \| (v,w,\mathbf{{\tilde A}},
\mathbf{{\tilde E}})\| _{\cal H}^2\leq c \widetilde{W}.
$$
Also since $W=O(e)$ the bound \eqref{bv} holds, and will be
used in the course of the proof. The estimate for $\widetilde{W}$
will be obtained as a consequence of the energy identity for
\eqref{Eq:d/dt(v,q)},\eqref{Eq:d/dt(w,s)}, so the next
stage is to write that identity down and separate the terms out
in a way that allows them to be usefully estimated.

\subsubsection{The energy identity for \eqref{de1}-\eqref{de4}}

\ba
\frac{d}{dt}\widetilde{W}&=&\left\langle 
\partial_t\bigl(
w,
\mathbf{{\tilde E}}
\bigr),
\bigl(
w ,
\mathbf{{\tilde E}}
\bigr) +\mathbf{u}\cdot\nabla \bigl(
v,
\mathbf{{\tilde A}}
\bigr) -i\gamma \omega \bigl(
v,
0
\bigr)  \right\rangle _{L^{2}}\la{dwtp}\\
&\;-&\left\langle \partial_t\bigl(
v ,
\mathbf{{\tilde A}}
\bigr) ,
\bigl(-D_v \tilde H,
-D_{\tilde A} \tilde H\bigr)
-\mathbf{u}\cdot\nabla \bigl(
w ,
\mathbf{{\tilde E}}
\bigr) +i\gamma \omega \bigl(
w,
0
\bigr) \right\rangle _{L^{2}} \notag\\
&\;\;+&\int \partial_t\tilde h(v,{\mathbf{\tilde A}}) dx
-\partial_t\bigl( \gamma \omega \bigr) \left\langle iv,w\right\rangle
_{L^{2}} 
+\left\langle \partial_t\mathbf{u}\cdot\nabla \bigl(
v ,
\mathbf{{\tilde A}}
\bigr) ,\bigl(
w,
\mathbf{{\tilde E}}
\bigr) \right\rangle_{L^2} \text{.}\notag
\ea
Here we have introduced a notation $\tilde h(v,{\mathbf{\tilde A}})$
for the integrand defining $\tilde H$, i.e.
\ba
\tilde H(v,{\mathbf{\tilde A}})&=&
\int \tilde h(v,{\mathbf{\tilde A}})\, dx\la{thd}\\
&=&
-\frac{1}{2}\int \sum_{n=2}^4
\left\langle(
v ,
\mathbf{{\tilde A}}),
{\mathbb{B}^{(n-1)}}(
v,
\mathbf{{\tilde A}}) 
\right\rangle \,dx.
\notag
\ea
Explicit expressions for the nonlinear operators ${\mathbb{B}^{(n-1)}}(
v, \mathbf{{\tilde A}}) $ show that they depend on $t,x$, and the 
$\partial_t\tilde h$ in the final line of \eqref{dwtp} 
refers to differentiation with 
$(v, \mathbf{{\tilde A}})$ held fixed; similar conventions will be
understood below.

Substituting for the time derivatives from \eqref{Eq:d/dt(v,q)} and
\eqref{Eq:d/dt(w,s)}, and noting the usual cancellations which occur in
the derivation of the energy identity, we obtain the following expression:
\ba
\frac{d}{dt}\widetilde{W}&=&
Q_1+Q_2+Q_3-\langle D_v\tilde H,ihv\rangle_{L^2}+
\int(\partial_t+{\mathbf u}\cdot\nabla)\tilde h(v,{\mathbf{\tilde
A}})\, dx
\la{dwt}\\
&\;-&
\langle iv,({\mathbf u}\cdot\nabla h)w
\rangle_{L^2}
-\partial_t\bigl( \gamma \omega \bigr) \left\langle iv,w\right\rangle
_{L^{2}} 
+\left\langle \partial_t\mathbf{u}\cdot\nabla (
v ,
\mathbf{{\tilde A}}
) ,
(w,
\mathbf{{\tilde E}}
) \right\rangle_{L^2} \text{,}
\notag
\ea
where
\ba
Q_1&=&
\left\langle 
\bigl(
\Phi_{21},\Phi_{22}
\bigr),
\bigl(
w ,
\mathbf{{\tilde E}}
\bigr) +\mathbf{u}\cdot\nabla \bigl(
v,
\mathbf{{\tilde A}}
\bigr) -i\gamma \omega \bigl(
v,
0
\bigr)  \right\rangle _{L^{2}} \notag\\
&&\quad -\left\langle \bigl(
\Phi_{11},0
\bigr) ,
\bigl(-D_v \tilde H,
-D_{\tilde A} \tilde H\bigr)
+\mathbf{u}\cdot\nabla \bigl(
w ,
\mathbf{{\tilde E}}
\bigr) -i\gamma \omega \bigl(
w,
0
\bigr) \right\rangle_{L^{2}}, \notag
\ea
\ba
Q_2&=&
\left\langle 
\bigl(
\jmath_2^0,
0
\bigr),
\bigl(
w ,
\mathbf{{\tilde E}}
\bigr) +\mathbf{u}\cdot\nabla \bigl(
v,
\mathbf{{\tilde A}}
\bigr) -i\gamma \omega \bigl(
v,
0
\bigr)  \right\rangle _{L^{2}} \notag\\
&&\quad -\left\langle\bigl(
\jmath_1^0,
\jmath_3^0
\bigr) ,
\bigl(-D_v \tilde H,
-D_{\tilde A} \tilde H\bigr)
-\mathbf{u}\cdot\nabla \bigl(
w ,
\mathbf{{\tilde E}}
\bigr) +i\gamma \omega \bigl(
w,
0
\bigr) \right\rangle _{L^{2}} \notag
\ea
and $Q_3=-\bigl(\partial_t\lambda-V_0(\lambda)\bigr)\cdot\tilde Q_3$, where
\ba
&&\tilde Q_3=
\left\langle 
\bigl(
\widetilde{\partial_{\lambda}}\psi _{SC,e},
\partial_{\lambda}\mathbf{E}_{SC,e}
\bigr),
\bigl(
w ,
\mathbf{{\tilde E}}
\bigr) +\mathbf{u}\cdot\nabla \bigl(
v,
\mathbf{{\tilde A}}
\bigr) -i\gamma \omega \bigl(
v,
0
\bigr)  \right\rangle _{L^{2}}
\notag\\
&&\; 
-\left\langle 
\bigl(
\widetilde{\partial_{\lambda}}\phi _{SC,e},
\partial_{\lambda}\mathbf{A}_{SC,e}
\bigr) ,
\bigl(-D_v \tilde H,
-D_{\tilde A} \tilde H\bigr)
-\mathbf{u}\cdot\nabla \bigl(
w ,
\mathbf{{\tilde E}}
\bigr) +i\gamma \omega \bigl(
w,
0
\bigr) \right\rangle _{L^{2}}.
\notag
\ea
We control $Q_1,Q_2,Q_3$ in the next three subsections before
completing the proof of lemma \ref{lem:dWtildedt}. 
In the course of estimating the various terms we will use bounds
for ${\mathcal N},h$ and the $\Phi's$ (which may be read off
from those in \S\ref{boundin}), and the bounds
for ${\tilde A}_0$ in \S\ref{a0est}.

\subsubsection{Estimation of $Q_1$}
The following proposition is the main result about $Q_1$ needed for the
basic growth estimate:

\begin{proposition}
\label{lem:<w,Phi2>-<A(v,q),Phi1>}
In the situation of lemma \ref{lem:dWtildedt}

$$
Q_1= 
\partial_t\left( o( \widetilde{W})+O(
e\widetilde{W}^{\frac{1}{2}}) \right) +
O\left(e^4+e\widetilde{W} 
+e^2 \widetilde{W}^{\frac{1}{2}}
+e{\delta} \widetilde{W}^{\frac{1}{2}}\right) \text{.}
$$

\end{proposition}

\noindent\proof
Substituting from \eqref{Eq:d/dt(v,q)} and \eqref{Eq:d/dt(w,s)}  we obtain:
\ba
Q_1&=& (\partial_t\lambda-V_0)\cdot\left\langle 
{\partial_{\lambda}}\mathbf{A}_{SC,e}
\Phi_{22}\right\rangle _{L^{2}}
\la{eq1}\\
&& \;+(\partial_t\lambda-V_0)\cdot\left[\left\langle (
\widetilde{\partial_{\lambda}}\phi _{SC,e} ,
\Phi_{21}\right\rangle _{L^{2}}
-\left\langle
\widetilde{\partial_{\lambda}}\psi _{SC,e},
\Phi _{11},\right\rangle _{L^{2}}\right]
\notag\\
&&\quad
+
\left\langle ie{{\tilde A}_0} 
(i\gamma ( \omega -e\alpha _{\omega ,e})
-\bfu\cdot\nabla )f_{\omega ,e},\Phi _{11}\right\rangle
_{L^{2}}
\notag\\
&&\qquad-\left\langle 
(ie{{\tilde A}_0} f_{\omega ,e},
\nabla {{\tilde A}_0})
 ,(\Phi _{21},\Phi_{22})\right\rangle _{L^{2}}\notag\\
&&\qquad\quad 
+
\left\langle ( \partial_t+\bfu\cdot\nabla ) 
v ,\Phi _{21}\right\rangle _{L^{2}}
+\left\langle ihv,\Phi _{21}\right\rangle _{L^{2}}-\left\langle ihw,\Phi
_{11}\right\rangle _{L^{2}}
\notag\\
&&\qquad\quad\quad 
+
\bigl\langle( \partial_t+\bfu\cdot\nabla ) 
\mathbf{{\tilde A}},
\Phi_{22}\bigr\rangle _{L^{2}}-\bigl\langle
( \partial_t+\bfu\cdot\nabla) w,
\Phi _{11}\bigr\rangle _{L^{2}}\notag
\ea
since $\Phi_{12}=0$.

\noindent{\it Estimation of the first line in $Q_1$}
The first line of $Q_1$ is easily seen to be
small, since $\Phi_{22}=-e^2{\bf
a}^{{\delta},\chi}f_{\omega,e}$ is $O(e^2)$ in every
$L^p$ by the bounds in \S\ref{boundin}. 
Together with the fact that,
$\|{\partial_{\lambda}}\mathbf{A}_{SC,e}\|_{L^p}=O(e)$
for $p>3$,
by  \eqref{lbs} and
the results of appendix \ref{Sec:alpha Estimates},
this implies that
$\langle {\partial_{\lambda}}\mathbf{A}_{SC,e}, 
\Phi_{22}\rangle _{L^{2}}=O(e^3)$, and so by \eqref{bv}
the first line is $O(e^4)$.

\noindent{\it Estimation of the second line in $Q_1$.}
The second line is
smaller than appears due to a cancellation which is a
consequence of the modulation equations, \eqref{r2m} or
\eqref{Eq:ModEq}. To see this, we refer to the decomposition
of the force on the right hand side of \eqref{Eq:ModEq} given in 
\S\ref{modcalc}, and using the definitions of the $\Phi_{IJ}$
in \eqref{Eq:d/dt(v,q)},\eqref{Eq:d/dt(w,s)}, we see that
\ba
\left\langle\widetilde{\partial_{\lambda_A}}\phi _{SC,e} ,
\Phi _{21}\right\rangle _{L^{2}}
&-&\left\langle 
\widetilde{\partial_{\lambda_A}}\psi _{SC,e},
\Phi _{11}\right\rangle _{L^{2}}
=-\mathbf{F}_{A}^{L}-\mathbf{F}_A^n+Err_A\notag\\
&&\qquad
=-\left( \mathbb{M}(e)_{AB}+\mathbb{j}_{AB}
\right)\bigl({\dot\lambda-V_0}\bigr)_B
+\mathbf{F}_{A}^p+Err_A\notag
\ea
where $$
Err_A=\left\langle\widetilde{\partial_{\lambda_A}}\phi _{SC,e} 
-\widetilde{\partial_{\lambda_A}}\phi _{S,0} ,
\Phi _{21}\right\rangle _{L^{2}}
-\left\langle 
\widetilde{\partial_{\lambda_A}}\psi _{SC,e}
-\widetilde{\partial_{\lambda_A}}\psi _{S,0},
\Phi _{11}\right\rangle _{L^{2}}.
$$
Using lemma \ref{lem:different},
the bound \eqref{estn} for ${\mathcal N}$, and the 
fact that from \S\ref{boundin} $\Phi_{11}=\jmath_1^{II}$ and 
$\Phi_{21}-\mathcal{N}=\jmath_2^{II}$ are $O(e)$,
we
deduce that $|Err_A|\leq c e^2(e+\widetilde{W}+\widetilde{W}^{5/2}
+e^2\widetilde{W}^{1/2})$.
Next notice that lemma \ref{lem:different} implies that 
$ \mathbb{M}(e)_{AB} -\mathbb{M}(0)_{AB}=O(e^2)$. Therefore since
$ \mathbb{M}(0)_{AB}=- \mathbb{M}(0)_{BA}$ the largest term drops out
and the second line of $Q_1$ can be rewritten as 
$$
\left( \mathbb{M}(e)_{AB}-
\mathbb{M}(0)_{AB}
+\mathbb{j}_{AB}
\right)\bigl({\dot\lambda-V_0}\bigr)_A\bigl({\dot\lambda-V_0}\bigr)_B
-\left(\mathbf{F}_{A}^p+Err_A\right)\bigl({\dot\lambda-V_0}\bigr)_A
$$
which, by the above and \eqref{bj},\eqref{bfp} is
$O(e^4+e^2\widetilde{W}^{1/2})$, for small $e$ and $\widetilde{W}$.

\noindent{\it Estimation of the third and fourth lines in $Q_1$.}Using 
lemma \ref{lem:rbarLp},\eqref{estn}, the bounds in
\S\ref{boundin} and the
properties of $f_{\omega,e}$ in appendix \ref{Sec:e=0soliton},
the third and fourth lines can be estimated immediately to be
$O\left(e^3\widetilde{W}^{1/2}+e^2\widetilde{W}^{3/2}\right)$.

\noindent{\it Estimation of the fifth and sixth line in $Q_1$.} This 
requires care because $h$ is unbounded as a function of
$x$.
This makes it essential to separate the nonlinear term ${\mathcal N}$
in $\Phi_{21}$ from the other terms (which are exponentially decreasing
in $x$ and can thus absorb the unboundedness of $h$). Therefore we
estimate first of all the quantity
\begin{equation}
\left\langle ihv,\Phi _{21}-{\mathcal N}(f_{\omega ,e},
f_{\omega},v)\right\rangle_{L^2}
-\left\langle ihw,\Phi _{11}\right\rangle_{L^{2}}=O\left({e^{2}}
\widetilde{W}^{\frac{1}{2}}\right) \text{,}
\end{equation}
by \eqref{bv} and the bounds for $h$ recorded in \S\ref{boundin}.
Next, write the first term on line five, together
with the missing piece $\langle ihv,{\mathcal N}\rangle_{L^2}$ 
from the previous estimation, 
as the sum of two quantities:
$$\left\langle \left( \partial_t+ih+\bf{u}\cdot\nabla \right)
v,{\mathcal N}
\right\rangle _{L^{2}}+Rem,$$ 
where
$Rem=\left\langle \left( \partial_t+\bf{u}\cdot\nabla \right)
v,\Phi_{21}-{\mathcal N}
\right\rangle _{L^{2}}$.
It is shown in lemma \ref{diff}
that the first of these quantities is
$
\partial_t( o(
\widetilde{W})) +O(e \widetilde{W}
+e^3\widetilde{W}^{\frac{1}{2}}).
$
To 
complete the proof of proposition \ref{lem:<w,Phi2>-<A(v,q),Phi1>} we need
to estimate the sixth line and the quantity
$Rem$ defined above. This is done by means of the integration
by parts identity \eqref{inme}, and taking advantage of the fact that
\beq
(\partial_t+\mathbf{u}\cdot\nabla) f_{\omega,e}
=
({{{\dot\lambda-V_0(\lambda)}}})\cdot 
\partial_{{\lambda}}f_{\omega,e},
\la{rfs}\eeq
is $O(e)$ by \eqref{bv}. Together with
\eqref{avs1}, this implies that 
\beq\la{ddd}
\|( \partial_t+\mathbf{u}\cdot\nabla)\Phi_{IJ}\|_{L^p}=O\bigl(e(e+{\delta})\bigr)
\eeq
for all $p$ and all $IJ$ except for $IJ=21$; but in that case \eqref{ddd} holds
instead for $\Phi_{21}-{\mathcal N}=\jmath_2^{II}$, (which is what is
actually needed to estimate $Rem$).
Putting this information into \eqref{inme}, we infer that the sixth
line and $Rem$ are
$
\partial_t( O(
e\widetilde{W}^{\frac{1}{2}})) +O(e (e
+{\delta})\widetilde{W}^{\frac{1}{2}}),
$
which is sufficient to complete
the proof of the proposition.
\myqed

\subsubsection{Estimation of $Q_2$}

The terms in $Q_2$ arising from 
$\jmath_1^0,\jmath_2^0$
can be estimated in a straightforward way
by the H\"older and Sobolev inequalities, 
because of the exponential decay of $f_{\omega,e}$,
and 
using lemma \ref{lem:rbarLp} to bound ${\tilde A}_0$. For example,
\beq
\left\langle w-i\gamma \omega v+\bfu\cdot\nabla v,
ie{\tilde A}_0\left( i\gamma \left( \omega
-e\alpha _{\omega ,e}\right) -\mathbf{u}.\nabla \right) f_{\omega
,e}\right\rangle _{L^{2}}=O\left(e^{2} \widetilde{W}\right)
\text{}
\end{equation}
by H\"older's inequality, 
since $f_{\omega,e}$ and $\nabla f_{\omega,e}$ are bounded in every
$L^p$ norm and $\|{\tilde A}_0\|_{L^p}=
O\left(e\widetilde{W}^{\frac{1}{2}}\right)$
for $3<p<\infty$. For the terms involving
$\jmath_3^0=\nabla{\tilde A}_0$ we can estimate,
\begin{equation}
\left\langle \mathbf{u}\cdot\nabla \mathbf{{\tilde E}},\nabla 
{\tilde A}_0\right\rangle
_{L^{2}}=\left\langle \dv\mathbf{{\tilde E}},\mathbf{u}\cdot
\nabla{\tilde A}_0\right\rangle_{L^{2}}
=O\left(e^2\widetilde{W}\right)
\text{,}
\end{equation}
since $\|\dv\mathbf{{\tilde E}}\|_{L^{3/2}}=
O\left(e\widetilde{W}^{\frac{1}{2}}\right)$ and
$\|\nabla{\tilde A}_0\|_{L^3}=O\left(e\widetilde{W}^{\frac{1}{2}}\right).$
Consider next the terms
$
\left\langle
\jmath_3^0,
-D_{\tilde A} \tilde H
\right\rangle _{L^{2}}.
$
Referring to the explicit expressions for $D_{\tilde A} \tilde H$ 
given in \S\ref{defin}, starting with \eqref{ml}, we see that
the resulting terms can all be estimated in a
straightforward way (using the bounds for $\nabla{\tilde A}_0$ in
appendix \ref{a0est})
to be $O(e^2\widetilde{W})$,
except for one, namely:
$$\langle\triangle\mathbf{\tilde A},\nabla{\tilde A}_0\rangle_{L^2},$$
but this vanishes by the Coulomb condition, and so 
$Q_2=O(e^2\widetilde{W})$.

\subsubsection{Estimation of $Q_3$}
The quantity ${\tilde Q}_3$ is smaller than it appears due to the
constraints. To see this first recall that, as used above already,
$\|{\partial_{\lambda}}\mathbf{A}_{SC,e}\|_{L^p}=O(e)$
for $p>3$, and
$\|{\partial_{\lambda}}\mathbf{E}_{SC,e}\|_{L^p}=O(e)$
for $p>3/2$,
by  \eqref{lbs} and
the results of appendices \ref{Sec:alpha Estimates} and
\ref{different}
Referring to the expressions for $D_{\tilde A} \tilde H$
in \S\ref{defin},
this means that
the electromagnetic contributions to $\tilde Q_3$ can be bounded as
$O(e\widetilde{W}^{\frac{1}{2}})$. But also, the expressions for
$D_{v} \tilde H$ in \S\ref{defin} imply that
$$
\left\langle
\widetilde{\partial_{\lambda}}\phi _{SC,e},
-D_v \tilde H+M_{\lambda}v
\right\rangle_{L^2}=O\left(e\widetilde{W}^{\frac{1}{2}}\right).$$
Therefore, up to $O(e\widetilde{W}^{\frac{1}{2}})$,
we deduce that ${\tilde Q}_3$ is equal to
$$
\left\langle\bfu\cdot\nabla w-i\omega\gamma w-M_\lambda v,
\widetilde{\partial_{\lambda}}\phi _{SC,e},
\right\rangle_{L^2}-
\left\langle\bfu\cdot\nabla v-i\omega\gamma v+w,
\widetilde{\partial_{\lambda}}\psi _{SC,e}\right\rangle_{L^2}.
$$
Now the identities in appendix \ref{Sec:Identities} 
and the constraints \eqref{Eq:Consa} imply that
this expression vanishes  if
$\phi_{SC,e},\psi _{SC,e}$ are replaced by 
$\phi_{S,0},\psi _{S,0}$. But
by lemma \ref{lem:different} this can be done at the expense of 
an $O\left(e^2\widetilde{W}^{\frac{1}{2}}\right)$ error. Therefore,
since $(\dot\lambda-V_0)=O(e)$ by \eqref{bv}, we deduce that 
$Q_3=O\left(e^2\widetilde{W}^{\frac{1}{2}}\right)$.

\subsubsection{Completion of proof of lemma \ref{lem:dWtildedt}}

The previous subsections have provided the requisite information
on the $Q's$, and so it now suffices to control the remaining 
quantities in \eqref{dwt} appearing after the $Q's$.
The following
two propositions treat the two quantities on the first line of 
\eqref{dwt}.

\begin{proposition}
\label{lem:(d/dt + u.nabla)A}\label{lem:(d/dt+u.nabla)A}Assume the
hypotheses of  lemma \ref{lem:dWtildedt}.
It follows that,
\ba
\int 
(\partial_t+\mathbf{u}\cdot\nabla)\tilde h(v,{\mathbf{\tilde A}})\, dx
&=&
-\frac{1}{2}\int \sum_{n=2}^4
\left\langle(
v ,
\mathbf{{\tilde A}}),
( \partial_t+\mathbf{u}\cdot\nabla){\mathbb{B}^{(n-1)}}
(v,
\mathbf{{\tilde A}}) 
\right\rangle \,dx \notag\\
&=&
O\left( e\widetilde{W} +\frac{e^{2}}{{\delta}}\widetilde{W}\right).
\notag
\ea
\end{proposition}

\noindent\proof
Observe
\begin{itemize}
\item
the fact that 
$\mathbf{a}^{{\delta},\chi}$ is
pointwise $O(\frac{1}{{\delta}})$, but its derivatives are $O(1)$,
in particular 
$
\|( \partial_t+\mathbf{u}\cdot\nabla)\mathbf{a}^{{{\delta}},\chi}
\| _{L^{\infty }}\leq 2(\|
\overset{.}{\mathbf{a}}\| _{L^{\infty }}
+\| \nabla \mathbf{a}\|_{L^{\infty }}) \text{.}
$
\item
the identity
\begin{equation*}
( \partial_t+\mathbf{u}\cdot\nabla) f_{\omega ,e}=
(\overset{.}{{{\lambda}}}-V_0)\cdot\partial_{{\lambda}}
f_{\omega,e},
\end{equation*}
which shows that the left hand side is $O(e)$ 
in every $L^p$, by \eqref{bv} and the
exponential decay properties in appendix \ref{Sec:Sol}.
Similarly,
$\|( \partial_t+\mathbf{u}\cdot\nabla)\alpha_{\omega,e}\|_{W^{1,\infty}}$
is $O(e^2)$ by \eqref{bv} and the bounds for $\alpha_{\omega,e}$
in appendix \ref{Sec:alpha Estimates}.
\end{itemize}
To prove the proposition now, just use these observations to estimate 
with H\"{o}lder's inequality 
each of the terms arising from differentiation of the expressions
for ${\mathbb{B}^{(n)}}$ in \S\ref{defin}.
\myqed

\begin{proposition} Assume the
hypotheses of  lemma \ref{lem:dWtildedt}.
It follows that
\beq
\langle D_v\tilde H, ihv\rangle_{L^2}=
O\left( e\widetilde{W} +\frac{e^{2}}{{\delta}}\widetilde{W}\right).
\notag
\eeq
\end{proposition}

\noindent\proof
Using the notation in \eqref{ml} for the Frechet derivative
$D_v\tilde H$,
we have
\beq
|\langle D_v\tilde H, ihv\rangle_{L^2}|
=
|\langle\mathbb{B}(v,{\mathbf{\tilde A}}),(ihv,0)\rangle_{L^2}|
\la{tbt}\eeq
and we can estimate term by term, but some care is needed since
$h$ is unbounded as a function of $x$, see \eqref{Eq:h}.
In addition to the first point in 
the proof of the previous proposition, 
we use the bounds for $h$ recorded in \S\ref{boundin}.
Those terms in \eqref{tbt} arising from
$\mathbb{B}^{(3)}$ vanish identically, while of those arising from
$\mathbb{B}^{(2)}$ the only non-zero ones are proportional to
$e\langle hv,\nabla v\mathbf{\tilde A}\rangle_{L^2}$. By the
Coulomb condition and the bound for $\nabla h$ from
\S\ref{boundin},
this term is  $O(e^2\widetilde{W}^{3/2})$.
It remains to bound those terms arising from $\mathbb{B}^{(1)}$.
Of these, it is straightforward to bound those 
arising from $\mathbb{B}_{12}$ as $O(e\widetilde{W})$ by the
second fact just mentioned, and the same goes for those arising
from $M_\lambda$ in $\mathbb{B}_{11}=-M_\lambda+e\mbR+\mbS$. 
However, there is a single non-zero term
arising from $e\mbR v$ which is proportional to
$$
\langle hv,\mathbf{a}^{{\delta},\chi}\cdot\nabla v\rangle
$$
which, with an integration by parts, 
can be bounded as $O(\frac{e^2}{{\delta}}\widetilde{W})$,
but, again, only after taking into account the Coulomb condition
$\nabla\cdot\mathbf{a}^{{\delta},\chi}=0$. Finally for the terms
arising from $\mbS$ we see from \eqref{Eq:E0} that
$$
\langle ihv,\mbS v\rangle_{L^2}=e\gamma\int [2\alpha_{\omega,e}
\langle v
\bfu\cdot\nabla v\rangle+\bfu\cdot\nabla\alpha_{\omega,e}|v|^2]\,dx=0,
$$
so that $\langle ihv,\mbS v\rangle_{L^2}=0$, and the proof of the
proposition is completed.
\myqed

The remaining terms on the second line of formula \eqref{dwt} are easily estimated
as $O(e\widetilde{W})$ by \eqref{bv}, and the proof of lemma
\ref{lem:dWtildedt} is completed.

\appendix

\section{Appendices}

\subsection{Further properties of the solitons}
\label{Sec:Sol}

\subsubsection{Exponential decay properties of the solitons}
\label{Sec:e=0soliton}

The $e=0$ solitons in the nonlinear Klein-Gordon equation 
\eqref{Eq:HNLW} are exponentially localized: to be precise
we have the following estimates for the profiles functions $f_\omega,g_\omega$:
\begin{equation}  \label{Eq:3 Derivatives of f}
\lim_{|x|\rightarrow \infty }\sup \sum_{|\alpha |\leq 3}\nabla ^{\alpha
}f_{\omega}Exp[|x|\left( \sqrt{m^{2}-\omega ^{2}}-\varepsilon \right)
]<\infty \text{ \ \ }\forall \varepsilon \in (0,\sqrt{m^{2}-\omega ^{2}})%
\text{,}
\end{equation}
together with
\begin{equation}  \label{eq:3 Derivatives of dfdomega}
\lim_{|x|\rightarrow \infty }\sup \sum_{|\alpha |\leq 3}\nabla ^{\alpha
}g_{\omega}Exp[|x|\left( \sqrt{m^{2}-\omega ^{2}}-\varepsilon \right)
]<\infty \text{ \ \ }\forall \varepsilon \in (0,\sqrt{m^{2}-\omega ^{2}})%
\text{,}
\end{equation}
and
\begin{equation}
\lim_{|x|\rightarrow \infty }\frac{f_{\omega}^{\prime }}{f_{\omega}}=-%
\sqrt{m^{2}-\omega ^{2}}\text{,}
\end{equation}
while $\forall \varepsilon >0$, there exists $c(\varepsilon )>0$ such that
\begin{equation}  \label{Eq:LowerExpBoundonf}
f_{\omega}(|x|)>c(\varepsilon )Exp[-|x|\left( \sqrt{m^{2}-\omega ^{2}}%
+\varepsilon \right) ]\text{.}
\end{equation}
(See theorem 1.4 in \cite{Stuart}).
Exponential decay also holds for the solitons coupled to
electromagnetism for small $e$:

\begin{lemma}
\label{lem:Expdec} Suppose that $|e|<e_{1}$, for some $e_{1}>0$. Under
conditions (\ref{Eq:Exist1}-\ref{Eq:Exist4}) on $U$,
\begin{equation}
|D^{\alpha }f_{\omega ,e}(x)|\leq CExp[-\kappa |x|]  \label{Eq:eff}
\end{equation}
for positive constants $C$ and $\kappa $, 
and where $\alpha$ is any multi-index with $|\alpha |\leq 2$.
Furthermore, the constants $C$ and $\kappa $ are independent of the coupling
constant $e$.
\end{lemma}

\noindent\proof
See \cite{Paper1}.
\myqed

\subsubsection{Some estimates of the soliton electromagnetic potential $\protect\alpha $}
\label{Sec:alpha Estimates}

\begin{lemma}
\label{lem:alphaFrechet}\label{lem:alphdiff} For each $f\in H_{r}^{2}(%
{{\Real}}^{3})$, there exists a unique $\alpha \in \overset{.}{H}_{r}^{1}(%
{{\Real}}^{3})$ such that
\begin{equation}
-\triangle \alpha +e^{2}f^{2}\alpha =\omega ef^{2}\text{.}
\end{equation}
Furthermore, the map $A:H^{2}({{\Real}}^{3})\longrightarrow \overset{.}{H}%
^{1}({{\Real}}^{3})$ defined by $A(f)=\alpha $ is continuously
Frechet-differentiable.
\end{lemma}

\noindent\proof
This follows from standard arguments.
\myqed

\begin{lemma}
\label{lem:alphaest} Suppose that $f\in H^{1}({{\Real}}^{3})$. Suppose
further that $\alpha $ solves
\begin{equation}
-\triangle \alpha +e^{2}f^{2}\alpha =e\omega f^{2}\text{.}
\end{equation}
It follows that $\nabla \alpha ,\nabla ^{i}\nabla ^{j}\alpha \in L^{2}(%
{{\Real}}^{3})$ for any $i,j\in (1,2,3)$. Furthermore, $\left\| \nabla
^{i}\nabla ^{j}\alpha \right\| _{L^{2}}$, $\left\| \nabla \alpha \right\|
_{L^{2}}$, $\left\| \alpha \right\| _{L^{\infty }}=O(e)$
\end{lemma}

\noindent\proof
\begin{equation}
\int \left| \nabla \alpha \right| ^{2}+e^{2}f^{2}\alpha^{2}=e\omega \int
f^{2}\alpha
\end{equation}
from which it easily follows via Sobolev's inequality that
\begin{equation}
\left\| \nabla \alpha \right\| _{L^{2}}\leq ce\left\| f\right\|
_{L^{2}}\left\| f\right\| _{L^{3}}\text{.}
\end{equation}
Next, since $-\triangle \alpha =e\left( \omega -e\alpha \right) f^{2}$, we
have
\begin{equation}
\left\| \triangle \alpha \right\| _{L^{2}}\leq e\left( \omega \left\|
f\right\| _{L^{4}}^{2}+e\left\| \alpha _{\omega ,e}\right\| _{L^{6}}\left\|
f\right\| _{L^{6}}^{2}\right) \text{.}
\end{equation}
By the Calderon-Zygmund inequality, we have that for any $i,j\in (1,2,3)$,
\begin{equation}
\left\| \nabla ^{i}\nabla ^{j}\alpha \right\| _{L^{2}}=O(e)\text{.}
\end{equation}
By Sobolev's inequality, we have thus shown that $\alpha \in W^{1,6}$ and
hence by Morrey's inequality, $\left\| \alpha \right\| _{L^{\infty }}=O(e)$.
\myqed

\begin{corollary}
Suppose that $f_{\omega ,e}\in H^{2}({{\Real}}^{3})$ solves
\begin{equation}
-\triangle f_{\omega ,e}+m^{2}f_{\omega ,e}-\left( \omega -e\alpha _{\omega
,e}\right) ^{2}f_{\omega ,e}=\beta (f_{\omega ,e})f_{\omega ,e}\text{,}
\end{equation}
where $\alpha _{\omega ,e}\in \overset{.}{H}_{r}^{1}({{\Real}}^{3})$ is a
non-local function of $f_{\omega ,e}$ uniquely determined by
\begin{equation}
-\triangle \alpha _{\omega ,e}+e^{2}f_{\omega ,e}^{2}\alpha _{\omega
,e}=\omega ef_{\omega ,e}^{2}\text{.}
\end{equation}
Then, $f_{\omega ,e}\in H^{4}({{\Real}}^{3})$.
\end{corollary}

\noindent\proof
Differentiate the equation for $f_{\omega ,e}$ and apply the
Calderon-Zygmund inequality.
\myqed

This leads naturally to the following lemma.

\begin{lemma}
Suppose that $f\in H^{4}({{\Real}}^{3})$ and that $\alpha $ solves
\begin{equation}  \label{Eq:alpha}
-\triangle \alpha +e^{2}f^{2}\alpha =e\omega f^{2}\text{.}
\end{equation}
It follows that $\nabla \alpha \in W^{3,p}({{\Real}}^{3})$ for any $p\in
\left( \frac{3}{2},\infty \right) $.
\end{lemma}

\noindent\proof
Differentiate (\ref{Eq:alpha}), and apply the Calderon-Zygmund inequality
(using the H\"older and Sobolev inequalities if necessary) to get the result.
\myqed

\begin{lemma}
Suppose that $f\in H^{2}({{\Real}}^{3})$ and that $\alpha $ solves
\begin{equation}
-\triangle \alpha +e^{2}f^{2}\alpha =e\omega f^{2}\text{.}
\end{equation}
It follows that
\begin{equation*}
0\leq sgn\left( \frac{\omega }{e}\right) \alpha \leq \left| \frac{\omega }{e}%
\right| \text{,}
\end{equation*}
where $sgn(x)=x/|x|$ for $x\neq 0$ and $sgn(0)=0$.
\end{lemma}

\noindent\proof
\label{lem:MaxPrinc} Assume that $f$ in $C_{c}^{\infty }({{\Real}}^{3}%
\mathbb{)}$. Define $\alpha ^{+}=\max (\alpha ,0)$ and $\alpha ^{-}=\max
(-\alpha ,0)$. Suppose $\omega e>0$, then by a weak maximum principle
(theorem 8.1 in \cite{Gilbarg-Trudinger}), $\alpha >0$. Now, $A_{0}=\alpha -$
$\frac{\omega }{e}$ solves $-\triangle A_{0}+e^{2}\left| f\right|
^{2}A_{0}=0 $, therefore $A_{0}\leq 0$ by the same weak maximum principle.
Hence, $0\leq \alpha \leq \frac{\omega }{e}$. Similarly, if $-\omega e>0$,
then $0\geq \alpha \geq -\frac{\omega }{e}$ so that $\left\| \alpha \right\|
_{L^{\infty }}\leq \left| \frac{\omega }{e}\right| $. The lemma follows by
approximation.
\myqed

\begin{lemma}
\label{lem:DalphaLambda}Suppose that $f_{\omega ,e}$ and $\alpha _{\omega
,e} $ are as given in theorem \ref{thm:Exist}. Then,
\begin{equation}
\left\| \nabla ^{i}\nabla ^{j}\frac{d\alpha _{\omega ,e}}{d{{\lambda}}}%
\right\| _{L^{p}}=O(e)
\end{equation}
for $p\in (1,\infty )$, and $i,j=1,2,3$. In addition, $\left\| \nabla \frac{%
d\alpha _{\omega ,e}}{d{{\lambda}}}\right\| _{W^{2,p}}=O(e)$ for any $%
p\in \left( \frac{3}{2},\infty \right) $.
\end{lemma}

\noindent\proof
From lemma \ref{lem:alphaFrechet} and theorem \ref{thm:Exist}, $\frac{%
d\alpha _{\omega ,e}}{d{{\lambda}}}$ is a well-defined object. We note
that
\begin{equation}
\triangle \frac{d\alpha _{\omega ,e}}{d{{\lambda}_A}}+e^{2}f_{\omega
,e}^{2}\frac{d\alpha _{\omega ,e}}{d{{\lambda}_A}}=ef_{\omega
,e}^{2}\delta_{-1\, A}+2ef_{\omega ,e}(\omega -e\alpha _{\omega ,e})\frac{df_{\omega ,e}}{d%
{{\lambda}_A}}
\end{equation}
from which $\left\| \triangle \frac{d\alpha _{\omega ,e}}{d{{\lambda}_A}}%
\right\| _{L^{p}}=O(e)$ for $p\in (1,\infty )$ follows immediately. The
lemma follows trivially from repeated differentiation, the Calderon-Zygmund
inequality and the H\"{o}lder and Sobolev inequalities.
\myqed

Let
$\zeta(x;\lambda) $ be the unique
solution in $\hh$ of \eqref{cc0}, 
$-\triangle \zeta =-\gamma \mathbf{u}\cdot\nabla \alpha _{\omega
,e}(\mathbf{Z})$,
which takes the Lorentz transformed
solitons into Coulomb gauge. Then
\begin{lemma}
\label{lem:Laplacezeta}
\begin{eqnarray}
\left\| \nabla ^{i}\nabla ^{j}\zeta \right\| _{L^{p}} &=&O(e)\text{,} \\
\left\| \nabla ^{i}\nabla ^{j}\partial_{\lambda}\zeta \right\|
_{L^{p}} &=&O(e)\text{,}  \notag
\end{eqnarray}
for $p\in (\frac{3}{2},\infty )$ and $i,j=1,2,3$.
\end{lemma}

\noindent\proof
By \eqref{cc0}, and its derivative:
and
\begin{equation*}
-\triangle \frac{d}{d{{\lambda}_A}}\zeta =-\gamma \mathbf{u}\cdot\nabla
\frac{d}{d{{\lambda}_A}}\alpha _{\omega ,e}-\left( \frac{d}{d\mathbf{%
\lambda }_A}\gamma \mathbf{u}\right) \cdot\nabla \alpha _{\omega ,e}\text{.}
\end{equation*}
 the result follows by means of 
lemmas \ref{lem:alphaest} and \ref{lem:DalphaLambda}.
\myqed

\subsubsection{Differentiability}
\la{different}
\begin{lemma}
\label{lem:different} Let $f_{\omega ,e}\in H^2$ be given by theorem \ref{thm:Exist}%
. Then it is a differentiable function of $\omega$ and satisfies,
for small $e$:
\begin{equation}
\left\| f_{\omega ,e}-f_{\omega}\right\| _{H^{2}}
+\left\| \partial_\omega f_{\omega ,e}-\partial_\omega
f_{\omega}\right\| _{H^{2}}
=O(e^{2}).
\end{equation}
\end{lemma}

\noindent\proof
See \cite{Paper1}.
\myqed

\begin{lemma}
\label{lem:ddo} Let $\widetilde{h}_{\omega }=h_{\omega }-\omega q_{\omega }$%
, where $h_{\omega }=H(\Phi _{S,e}(0,\omega ,0,0))$ while $q_{\omega
}=Q(\Phi _{S,e}(0,\omega ,0,0))$. Then
\begin{equation}
\frac{d}{d\omega }\widetilde{h}_{\omega }=-q_{\omega }\text{.}
\end{equation}
\end{lemma}

\noindent\proof
Following the argument given in \cite{GSS1}, we note that
\begin{equation}
\frac{d}{d\omega }\widetilde{h}_{\omega }=-q_{\omega }+\left\langle
H^{\prime }(\Phi _{S,e}({{\lambda}}_{0}))-\omega Q^{\prime
}(\Phi _{S,e}({{\lambda}}_{0})),\frac{d}{d\omega }\Phi
_{S,e}({{\lambda}}_{0})\right\rangle_{L^2} \text{,}
\end{equation}
where ${{\lambda}}_{0}=(\omega,0 ,0,0)$. The result follows from the
fact that $H^{\prime }(\Phi _{S,e}({{\lambda}}_{0}))-\omega Q^{\prime
}(\Phi _{S,e}({{\lambda}}_{0}))=0$.
\myqed

\subsubsection{Some identities involving 
$\left( \widetilde{\partial_{\lambda}}{\phi }_{S,0},
\widetilde{\partial_{\lambda}}{\psi }_{S,0}\right) 
$}
\label{Sec:Identities}

The explicit calculation of the modulation equations 
can be carried out by making use of the following functions 
$(a_{A}(\mathbf{Z}(\mathbf{x},\lambda);\lambda),
b_{A}(\mathbf{Z}(\mathbf{x},\lambda);\lambda)$
from \cite{Stuart}:
\begin{eqnarray}
b_{-1}(\mathbf{Z};{{\lambda}}) &=&g_{\omega}-i\mathbf{u}.\mathbf{Z}%
f_{\omega},  \label{Eq:Consgen2} \\
b_{0}(\mathbf{Z};{{\lambda}}) &=&if_{\omega},  \label{Eq:Consgenb0}
\\
b_{i}(\mathbf{Z};{{\lambda}}) &=&\nabla _{\mathbf{Z}}^{i}f_{\omega}(%
\mathbf{Z}), \\
b_{3+i}(\mathbf{Z};{{\lambda}}) &=&\zeta _{ji}\nabla _{\mathbf{Z}%
}^{j}f_{\omega}(\mathbf{Z})-i\omega \gamma (\left( \gamma P_{\mathbf{u}%
}+Q_{\mathbf{u}}\right) \mathbf{Z})_{i}f_{\omega}(\mathbf{Z}),
\end{eqnarray}
while
\begin{eqnarray}
a_{-1}(\mathbf{Z};{{\lambda}}) &=&-\gamma ^{-1}b_{0}+\left( \gamma
\mathbf{u}.\nabla _{\mathbf{Z}}-i\gamma \omega \right) b_{-1},
\label{Eq:Consgen4} \\
a_{0}(\mathbf{Z};{{\lambda}}) &=&\left( \gamma \mathbf{u}.\nabla _{%
\mathbf{Z}}-i\gamma \omega \right) b_{0}  \label{Eq:Consgena0} \\
a_{i}(\mathbf{Z};{{\lambda}}) &=&\left( \gamma \mathbf{u}.\nabla _{%
\mathbf{Z}}-i\gamma \omega \right) b_{i}, \\
a_{3+i}(\mathbf{Z};{{\lambda}}) &=&\left( \gamma P_{\mathbf{u}}+Q_{%
\mathbf{u}}\right) \mathbf{Z})_{ij}b_{j}+\left( \gamma \mathbf{u}.\nabla _{%
\mathbf{Z}}-i\gamma \omega \right) b_{3+i}\text{,}
\end{eqnarray}
where $i,j=1,2,3$, $g_{\omega}=\frac{d}{d\omega }f_{\omega}$, and 
$$
\zeta _{ji}=\gamma^2({\mathbf u}\cdot {\mathbf Z})(P_{\mathbf{u}})_{ji}
+\frac{\gamma-1}{\gamma|{\mathbf u}|^2}({\mathbf u}\cdot {\mathbf Z})(Q_{\mathbf{u}})_{ji}
+\frac{\gamma-1}{|{\mathbf u}|^2}(Q_{\mathbf u}Z)_i{\mathbf u}_j.
$$
These are convenient for computation of the modulation equations because
the linear span of the 
$\widetilde{\partial_{{\lambda}_{A}}}
(\phi_{S,0},\psi_{S,0})$ is the same as the linear span of the
$\left(b_A,-a_A\right)$.(To be precise: 
except for $A=j\in\{1,2,3\}$, 
we have
$\widetilde{\partial_{{\lambda}_{A}}}
(\phi_{S,0},\psi_{S,0})=\left(b_A,-a_A\right)$, and for $A=j$
we have $\widetilde{\partial_{{\xi^j}}}
(\phi_{S,0},\psi_{S,0})=
-(\gamma P_{\mathbf{u}}+Q_{\mathbf{u}})_{jk}\left(b_k,-a_k\right)
+\omega\gamma u^j(b_0,a_0)$.)

The following identities are equivalent to lemma 2.2 in
\cite{Stuart}, and can be obtained by differentiating the 
Euler-Lagrange equation $F_0'=0$, where $F_0$ is the augmented Hamiltonian
\eqref{augh}:
\begin{eqnarray}
\left( i\gamma \omega -\mathbf{u}\cdot \nabla \right) 
\widetilde{\partial_{{\lambda}_{0}}}{\phi }_{S,0}
- \widetilde{\partial_{\lambda_{0}}}
{\psi_{S,0}} &=&0, \\
\left( i\gamma \omega -\mathbf{u}\cdot \nabla \right) 
\widetilde{\partial_{\lambda_{0}}}\psi_{S,0}}
-M_{{{\lambda}}}\widetilde{\partial_{\lambda_{0}}}{\phi_{S,0}
&=&0, \\
\left( i\gamma \omega -\mathbf{u}\cdot \nabla \right) 
\widetilde{\partial_{\lambda_{j}}}\phi_{S,0}
-\widetilde{\partial_{\lambda_{j}}}{\psi_{S,0}} &=&0, \\
\left( i\gamma \omega -\mathbf{u}\cdot \nabla \right) 
\widetilde{\partial_{{\lambda}_{j}}}\psi_{S,0} 
-M_{{{\lambda}}}\widetilde{\partial_{{\lambda}_{j}}}{\phi }_{S,0}&=&0, \\
\left( i\gamma \omega -\mathbf{u}\cdot \nabla \right) 
\widetilde{\partial_{\lambda_{-1}}}{\phi_{S,0}}
-\widetilde{\partial_{\lambda_{-1}}}\psi_{S,0} 
&=&-\frac{1}{\gamma }
\widetilde{\partial_{\lambda_{0}}}
{\phi_{S,0}}, \\
\left( i\gamma \omega -\mathbf{u}\cdot \nabla \right) 
\widetilde{\partial_{\lambda_{-1}}}\psi_{S,0}
-M_{{{\lambda}}}\widetilde{\partial_{\lambda_{-1}}}{\phi_{S,0}}&=&
-\frac{1}{\gamma}
\widetilde{\partial_{\lambda_{0}}}{\psi_{S,0}}, \\
\left( i\gamma \omega -\mathbf{u}\cdot \nabla \right) 
\widetilde{\partial_{\lambda_{3+j}}}\phi_{S,0}
-\widetilde{\partial_{\lambda_{3+j}}}
\psi_{S,0}
&=&-\widetilde{\partial_{\lambda_{j}}}
\phi_{S,0}-\gamma \omega{u}_{j}
\widetilde{\partial_{\lambda_{0}}}\phi_{S,0}
, \\
\left( i\gamma \omega -\mathbf{u}\cdot \nabla \right) 
\widetilde{\partial_{\lambda_{3+j}}}\psi_{S,0}
-M_{{{\lambda}}}\widetilde{\partial_{\lambda_{3+j}}}
{\phi_{S,0}}
&=&-\widetilde{\partial_{\lambda_{j}}}{\psi_{S,0}}
-\gamma\omega{u}_{j}
\widetilde{\partial_{\lambda_{0}}}{\psi_{S,0}}
\end{eqnarray}
where the index $j$ runs from $1$ to $3$.

\subsection{Some estimates}

\subsubsection{Estimates related to the external field}
\la{eref}

\begin{lemma}
\la{err}
Let $f$ be a measurable function with $(1+|x|)f\in L^1$. Then
if $a^{{\delta},\chi}$ is as in \eqref{da}
\begin{equation}
\left\| ea_{0}^{{{\delta}},\chi } f
\right\| _{L^{p}}\leq ce
L_1\left\|(1+ |x-\xi |)f\right\|_{L^{p}}
\text{,\label{Eq:a0mu}}
\end{equation}
and
\begin{equation}
\left\| e\mathbf{a}^{{{\delta}},\chi }f
\right\| _{L^{p}}\leq ce L_1
\left\| (1+ |x-\xi |)f\right\| _{L^{p}}  \label{Eq:amu}
\end{equation}
for $p\in \lbrack 1,\infty ]$. If in addition $f_{even}$ is an 
even function of $(\mathbf{x}-\mbxi)$ 
and  $(1+|x|)^2 f_{even}\in L^1$ then
\begin{equation}
\left|\int a_\mu^{{{\delta}},\chi }f_{even}
d^3 x\right|\leq cL_2{\delta} 
\left\| (1+ |x-\xi |)^2f_{even}\right\| _{L^{1}}  \label{Eq:amueven}
\eeq
with $L_1,L_2$ as in \eqref{Eq:ExternalAssumption2}.
\end{lemma}
\noindent\proof
Recall \eqref{da} and \eqref{dad}.
Writing \beq
a_0^{\delta} (t,x)-a_0^{\delta} (t,\xi)=(\mathbf{x}-\xi)\cdot\int\nabla a_0^{\delta}
(t,\xi+s(x-\xi))ds
\la{rep}\eeq
etc, by the fundamental theorem of calculus,
the result then follows, using the fact that the gradients of 
$a_0^{\delta},\mathbf{a}^{\delta}$ are bounded independent of ${\delta}$ by
assumption (see \S\ref{Sec:ExternalField}). For the proof of 
\eqref{Eq:amueven}, it suffices to use the
 identity for $\nabla a_\mu^{\delta}$  corresponding to \eqref{rep},
and then substitute this back into \eqref{rep} and use the fact
that $\int (\mathbf{x}-\xi) f_{even}=0$.
\myqed

Similarly, we have the following bounds:
\begin{lemma}
\la{absurd}

\ba
\left\|(1+|x-\xi|)^{-1} ( \partial_t+\mathbf{u}\cdot\nabla)
a_{\mu}^{{{\delta},\chi} }
\right\| _{L^{\infty }} &\leq &C_1(|{\delta}|+|e|)
\la{avs1}\\
\int_{{{\Real}}^{3}}
f(x)\bigl|\nabla_{t,x} a_{\mu}^{{{\delta}} }(t,%
\mathbf{x})
-(\nabla_{t,x} a_{\mu}^{{{\delta}} })(t,\mbxi)
\bigr|
d\mathbf{x}&\leq &C_2|{\delta}|\text{,}
\la{avs3}\ea
where we use \eqref{bv},
\eqref{Eq:ExternalAssumption2},
and $C_1=C_1(L_1,L_2)$ and
$C_2=C_2(L_2,\|(1+|x|)f\|_{L^1})$.
\end{lemma}

\subsubsection{Estimates for the time component of the 
electromagnetic potential}
\la{a0est}
\begin{lemma}
\label{lem:rbarLp}
Given $(v,w)\in H^1\times L^2$ and $\lambda\in\widetilde{O}$
there exists a unique $\tilde{A}_0\in\hh$ solving \eqref{Eq:Gauss2m}
such that
\begin{equation}
\left\| \nabla \tilde{A}_0\right\| _{L^{p}}=
O\left(e
\| (v,w,\mathbf{{\tilde A}},
\mathbf{{\tilde E}})\|_{\cal H}
+e
\| (v,w,\mathbf{{\tilde A}},
\mathbf{{\tilde E}})\|_{\cal H}^2
\right) \text{,}
\end{equation}
for $p\in (\frac{3}{2},3]$. Consequently $\left\| \tilde{A}_0\right\|
_{L^{q}}$ satisfies the same bound for $3<q<\infty$ by Sobolev's
inequality.
\end{lemma}
\noindent\proof
From Gauss's law \eqref{Eq:Gauss2m}, we have explicitly
\begin{equation}
-\triangle \tilde{A}_0=e\left\langle if_{\omega ,e},w\right\rangle
+e\left\langle iv,\left( i\gamma (\omega -e\alpha _{\omega ,e})-\mathbf{u}%
.\nabla \right) f_{\omega ,e}+w\right\rangle \text{.}
\end{equation}
By Sobolev's and H\"{o}lder's respective inequalities,
\begin{equation*}
\left\| \triangle \tilde{A}_0\right\| _{L^{q}}=O\left(e
\| (v,w,\mathbf{{\tilde A}},
\mathbf{{\tilde E}})\| _{\cal H}
+e
\| (v,w,\mathbf{{\tilde A}},
\mathbf{{\tilde E}})\|_{\cal H}^2
\right) \text{.}
\end{equation*}
for $q\in \lbrack 1,\frac{3}{2}]$. The lemma follows from the Sobolev
inequality and from the Calderon-Zygmund inequality, 
\cite[section 9.4]{Gilbarg-Trudinger}.
\myqed

\subsubsection{Integration by parts and simple averaging}
First we recall the phenomenon of averaging in the context of
ordinary differential equations, in the simplest possible 
case of the perturbed harmonic oscillator.
Let $g$ be a $C^1$ function of $t\in\Real$, with
$|g|\le M$ and $|\dot g|\le N$. For $0<\epsilon <<1$ let
$y^\epsilon$ be the solution of $\ddot y+y=\epsilon g(\epsilon t)$
with initial data $y^\epsilon(0)=y_0,\dot y^\epsilon(0)=y_1$
(fixed independent of $\epsilon$). Then $y^\epsilon-y^0$ 
is $O(\epsilon)$ in $C^1([-T,T])$ norm for times of 
$T=O(\frac{1}{\epsilon})$. One way to prove this is to 
define $f=\epsilon^{-1}(y^\epsilon-y^0)$, which solves
$\ddot f+f=g(\epsilon t)$ with zero initial data. Let 
$E(t)=(f^2+\dot f^2)/2$ be the energy; it satisfies $E(0)=0$
and $\dot E(t)=\dot f(t)g(\epsilon t)$. Now an integration by parts
gives
\ba
|\int_0^T \dot f(t)g(\epsilon t)dt|&\leq & M|f(T)|
+\epsilon N\int_0^T|f(t)|dt\notag\\
&\leq&
M|f(T)|+\frac{\epsilon TN^2}{2}
+\epsilon\int_0^T\frac{|f(t)|^2}{2}dt,\notag
\ea
which, by Gronwall's inequality, implies $E(t)=O(1)$ for 
$t=O(\frac{1}{\epsilon})$ as claimed.
To conclude, this simple fact - that a small slowly varying inhomogeneous 
$\epsilon g(\epsilon t)$ term only 
influences a simple harmonic oscillator  to $O(\epsilon)$ on 
time scales of $O(\frac{1}{\epsilon})$ -
expresses a weak averaging effect, and can be proved by integration
by parts. Of course, this argument can be modified to give information
about perturbed oscillators on longer times scales of
$O(\frac{1}{\epsilon^a}),\,a<2$, and many
different generalizations are possible.

A simple
generalization, which is usful 
for the study of slow motion of
solitons, can be obtained by integrating the identity
\beq
\la{inme}
\langle(\partial_t+\bfu\cdot\nabla)F,G\rangle_{L^2}=
\partial_t\langle F,G\rangle_{L^2}-
\langle F,(\partial_t+\bfu\cdot\nabla)G\rangle_{L^2}
\eeq
where $F,G$ are sufficiently regular functions of 
$t,x$ but $\bfu=\bfu(t)$ depends on $t$ only and the inner product
is $L^2(dx)$. This is often useful because in perturbation theory for solitons 
functions often arise with $(\partial_t+\bfu\cdot\nabla)G$ small - see
\eqref{ddd}.

The following result, used in the 
proof of proposition \ref{lem:<w,Phi2>-<A(v,q),Phi1>}, is a more
complicated version of this idea:

\begin{proposition}
\la{diff}
In the situation of lemma \ref{lem:dWtildedt},
$$
\left\langle \left( \partial_t+ih+\mathbf{u}\cdot\nabla \right) v,{\mathcal N}(f_{\omega
,e},f_{\omega},v)\right\rangle _{L^{2}}=\partial_t\left( o\left(
\widetilde{W}\right) \right) +O\left(e \widetilde{W}
+e^3\widetilde{W}^{\frac{1}{2}}\right) \text{,}
$$
where a function $f$ satisfies $f=d/dt\left( o\left( \widetilde{W}\right)
\right) $ if there exists a $C^1$ function 
$g=o\left( \widetilde{W}\right) $ and $f=\frac{d}{dt}g$.
\end{proposition}

\noindent\proof
We work mostly with the potential $\mcn_1(\phi)=-U(|\phi|)$ 
which determines ${\mathcal N}$: recall that 
$\mcn_1'(\phi)=-\beta(|\phi|)\phi$, and (being slightly cavalier with
notation) \eqref{Eq:N(f,f,v)} can be rewritten
$$
{\mathcal N}(f_{\omega ,e},f_{\omega},v)
=-\mcn_1^{'}(f_{\omega ,e}+v)
+\mcn_1^{'}(f_{\omega ,e})+\mcn_1^{''}(f_{\omega})(v)
$$

Define
\begin{equation}
\overline{\Theta }=\int_{0}^{t}hds\text{,}
\end{equation}
\begin{equation}
f_{\omega, e }^{\ast }=Exp[i\overline{\Theta }]f_{\omega}
\end{equation}
and
\begin{equation}
v^{\ast }=Exp[i\overline{\Theta }]v\text{.}
\end{equation}
Then, as with \eqref{rrr}, and using the fact that
$\|\partial_t v^{\ast}\|_{L^2}=O(e+\widetilde{W}^{\frac{1}{2}})$ by
\eqref{de1},
we have
\begin{multline}
\left\langle \partial_t v+ihv,{\mathcal N}(f_{\omega ,e},f_{\omega},v)\right\rangle_{L^2}
= \\
-\left\langle \partial_tv^{\ast },\mcn_1^{\prime }(f_{\omega, e }^{\ast }+v^{\ast
})-\mcn_1^{\prime }(f_{\omega, e }^{\ast })-\mcn_1^{\prime \prime }(f_{\omega, e }^{\ast
})[v^{\ast }]\right\rangle_{L^2} +O\left(e^{3}\widetilde{W}^{\frac{1}{2}}
+e^{2}\widetilde{W}\right) \text{.}
\end{multline}
But,
\begin{multline}
\left\langle \partial_tv^{\ast },\mcn_1^{\prime }(f_{\omega, e }^{\ast }+v^{\ast
})-\mcn_1^{\prime }(f_{\omega, e }^{\ast })-\mcn_1^{\prime \prime }(f_{\omega, e }^{\ast
})[v^{\ast }]\right\rangle _{L^{2}}= \\
\partial_t\int\left( \mcn_1(f_{\omega, e }^{\ast }+v^{\ast })-\mcn_1(f_{\omega, e }^{\ast
})-\mcn_1^{\prime }(f_{\omega, e }^{\ast })[v^{\ast }]-\frac{1}{2}\mcn_1^{\prime \prime
}(f_{\omega, e }^{\ast })[v^{\ast }]^{2}\right) dx\\
-\left\langle \partial_tf_{\omega, e }^{\ast },\mcn_1^{\prime }(f_{\omega, e }^{\ast
}+v^{\ast })-\mcn_1^{\prime }(f_{\omega, e }^{\ast })-\mcn_1^{\prime \prime }(f_{\omega,e}^{\ast })[v^{\ast }]-\frac{1}{2}\mcn_1^{(3)}(f_{\omega, e }^{\ast })[v^{\ast
}]^{2}\right\rangle _{L^{2}}\text{.}
\end{multline}
Hence,
\begin{multline}
\left\langle \partial_tv^{\ast },\mcn_1^{\prime }(f_{\omega, e }^{\ast }+v^{\ast
})-\mcn_1^{\prime }(f_{\omega, e }^{\ast })-\mcn_1^{\prime \prime }(f_{\omega, e }^{\ast
})[v^{\ast }]\right\rangle _{L^{2}}= \\
\partial_t\int\left( \mcn_1(f_{\omega,e}+v)-\mcn_1(f_{\omega,e})-\mcn_1^{\prime
}(f_{\omega,e})[v]-\frac{1}{2}\mcn_1^{\prime \prime }(f_{\omega
,e})[v]^{2}\right) dx\\ - \left\langle \left( \partial_t+ih\right)
f_{\omega,e},\mcn_1^{\prime }(f_{\omega,e}+v)-\mcn_1^{\prime }(f_{\omega
,e})-\mcn_1^{\prime \prime }(f_{\omega
,e})[v]\right\rangle _{L^{2}} - \\
\left\langle \left( \partial_t+ih\right) f_{\omega
,e},\frac{1}{2}\mcn_1^{(3)}(f_{\omega,e})[v]^{2}\right\rangle _{L^{2}}
\text{.}
\end{multline}
Now,
\begin{equation}
\left\langle ihf_{\omega,e},\mcn_1^{(3)}(f_{\omega,e})[v]^{2}\right\rangle
_{L^{2}}\leq c\int |f_{\omega,e}h|\left( 1+|f_{\omega,e}|^{3}\right)
|v|^{2}dx\text{,}
\la{em1}
\end{equation}
by condition (\ref{Eq:U2prime2}). Additionally,
\ba
&&\left\langle ihf_{\omega,e},\mcn_1^{\prime }(f_{\omega,e}+v)-\mcn_1^{\prime}
(f_{\omega,e})-\mcn_1^{\prime \prime }(f_{\omega,e})[v]\right\rangle _{L^{2}}
\la{em2}\\
&&\qquad\qquad=\int_{0}^{1}(1-s)\left\langle ihf_{\omega,e},\left( \mcn_1^{\prime \prime}
(f_{\omega,e}+sv)-\mcn_1^{\prime \prime }(f_{\omega,e})\right)
[v]\right\rangle _{L^{2}}\text{,}
\notag\\
&&\qquad\qquad\quad\leq c\int |f_{\omega,e}h|\left( 1+|f_{\omega,e}|^{3}\right) \left(
|v|^{2}+|v|^{5}\right) dx\notag\text{,}
\ea
by condition (\ref{Eq:U2prime2}). Therefore, by the exponential decay 
of $f_{\omega,e}$ and the fact that $|f_{\omega,e}h|_{L^p}=O(e)$ by
the bounds of \S\ref{boundin},
$$
\left\langle ihf_{\omega,e},\mcn_1^{\prime }(f_{\omega,e}+v)-\mcn_1^{\prime
}(f_{\omega,e})-\mcn_1^{\prime \prime }(f_{\omega,e})[v]-\frac{1}{2}%
\mcn_1^{(3)}(f_{\omega,e})[v]^{2}\right\rangle _{L^{2}} \\
=O(e\widetilde{W})\text{.}
$$
Integration by parts and lemma \ref{lem:different} imply that
\begin{multline}
\left\langle (\mathbf{u}\cdot\nabla v,{\mathcal N}(f_{\omega ,e},f_{\omega
,0},v)\right\rangle =O(e^{2}\widetilde{W})+ \notag\\
\left\langle \mathbf{u}\cdot\nabla f_{\omega,e},\mcn_1^{\prime }(f_{\omega
,e}+v)-\mcn_1^{\prime }(f_{\omega,e})-\mcn_1^{\prime \prime }(f_{\omega,e})[v]-\frac{%
1}{2}\mcn_1^{(3)}(f_{\omega,e})[v]^{2}\right\rangle \text{.}
\notag\end{multline}
Next notice that the quantity
$$
\left\langle (\partial_t+\mathbf{u}\cdot\nabla) f_{\omega,e},\mcn_1^{\prime }(f_{\omega
,e}+v)-\mcn_1^{\prime }(f_{\omega,e})-\mcn_1^{\prime \prime }(f_{\omega,e})[v]-\frac{%
1}{2}\mcn_1^{(3)}(f_{\omega,e})[v]^{2}\right\rangle 
$$
can be estimated to be $O(e\widetilde{W})$ 
in the same way as the bounds \eqref{em1},\eqref{em2}
once we note that, for every $p\in[1,\infty]$,
\begin{equation}
\left\| \partial_tf_{\omega,e}+\mathbf{u}\cdot\nabla f_{\omega,e}\right\|
_{L^{p }}=\left\|({{{\dot\lambda-V_0(\lambda)}}})\cdot 
\partial_{{\lambda}}f_{\omega,e}\right\|
_{L^{p}}=O(e)\text{,}
\end{equation}
by \eqref{bv}.
The proof is now completed by noticing that Taylor's theorem and
\eqref{Eq:U2prime2} imply that the quantity
$$
\int\left( \mcn_1(f_{\omega, e }^{\ast }+v^{\ast })-\mcn_1(f_{\omega, e }^{\ast
})-\mcn_1^{\prime }(f_{\omega, e }^{\ast })[v^{\ast }]-\frac{1}{2}\mcn_1^{\prime \prime
}(f_{\omega, e }^{\ast })[v^{\ast }]^{2}\right)
dx$$
is
$O(\widetilde{W}^{3/2})
+O(\widetilde{W}^3)=o(\widetilde{W}),
$
since $\widetilde{W}$ is small by assumption.
\myqed

\end{document}